\documentclass[fleqn,12pt,a4paper]{article}
\usepackage{amsthm}
\usepackage{amsmath}
\usepackage{amssymb}
\usepackage{amsopn}
\usepackage{latexsym}
\usepackage{setspace}
\usepackage{dsfont}
\usepackage{datetime}
\usepackage[allnumber,warning,easyold,math]{easyeqn}
\usepackage{xcolor}
\usepackage[12pt]{moresize}
\usepackage{natbib}
\usepackage[bookmarks=true,bookmarksnumbered=true,pdfpagemode=None,
pdfstartview=FitH,hidelinks]{hyperref}
\usepackage{fix-cm}
\usepackage[calcwidth]{titlesec}
\usepackage{fix-cm}
\usepackage{graphicx}
\graphicspath{{./images/}}
\usepackage{sgame}
\usepackage[utf8]{inputenc}
\usepackage[english]{babel}
\usepackage{amssymb}
\usepackage{braket}
\usepackage{upgreek}
\usepackage[font=footnotesize,skip=-2pt]{caption}
\usepackage[font=footnotesize,skip=-2pt]{subcaption}
\usepackage{float}
\usepackage{multirow}
\usepackage[normalem]{ulem}
\usepackage{import}

\theoremstyle{break}

\newtheorem*{thm*}{Theorem}

\newcommand{\Prests}{\raisebox{-0.7pt}{%
\includegraphics[height=0.35cm]{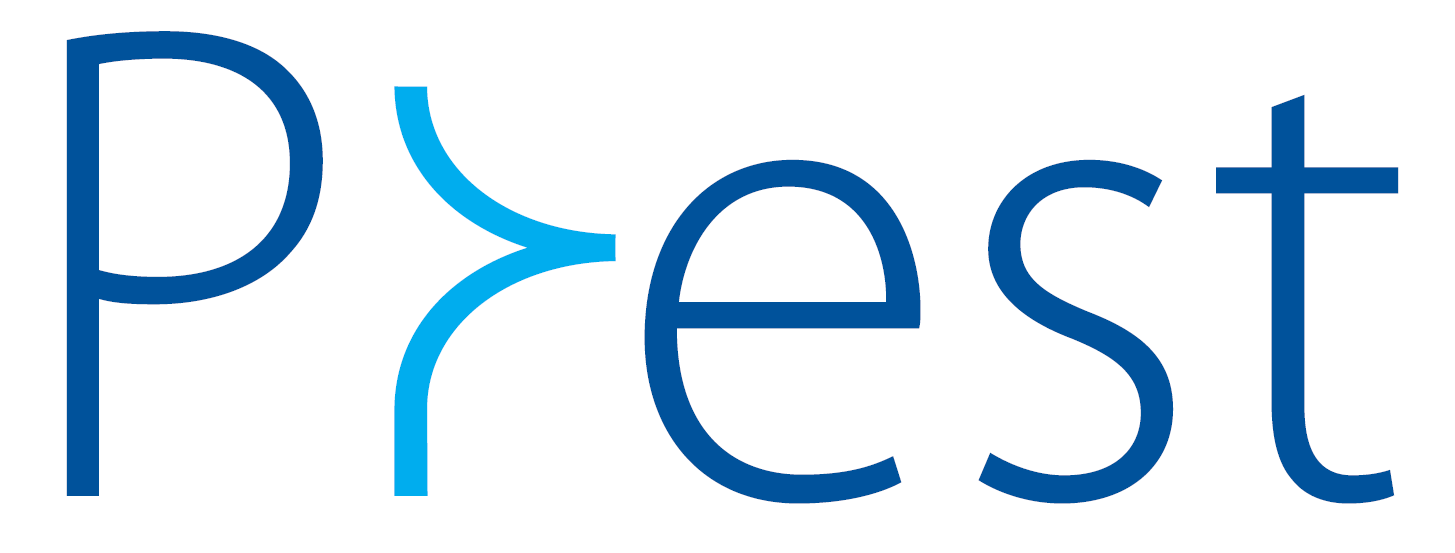}\;}%
}

\newcommand{\PrestsFootnote}{\raisebox{-0.7pt}{%
\includegraphics[height=0.30cm]{prest.png}\;}%
}

\titleformat{\section}[hang]{\rmfamily\bfseries}{\color{black}\fontsize{14}{14}
\selectfont\thesection}{30pt}{\fontsize{14}{14}\selectfont}
\titleformat{\subsection}[hang]{\rmfamily\bfseries}{\color{black}\fontsize{13}{13}
\selectfont\thesubsection}{30pt}{\fontsize{13}{13}\selectfont}
\titleformat{\subsubsection}[hang]{\rmfamily\bfseries}{\color{black}\fontsize{12}{12}
\selectfont\thesubsubsection}{30pt}{\fontsize{12}{12}\selectfont}

\begin{document}

\parskip=5pt

\author{Georgios Gerasimou\thanks{Email: 
\href{mailto:Georgios.Gerasimou@glasgow.ac.uk}{\color{blue}
Georgios.Gerasimou@glasgow.ac.uk}.
\color{black} I cordially thank Carlos Cueva, Miguel Costa-Gomes 
and Matúš Tejiščák for our previous joint work, as well as 
Elias Bouacida for comments on this paper. 
I also thank Nathan Jones, Iravati Kashyap, Kanishk Khanna 
and Jacob Beaddie for excellent research assistance with Python 
and constraint-programming computational tasks. The paper was presented at 
SAET 2021 (online), SPUDM 2021 (online), DETC 2022 (Durham), FUR 2022 (Ghent), 
SAET 2022 (online), EEA-ESEM 2022 (Milan), whose audiences I thank 
for helpful feedback. 
The scores reported in Sections 4 and 5 were computed in \Prests \citep{prest}.
All statistical analyses and figures were produced in R with RStudio 
\citep*{Rcitation,Rstudio}. 
The project was supported by British Academy and University of St Andrews funds
and received approval by the University of St Andrews Research Ethics Committee 
(EC9805/EC16707).
Earlier versions were circulated as 
\textit{``Towards eliciting weak or incomplete preferences in the lab: 
a model-rich approach''} and \textit{``Model-rich approaches to eliciting 
weak preferences: evidence from a multi-valued choice experiment''}. 
Any errors are my own.} 
\vspace{-7pt}
\\  \footnotesize University of Glasgow \vspace{5pt} 
\\ \scriptsize This version: \!\! \today \vspace{-10pt} \\  
\scriptsize \ First version: 
\href{https://arxiv.org/abs/2111.14431v1}{\color{blue}{November 29, 2021}}\vspace{-20pt}}

\title{\vspace{-40pt} \large \textbf{Eliciting and Distinguishing Between Weak 
and Incomplete Preferences: Theory, Experiment and Computation}}

\parindent=0pt

\date{}

\maketitle

\vspace{-5pt}

\begin{abstract}\footnotesize
\noindent 
Recovering and distinguishing between the strict-preference, indifference 
and/or indecisiveness parts of a decision maker's preferences 
is a challenging task but also important for testing theory and conducting 
welfare analysis.
This paper contributes towards this goal by reporting on data 
from a lab experiment on riskless choice that were analyzed with novel 
theory-guided computational methods.
The experiment included both Forced- and Free-Choice 
treatments. Its main novelty consisted of allowing 
subjects to select multiple alternatives at each menu. 
Based on a new non-parametric goodness-of-fit
criterion that we introduce, which generalizes 
a widely used pre-existing method to environments 
of multi-valued choices,
each subject's decisions were tested against 
three structured general choice models that feature maximization 
of stable but potentially weak and/or incomplete preferences. 
Nearly 60\% of all subjects' are well-explained by one of these
models, typically with a unique model-optimal preference relation per subject. 
Importantly, revealed preferences typically have a non-trivial indifference part 
that, on average, accounts for up to 19\% of all possible comparisons. 
In addition, 22\% of all subjects are best explained by models of incomplete-preference 
maximization and reveal preferences that typically exhibit
the distinctions between indifference and indecisiveness that these models afford or predict.
These distinctions are documented empirically for the first time.\\

\end{abstract}

\setcounter{page}{0}

\thispagestyle{empty}

\vfill

\pagebreak

\parindent=15pt

\renewcommand*{\thepage}{\footnotesize{\arabic{page}}}

\section{Introduction}

\subsection{Motivation}

The state in which a decision maker is indifferent between two or more 
choice alternatives in the sense of considering them to be equivalent 
plays a prominent role in many domains of economic analysis. 
As such, it is important to understand what kinds of observable decision 
environments and data-analytic methods can in principle allow for 
extracting an individual's potentially weak preferences and distinguishing 
between their strict-preference and indifference parts. 
This is particularly pertinent if one also accepts that the individual 
in question may be indecisive, with preferences that are transitive 
but potentially incomplete. In this situation, 
it is not obvious how such observable 
decision data could be used to separate indifference 
from incomparability/indecisiveness. 

These problems are important, both from a 
theoretical and a policy perspective. 
First, since many economic models of individual, 
collective or interactive decisions 
assume that agents have preferences with non-trivial indifference parts, 
elicitation of the agents' weak preference relations 
would allow for testing 
these models' descriptive relevance more accurately 
than if indifferences were assumed away. 
Additionally, knowing, for example, how many decision makers within 
a community consider e.g. a tree-planting 
program to be equally good to the development of a playground 
and how many of them have a strict preference instead would generally 
allow the community leader to 
arrive at a better decision than if all preferences were mistakenly 
interpreted to be strict. 
This example is made more concrete with the possibilities laid out
in the table below,  
which shows how eliciting single choices 
in the presence of indifferences may lead to a Pareto inefficient 
social outcome while eliciting 
multiple choices would not:

\begin{table}[!htbp]
\scriptsize
\centering
\setlength{\tabcolsep}{5pt} 
\renewcommand{\arraystretch}{1.3} 
\begin{tabular}{|l|c|c|c|c|}
\hline
& \multirow{2}{*}{\textbf{Anna}}  
& \multirow{2}{*}{\textbf{Basel}} 
& \multirow{2}{*}{\textbf{Cora}} 
& \multirow{1.5}{*}{\textbf{Majority-rule}}\\
&&&& \multirow{1}{*}{\textbf{outcome}} \\
\hline
\multirow{3}{*}{\textbf{True preferences}}	
& $Tree Planting$ 
& $TreePlanting$  
& $TreePlanting$ 
& \multirow{3}{*}{} \\
& $\succ_A$  
& $\sim_B$  
& $\sim_C$  
&  $TreePlanting$\\
& $PlayGround$ 
& $PlayGround$ 
& $PlayGround$ 
& \\
\hline
\multirow{1.5}{*}{\textbf{Possible revealed preferences}} 
& \multirow{2.2}{*}{$TreePlanting$} 
& \multirow{2.2}{*}{$PlayGround$} 
& \multirow{2.2}{*}{$PlayGround$} 
& \multirow{2.2}{*}{$PlayGround$} \\
\multirow{1}{*}{\textbf{when only single choices allowed}}
&&&&\\
\hline
\multirow{2.5}{*}{\textbf{Revealed preferences when}}	
& \multirow{3}{*}{$TreePlanting$} 
& \multirow{1.5}{*}{$TreePlanting$} 
& \multirow{1.5}{*}{$TreePlanting$} 
& \multirow{3}{*}{$TreePlanting$}\\
\multirow{2}{*}{\textbf{multiple choices allowed}}
&&&&\\
&
&$PlayGround$ 
&$PlayGround$ 
&\\
\hline
\end{tabular}
\end{table}

\noindent Similarly, understanding when agents are indifferent 
and when they are indecisive can enable the 
policy maker to understand whether they should immediately make an 
active choice that will impact everyone 
in the group or delay such a choice and aim instead to 
inform the group's members towards 
resolving their indecisiveness, 
e.g. via targeted information campaigns.

The relevance of indecisiveness as a concept distinct from indifference 
also manifests itself in large-stake real-world domains such as 
high-court judicial decision making. 
Owing to the fact that justices in such courts often 
have the luxury to choose which cases to hear --and make a decision on-- 
and which to ignore, it is conceivable that those binary 
hear/no-hear choices are not random but, instead, driven by some pattern. 
As a case in point, we cite the work of \cite*{hitt2019} 
who notes in relation to the US Supreme Court (pp. 6-7) that  
\textit{``the Court could protect the logical consistency and quality 
of its opinions by ignoring complex and multifaceted cases. [...] 
Essentially, [the Supreme Court Case Selections Act of 1988] gave 
the justices almost total freedom to opt \textnormal{not} to decide most disputes. 
As such, whether consciously or not, the modern Supreme Court actively 
evolved away from decisiveness''.} 
Hitt further argued that such potentially very impactful decisions may have 
traceable and behaviourally intuitive underlying patterns, noting that 
\textit{``prioritizing consistency means that the Court will leave numerous 
important questions and conflicts unresolved''} because \textit{``a desire 
to produce `good' (consistent) law may induce 
the justices to prioritize consistency over decisiveness.''}
Such prioritization of consistency over decisiveness might be viewed 
by the economic analyst as defining a particular kind of partial 
preference relation over judicial cases whereby one is preferred to another 
if and only if the former is less complex/multifaceted 
\textit{and} more likely to result in the production of ``good''/consistent 
law if it was taken up by the judge. 
When confronted with two cases that generate trade-offs across these 
two dimensions, such a judge would be indecisive between them, with this 
mental state potentially manifesting itself in delayed hear/no-hear decision. 
In contrast, if the judge was instead indifferent between such trade-off 
generating cases, then one would expect both types of cases to be heard 
in more or less equal frequency, contrary to the legal scholarly 
arguments cited above.

The relevance of understanding indecisiveness and, where necessary, 
distinguishing it from indifference/equivalence, also manifests itself in the 
the rapidly evolving landscape of Artificial-Intelligence (AI) 
Large Language Models (LLMs) that are currently in daily use 
by a sizeable and growing fraction of the global population. 
Specifically, in response to the 
persistent ``hallucinations'' problem\footnote{This term refers to LLMs' 
occasional answers to user questions that are confident- 
and authoritative-sounding but contain misleading, false or 
completely invented information.} facing all LLMs that have become 
mainstream in the past few years, and which erodes public trust in these models, 
a segment of the machine-learning literature seeks to 
find ways in which this problem can be mitigated. 
Notably, in the very recent study that is co-authored
by research scientists of a leading LLM provider, \cite*{hallucinations25}
attribute the problem to the models' incentives in their training 
and, especially, evaluation stages, which 
\textit{``reward guessing over acknowledging uncertainty''}.
Motivated by prior literature and pre-existing test-grading systems 
that penalize incorrect answers to dis-incentivize guessing but 
do not penalize \textit{``I don't know''}/abstaining answers,  
and where the relevant rules are known to test-takers in advance,  
the authors propose that LLM evaluation instructions at 
the models' post-training stage explicitly state confidence 
targets within user prompts. For a confidence parameter 
$t$ between 0 and 1, for example, they suggest  
including the following at the end of every prompt:
\textit{``Answer only if you are $>t$ confident, since mistakes 
are penalized $\dfrac{t}{1-t}$ points, while correct answers receive 
1 point, and an answer of \textnormal{``I don't know''} receives 
0 points.''} The potential enforcement of such a rule with a relatively 
high value of $t$ in an LLM's post-training evaluation stage would clearly 
aim to prioritize accuracy and consistency over decisiveness, in partial
analogy to Hitt's (\citeyear{hitt2019}) analysis of decision-making in the 
US Supreme Court.

\subsection{Contribution}

This paper extends and combines 
recent developments in choice theory, computational revealed preference 
analysis and experimental economics in an attempt to elicit human decision 
makers' potentially weak and/or incomplete preferences and, where relevant, to 
separate their revealed strict preference, indifference and 
incomparability/indecisiveness components. 
The paper contributes in this direction with a combination of 
methodological and empirical innovations.
First, it proposes and reports on an implementation of 
a lab experimental design which elicits observable behavioural data aiming
at recovering an individual's possibly 
weak preferences. 
The main feature of the design is that it allows subjects in both 
its Forced- and Free-Choice treatments  
to make multiple choices from every menu of alternatives 
they see by introducing incentives that 
help them make such possibly multi-valued choices in a 
preference-guided way. Subjects in the Free-Choice treatment 
can additionally avoid/delay 
making a choice from any menu at a 
small cost, without receiving any additional information 
about the alternatives. Subjects in 
the otherwise standard Forced-Choice treatment on the other 
hand are always required to choose at least one option. 
This experimental design was implemented in a lab environment 
of riskless choice from 50 distinct menus 
that were derived from 6 pairs of popular \pounds 10 gift cards, 
and data from 273 subjects are reported on. 
The main novelty of this design relative 
to existing ones is that it enables 
the elicitation of multiple choices per subject in both 
its treatments. Evidently, this is 
crucial for testing theories of choice that 
predict multi- and/or empty-valued choice
correspondences as the outcome of preference maximization.
While the design's incentives are more closely aligned 
with testing the benchmark model of 
Rational Choice/Utility Maximization --importantly, 
allowing for possible indifferences-- in this riskless 
environment, we also consider two distinct models of 
\textit{incomplete} preference maximization to describe 
subjects whose behaviour is incompatible with that benchmark.

Indeed, the paper's second methodological contribution consists 
of analysing these data by applying a novel computational method 
that is based on combinatorial optimization and allows for a model-rich approach 
towards recovering an individual's potentially weak and/or incomplete preferences
that pays particular attention to the fact that choices are possibly multi-valued. 
This method builds on and extends in the direction of multi-valued choice 
--and in a model-based way-- the classic \cite*{houtman&maks} technique that 
is routinely used in empirical revealed preference tests of the 
rational choice/utility maximization hypothesis. 
More specifically, the original Houtman-Maks technique 
computes the maximal subset of a subject's dataset that is consistent 
with rational choice, essentially allowing for an intuitive quantification 
of the subject's behavioural proximity with that model. 
The extension that we use, which we refer to as the 
\textit{Jaccard-}Houtman-Maks method, accounts 
for the general multi-valuedness of choice data in these computations
and applies this approximation principle 
\textit{simultaneously} to the model 
of (indifference-permitting) rational choice/utility maximization 
and two additional models of 
\textit{incomplete-}preference maximization: 
(i) \textit{undominated} choice, whereby an individual chooses the 
feasible alternative(s) that are not worse than anything else; 
(ii) \textit{dominant} choice, whereby they choose the most 
preferred feasible alternative(s) 
if and only if those exist and defer otherwise.

Using the data collected from an experiment that was designed in the 
way described previously and analyzing them with the Jaccard-modified 
Houtman-Maks method outlined a few lines above, 
the paper makes several novel empirical contributions. 
First, despite the relatively large number of experimental 
decisions, 56\% of all subjects are well-approximated 
(specifically, they are no more than 10\% away from being perfectly 
explained) by one of the above three models of preference maximization. 
Furthermore, and perhaps surprisingly, 
the model-optimally recovered preferences of 77\% of all 
subjects in this group feature at least one indifference 
comparison between distinct alternatives and, on average, almost 2.5
such comparisons, accounting for 15\% of all possible binary 
comparisons between distinct alternatives.
The prevalence of indifference suggested by these finding is 
important and merits further investigation in future studies. 
As far as the decomposition of preference-maximization types is concerned, 
34\% and 22\% of all subjects are best matched by utility 
maximization and the two models of incomplete-preference maximization, 
respectively. 
In addition, a \textit{single} optimal preference relation 
is recovered from 86\% of all these subjects, highlighting the 
strengths of this method in sharp preference identification 
when applied on relatively rich --if incomplete-- data sets 
and in relation to models that are in principle uniquely identifiable.

Importantly, the preferences elicited from nearly two thirds of all 
incomplete-preference maximizers document empirically --and for the first time-- 
the distinct theoretical separations between 
\textit{revealed indifference and revealed indecisiveness} that are afforded 
by the two relevant models. Both these separations assume the analyst 
has access to the kind of multi-valued choice 
data that we elicit in this paper. Recalling also the motivational remarks 
earlier in this Introduction, we view this documentation as 
a significant empirical contribution of the paper because it demonstrates 
the relevance of theories of incomplete weak preferences,  
confirms that their predictions can indeed be tested by observable choices,
and shows how this can be done.
Together, the above findings highlight the potential usefulness 
of multi-valued choice data, 
indifference-permitting deterministic models of complete 
and incomplete preferences, and combinatorial-optimization 
goodness-of-fit methods of testing them.

The paper's next empirical contribution, this time from analysing data 
at the aggregate level, is the documentation of a significant, 
and nearly identical, negative correlation between subjects' 
choice consistency and their average response times. 
Although seemingly counter-intuitive, this interesting 
finding is broadly in line with a key prediction of the influential 
\textit{drift-diffusion} neuro-economic model in binary forced-choice tasks 
(see Ratcliff and McCoon, \citeyear{ratcliff&mckoon08}, 
Al\'{o}s-Ferrer et al., \citeyear{alos-ferrer-fehr-netzer} and references
therein). 
The prediction might be interpreted as suggesting that 
shorter response times are more likely when the decision maker has a 
clear preference between the feasible options. 
Such clarity in turn would be expected to translate into more consistent 
active choices, which is indeed what we find. 
In addition, and despite the fact that subjects in our experiment were allowed to 
--and typically did-- make multi-valued 
instead of single-valued choices, those in the Free-Choice 
treatment behaved significantly more consistently. 
This provides a positive robustness check of the main finding 
in the study by \citet*{CCGT22}, which, unlike the present study, 
was predominantly designed to test for such potential differences 
in consistency, and did so with \textit{single-valued} choice data.
Finally, the present study exploits the richness and generality of the collected 
experimental data by also introducing and discussing the behaviour 
of new variables/statistics that are specific to multi-valued choice 
data, such as \textit{choice sizes} and \textit{choice proportions}, 
and analyses them relative to other relevant observables such as 
response times and choice-consistency indicators.

\subsection{Related Literature}

The core of this paper's experimental design extends 
in the multi-valued choice direction the design of 
\citet*{CCGT22}. That study also reported on 
forced- and non-forced choice treatments but only elicited 
single-valued choices. This made it impossible 
to raise and answer the novel questions that this paper 
is mainly concerned with. 
Although a pre-publication version of that paper from \citeyear{CCGT16}
had proposed a method of multi-valued choice construction, 
that method is considerably more restrictive than the one advanced 
here. In particular, contrary to this paper's unrestricted 
elicitation of subjects' multi-valued choices, that method 
used survey data at \textit{binary} menus in 
order to augment with (constructed) transitive indifference 
relations the subjects' single-valued choices 
\textit{at binary as well as non-binary} menus. 
In work that followed the above-mentioned study, but which predates 
the present one, \cite*{bouacida21} proposed 
a different design that elicits multi-valued choices directly, 
by paying subjects a fixed extra amount for every additional 
alternative they choose from a menu of real-effort tasks, 
assigning them one of these tasks at random as a reward.

The design proposed herein differs from both these pre-existing ones and, 
in our view, materially so. Specifically, like \cite*{bouacida21} 
but unlike \cite*{CCGT16}, we allow subjects to choose multiple 
alternatives at every menu directly, 
thereby avoiding the imposition of a priori consistency constraints 
on the collected multi-valued choice data. 
Yet, like \cite*{CCGT16} and unlike \cite*{bouacida21}, 
we penalize inconsistent choices towards incentivizing 
subjects to reveal their true preferences. 
Furthermore, \cite*{CCGT16} implemented their design on 
all menus derived from a set of 5 headsets 
and with a 1-out-of-4 subject reward frequency\footnote{A more complete list
of the main differences between this experiment 
and those of that study is as follows:
(i) (it) allows for potentially multi-valued choices 
vs requires necessarily 
single-valued ones; 
(ii) features 6 vs 5 choice alternatives (gift-card bundles 
vs headphones); 
(iii) presents 50 decisions vs 26; 
(iv) includes 273 subjects in the sample vs 161 \& 121;
(v) gives choice rewards to every subject vs 1 out of 4;
(vi) makes no provision of any information about the options 
in the randomly selected menu before the 2nd decision
vs provision of such information in the 1st experiment of the earlier study.}, 
while \cite*{bouacida21} implemented his on all menus 
derived from a set of 4 real-effort tasks where all subjects were rewarded. 
In addition, there are significant differences in the choice alternatives, 
number of choice problems, subjects' reward frequency, and most importantly,
in the data-analytic methods, research questions and conclusions of these papers.

Turning to the pre-existing literature on the 
elicitation of incomplete preferences, 
this has mainly revolved around choice over 
binary menus of lotteries and/or uncertain acts, 
and is reviewed in detail in \cite*{CCGT22}. More summarily here: 
to identify incompleteness in such environments those studies 
have used choice-deferral alongside preference-for-flexibility models 
\citep*{danan&ziegelmeyer06}; 
partially incentivized methods of imprecise-preference revelation from menu 
lists (see \cite*{cubitt&navarro-martinez&starmer15} and references therein); 
and incentivized methods where incompleteness 
can potentially be revealed via certain patterns of randomized choice 
\citep*{cettolin&riedl,agranov-ortoleva20}. 
In general choice environments over durable goods on the other hand, 
\textit{strict} incomplete preferences were documented 
in the non-forced-choice treatment of \cite*{CCGT22} via costly choice deferrals
and restricted application of the model-based Houtman-Maks method 
to the dominant-choice model with \textit{strict} incomplete preferences. 
By allowing for multi-valued choices in both Forced- and 
Free-Choice treatments, and by proposing and building the analysis upon 
the novel Jaccard-similarity modified variation of the Houtman-Maks method 
for multi-valued choice data, 
the design proposed in the present paper significantly 
improves upon \cite{CCGT22} by enabling the analyst to test 
the hypothesis of of incomplete-preference 
maximization in the most general 
choice environments possible. Applied in these data, 
and based on the distinct choice-theoretic separations of these 
relations that were proposed in \cite{ok_eliaz_2006} 
and \cite{gerasimou18}, the method allows to document 
empirically --and for the first time--
the uncoupling of revealed indifference and 
incomparability/incompleteness that is predicted by these models.

\section{Theoretical Background}

\subsection{Three Deterministic Models of Preference-Maximizing Choice}

As was mentioned in the Introduction, 
in the main part of our individual-level analysis 
we consider three simple general-choice models 
of deterministic preference maximization 
that impose a rich structure on observable behaviour:
\begin{enumerate}
\itemsep0em 
\item[I.] Rational Choice/Utility Maximization.
\item[II.] Undominated Choice with Incomplete Preferences.
\item[III.] Dominant Choice with Incomplete Preferences.
\end{enumerate}
We focus on these models for several reasons:
\begin{enumerate}
\item All three feature stable preferences and predict both single-valued and 
multi-valued choices under different preference orderings, 
with multi-valuedness potentially interpretable as revealing indifferences 
and single-valuedness as revealing strict preferences. 
\item They impose strong behavioural restrictions. 
In particular, all three models satisfy the fundamental 
\textit{Property $\alpha$} or \textit{Contraction Consistency} 
\citep*{sen71,sen97} principle, which requires an alternative 
to be chosen at a menu whenever it is feasible at that menu 
and also chosen at a larger menu. 
In addition, the first and third models predict active choices that 
are actually consistent with the Congruence/Strong Axiom 
of Revealed Preference principle, which rules out all 
forms of choice cycles. 
\item They are defined in terms of 
(and hence in principle allow the analyst to recover) 
a single preference relation, thereby making the welfare-relevant parts 
of the analysis unambiguous. 
\item All three models are uniquely identifiable. 
That is, if a decision maker's observable 
behaviour is perfectly compatible with one of these models and the 
available data are sufficiently rich (as is the case 
in our experiment), then there is a unique preference relation with 
which that model explains the individual's behaviour. 
\item They are sufficiently computationally tractable to allow for 
the behaviourally intuitive optimization-based 
goodness-of-fit test that we describe below. 
\item In light also of the preceding discussion about the structure 
and possible interpretation of the experimental design, 
the three models predict the kinds of active-choice and deferring 
behaviour that one might expect to observe in our data.
\end{enumerate}

To state the models formally we first define a decision maker's choice dataset 
$\mathcal{D}=\big(A_i,C(A_i)\bigr)_{i=1}^k$ 
on a finite grand choice set $X$ to be a collection of pairs that comprise 
a non-empty menu $A_i\subseteq X$ and 
a -possibly empty- set of alternatives that were chosen at this menu when the 
decision maker was presented with it. 
Thus, $\emptyset\subseteq C(A_i)\subseteq A_i$ holds for all $i\leq k$. 
Dataset $\mathcal{D}$ is explainable by 
\textit{rational choice/utility maximization} if there exists a complete 
and transitive preference relation 
$\succsim$ on $X$ such that, for all $i\leq k$,
\begin{equation}
\label{UM} C(A_i) = \{x\in A_i: x\succsim y \text{ for all } y\in A_i\}.
\end{equation}
If, instead, \eqref{UM} is true for all $i\leq k$ with respect to a 
reflexive and transitive but incomplete preference relation, 
then $\mathcal{D}$ is explainable by the model of 
\textit{dominant choice with incomplete preferences}. In that case we have
\begin{eqnarray*}
\label{mdc-1}	
C(A)\neq\emptyset 	
& \Longleftrightarrow 
& \text{ there is } x\in A \text{ such that } 
x\succsim y \text{ for all } y\in A,\\
\label{mdc-2}	
C(A) = \emptyset 	
& \Longleftrightarrow 
& \text{ for all } x\in A \text{ there is } 
y\in A \text{ such that } x\not\succsim y.
\end{eqnarray*}
Finally, $\mathcal{D}$ is explainable by the model of 
\textit{undominated choice with incomplete preferences} 
if there is a reflexive, transitive and incomplete preference relation 
$\succsim$ whose asymmetric part is $\succ$, such that, for all $i\leq k$,
\begin{equation}
\label{UC} 
C(A_i) = \{x\in A_i: y\not\succ x \text{ for all } y\in A_i\}.
\end{equation}

The model of rational choice/utility maximization was characterized by 
\cite*{richter66} in a general environment that encompasses the one 
within which we are operating here. 
The model of undominated choice with incomplete preferences was, 
to the best of our knowledge,  introduced by \cite*{schmeidler69} 
in a general equilibrium setting 
and was analyzed choice-theoretically under a variety of 
decision environments and preference structures by, most notably, 
\cite*{schwartz76}, \cite*{bossert&etal05}, \cite*{ok_eliaz_2006}, 
\cite*{bossert-suzumura10} and \cite*{stoye15}.\footnote{ 
This model is implicitly also used in stochastic-dominance applications 
of portfolio efficiency, along the lines studied in \cite*{levy16}, 
\cite*{linton-post-whang14} and \cite*{arvanitis-scaillet-topaloglou23}, 
for example. \cite{bouacida21} tests the no-indifference version of this 
model with the data collected under that study's experimental design.} 
Dominant choice with incomplete preferences was studied theoretically 
in \cite*{gerasimou18}\footnote{A related pre-dating study is
\cite*{dean08wp}, which proposed a class of 
\textit{decision-avoidance} models that 
focused on explaining the increasing prevalence of \textit{status quo bias} 
--a phenomenon that is distinct from \textit{choice deferral}--
as menu size increases. When the decision problem contains no natural status quo 
and no dominant alternative, these models predict a compensatory decision process 
via which active choices (not fully consistent in general) are always made. 
Dominant choice with incomplete preferences 
on the other hand focuses on decision problems
without a natural status quo option and features a non-compensatory decision 
process whereby the agent makes (fully consistent) active choices when 
and only when a most preferred alternative exists.} 
and empirically in \cite*{CCGT22}, 
with the latter analysis limited to an environment of single-valued 
non-forced choice experimental data. Those data, in particular, 
do not allow for testing the model 
of undominated choice or distinguishing between indifference and indecisiveness 
in either of the two models of choice with incomplete preferences, which is 
a primary goal of this paper.

The two models of incomplete-preference maximization are logically distinct. 
Moreover, if we replace the term ``incomplete'' with ``possibly incomplete'' 
in their respective statements, then these models generalize rational choice 
in different ways. 
The first does so by relaxing active-choice 
consistency while retaining the decisiveness 
(non-emptiness) assumption that requires $C(A_i)\neq\emptyset$ for all $i\leq k$. 
The second model does so by relaxing the decisiveness assumption while 
retaining active-choice consistency.

\subsection{Distinguishing Between Indifference and Incomparability/Indecisiveness}

For a decision maker with incomplete preferences who is also indifferent 
between some alternatives, a non-trivial question that 
emerges naturally is how one might use observable behavioural 
data in conjunction with some model in order to separate those 
pairs of alternatives between which the agent is indifferent 
from those where the agent is indecisive/unable to compare. 
\cite*{ok_eliaz_2006} were the first to raise and provide an 
answer to this question. Taking the model of undominated 
choice as their primitive, the authors focused on and characterized 
the special case where the model's rationalizing 
incomplete preference relation $\succsim$ is ``regular'' 
in the sense that whenever $x\not\succsim y$ and $y\not\succsim x$ 
are both true, then there is $z\in X$ such that either 
$x\not\succsim z$, $z\not\succsim x$ and $y\succ z$ or $z\succ y$, 
or $y\not\succsim z$, $z\not\succsim y$ and $x\succ z$ or $z\succ x$ 
(Table \ref{enumeration} shows how the number 
of indifference-permitting regular incomplete preorders varies when 
$|X|\in \{3,4,5,6,7\}$). The authors' proposed 
distinction can then be summarized as follows:

\vspace{10pt}

\begin{minipage}{50em}
\noindent An agent whose incomplete preferences 
are captured by a regular preorder and\\ 
who maximizes these preferences according to the undominated-choice model\\ 
is revealed to be:

\vspace{4pt}

\noindent \textit{indifferent} between $x$ and $y$ \textit{\textbf{only if}} 
$\big[$$x,y\in A$, $y\in C(A)$$\bigr]$ $\Rightarrow$ $x\in C(A)$;

\vspace{4pt}

\noindent \textit{indecisive} between $x$ and $y$ \textit{\textbf{only if}} 
$x\in C(A)$, $y\in A\setminus C(A)$, $y\in C(B)$, $x\in B$\\ 
\hspace*{188pt}{for distinct menus $A$ and $B$.}
\end{minipage}

\vspace{10pt}

\begin{table}[!htbp]
\centering
\footnotesize
\caption{Enumeration of regular incomplete preorders that 
allow for non-trivial indifferences.}
\setlength{\tabcolsep}{4pt} 
\renewcommand{\arraystretch}{1.3} 
\makebox[\textwidth][c]{
\begin{tabular}{|c|c|c|c|}
\hline
& \textbf{All} 	&  \textbf{Regular}  	&  \\
$|X|$ & \textbf{indifference-permitting} & \textbf{indifference-permitting}&\textbf{\%}\\
& \textbf{incomplete preorders}		&  \textbf{incomplete preorders} 	&  \\
\hline
3 	& 3 			& 0 				& 0.00\% \\
\hline
4  	& 85 			&  54 				& 63.53\% \\
\hline
5  	& 2,290 		&  1,705 			& 74.45\% \\
\hline
\textbf{6}  & \textbf{75,541}&  \textbf{60,455} & \textbf{80.03\%} \\
\hline
7  & 3,363,129 		& 2,799,615 		& 83.24\%  \\
\hline
\end{tabular}
}
\caption*{\centering \scriptsize Source: Output from constraint-satisfaction 
problems written in the Essence$^\prime$ 
language and solved by the MINION solver \citep*{MINION} using the 
Savile Row modelling assistant (\url{https://savilerow.cs.st-andrews.ac.uk/}).}
\label{enumeration}
\end{table}

\noindent In words, the agent is indifferent only 
if the two options are either chosen 
or rejected together when both are feasible, 
while they are indecisive only if one is 
chosen over the other in some menu and the 
latter is chosen in the presence of the 
former in another menu. Importantly, the revealed indifference relation 
here is transitive, whereas the revealed indecisiveness one is not 
(see also Mandler, \citeyear{mandler09}). 
Also importantly, although this choice-reversal-based 
distinction between the two notions is intuitive, it is not robust. 
Indeed, any behaviour that is 
compatible with such an indifference-permitting instance of that model 
is observationally equivalent 
to the same instance of the model where the decision maker is 
simply indecisive between any 
two alternatives that are not ranked by strict preference 
(see also Theorem 1 in Bossert et al., \citeyear{bossert&etal05} and 
Theorem 3.3 in Bossert and Suzumura, \citeyear{bossert-suzumura10}).

More recently, \cite*{gerasimou18} noted a distinct and robust behavioural 
separation between indifference and indecisiveness 
that is afforded by the dominant-choice model. 
This can be summarized as follows: 

\vspace{10pt}

\begin{minipage}{50em}
\noindent A decision maker whose incomplete 
preferences are captured by a preorder and\\ 
who maximizes these preferences according to the dominant-choice model\\ 
is revealed to be:

\vspace{4pt}

\noindent \textit{indifferent} between $x$ and $y$ 
\textit{\textbf{if and only if}} 
$\big[$$x,y\in A$, $y\in C(A)$$\bigr]$ $\Rightarrow$ $x\in C(A)$;

\vspace{4pt}

\noindent \textit{indecisive} between $x$ and $y$ 
\textit{\textbf{if and only if}} 
$x,y\in A$ $\Rightarrow$ $x,y\not\in C(A)$.	
\end{minipage}

\vspace{10pt}

\noindent In words, the agent is indifferent iff both options 
are either chosen or rejected together when both are feasible, 
and they are indecisive between these options iff neither is 
ever chosen in the presence of the other. This distinction is 
obviously robust and not subject to alternative interpretations 
because reducing a relation's indifferences to incomparabilities 
in this case generates different active-choice 
and deferring behaviour. 

\subsection{Measuring a Choice Correspondence's Proximity to A Model: \newline
The \textit{Jaccard}-Houtman-Maks Index}

Our non-parametric model-fitting analysis 
extends the classic Houtman-Maks (HM) \citep{houtman&maks} method 
that was outlined in the introduction by accounting for: 
(i) multi-valued (as well as empty-valued) choice data, and (ii) 
the possibility that the decision maker generating such data 
behaves under a model that is more general than utility maximization.
Indeed, by finding which model is closest to explaining each subject's behaviour, 
this method effectively allows for recovering --possibly approximately, 
and with the standard \textit{as if} qualifications in place-- 
both the individual's deterministic \textit{decision rule} 
and their \textit{preferences conditional on that rule}.
This explains the ``model-based'' part of the claimed extension.
In order to also account for the general multi-valuedness 
of subjects' choices and of the model-predicted optimal choices, 
our proposed extension of the HM method computes the (dis-)similarity 
between these sets --which intuitively captures the proximity of observed 
and model-predicted choices-- by means of 
the classic \textit{Jaccard (dis-)similarity metric} between finite sets 
of discrete objects 
\citep{jaccard,marczewski-steinhaus,levandowsky-winter}. 
This, in particular, allows one to have a more nuanced understanding 
of when a deviation of a subject's set of chosen 
alternatives from the model-predicted ones is a ``severe'' mistake or 
a relatively ``mild'' one, contrary to the binary ``pass/fail'' approach 
of the straightforward extension of the HM method in this environment. 
Hence, we will refer to this as the \textit{Jaccard-}Houtman-Maks (JHM) method.

More formally, recall that a dataset $\mathcal{D}$ consists of a finite collection  
of observations $\big(A,C(A)\bigr)$, where 
$\emptyset\neq A\subseteq X$ and $\emptyset\subseteq C(A)\subseteq A$.
Denote by $\mathbf{D}$ the collection of all such datasets. 
In the important special case where $|C(A)|=1$ is true 
for every pair $\big(A,C(A)\bigr)$ in $\mathcal{D}\in\mathbf{D}$ 
one can measure the proximity of $\mathcal{D}$ to the model of 
utility maximization with strict preferences with 
the classic HM index of rationality due to \cite{houtman&maks}\footnote{An 
axiomatization of this index in a rich-dataset environment is given in  
\cite{apesteguia&ballester15}.}, 
defined by the function $\mathtt{HM}:\mathbf{D}\rightarrow\mathbb{N}_0$
such that
\begin{eqnarray}
\label{hm} \mathtt{HM}(\mathcal{D}) & := & 
\min_{\succ \in \mathcal{P}} 
\left|\big(A,C(A)\bigr)\in\mathcal{D}:C(A)\neq C_{\succ}(A)\right|,
\end{eqnarray}
where $\mathcal{P}$ is the set of strict linear orders on $X$
and $C_{\succ}(A)$ the unique $\succ$-maximizer at menu $A$. 
In words, $\mathtt{HM}$ associates with each dataset a natural 
number that captures the smallest number of choices that need 
to be dropped or modified in that dataset in order for it to be
rationalizable by utility maximization under some strict preference
relation. 

Now suppose $|C(A)|\geq 1$ for every $\big(A,C(A)\bigr)$ 
in $\mathcal{D}$. Replacing in \eqref{hm} the set 
$\mathcal{P}$ with the collection of 
\textit{complete} preorders on $X$ --denoted by $\mathcal{R}$-- and $C_{\succ}(A)$ 
with $C_{\succsim}(A)$ for $\succsim\in \mathcal{R}$ provides a direct 
translation of the HM rationality index with strict preferences 
to its weak-preference counterpart. 
There is, however, a conceptual subtlety that is lost in this translation: 
while there is a unique way in which a single-valued choice
$C(A)$ could differ from the $\succ$-optimal choice $C_{\succ}(A)$ (i.e. 
either the two coincide or they do not), there are 
many ways in which a \textit{multi-valued} choice $C(A)$ can differ from 
the $\succsim$-optimal choice $C_{\succsim}(A)$. 
We illustrate this with the following example that 
presents three hypothetical subjects' choices at the same menu:

\begin{tabular}{llllllll}
$\succsim$ : & \multicolumn{7}{c}{$a \sim b \succ c \succ d$} \\
$A$ & = & \multicolumn{6}{c}{$\{a, b, c \}$}  \\
$C_\succsim(A)$ & = & \multicolumn{6}{c}{$\{a,b\}$} \\
$C_1(A)$ & = & \multicolumn{6}{c}{$\{a\}$}  \\
$C_2(A)$ & = & \multicolumn{6}{c}{$\{c,d\}$} \\
$C_3(A)$ & = & \multicolumn{6}{c}{$\{a,b\}$} \\
\end{tabular}

\noindent Here, the $\succsim$-optimal choices at $A$ are in the set 
$\{a,b\}$ and coincide with the choices made by subject 3. 
Those made by subjects 1 and 2 on the other hand both deviate
from the optimal choice. However, there is a natural sense in which the 
former subject's choice is ``closer'' to being optimal than that of subject 2:
it includes one of the two $\succsim$-best alternatives at the menu, 
whereas the choice of subject 2 contains neither. 
In the language of mistakes, one might label the first subject's 
as a ``partially'' while that of the second subject as a ``fully'' 
mistaken decision.  
The straightforward extension of the HM index presented above 
does not account for this nuance and 
``penalizes'' both $C_1(A)$ and $C_2(A)$ in the same way. 

For this reason, instead of generalizing \eqref{hm} to the multi-valued 
choice case in what appears to be a counter-intuitively punitive way, 
we propose a generalization that explicitly 
accounts for the \textit{content-similarity} of sets $C(A)$ and 
$C_{\succsim}(A)$ across all pairs $\big(A,C(A)\bigr)\in\mathcal{D}$
and every weak preference relation $\succsim\in\mathcal{R}$. 
More specifically, to assess how dissimilar is the observed choice $C(A)$ 
to the choice $C_\succsim(A)$ that is optimal under some $\succsim\in\mathcal{R}$
we use the classic \textit{Jaccard} dissimilarity metric, 
whereby 
\begin{eqnarray}
\nonumber J\big(C(A),C_\succsim(A)\bigr) & := & 
\dfrac{|C(A)\cup C_\succsim(A)|-|C(A)\cap 
C_\succsim(A)|}{|C(A)\cup C_\succsim(A)|} \\
\label{jaccard} & = & 1-\dfrac{|C(A)\cap 
C_\succsim(A)|}{|C(A)\cup C_\succsim(A)|}\\
& \in & [0,1].
\end{eqnarray}
This function, which is widely applied in the life sciences and elsewhere, 
is a proper metric \citep{marczewski-steinhaus,
levandowsky-winter,gerasimou24} 
and identifies the dissimilarity 
between two sets as the proportion of their 
non-common elements relative to the total number of distinct 
elements between them.
With this definition in place, we can now formally introduce our 
proposed generalization of \eqref{hm}
for the case of possibly multi-valued choice data as the function 
$\mathtt{JHM}:\mathbf{D}\rightarrow\mathbb{Q}_+$ where  
\begin{eqnarray}
\mathtt{JHM}(\mathcal{D}) & := & \min_{\succsim\in\mathcal{R}} 
\sum_{\big(A,C(A)\bigr)\in\mathcal{D}}J\big(C(A),C_\succsim(A)\bigr)
\end{eqnarray}
In words, $\mathtt{JHM}$ differs from the cruder multi-valued choice 
extension of $\mathtt{HM}$ in that it distinguishes between --and penalizes--
``partially'' and ``fully'' mistaken decisions at a menu according to the Jaccard
similarity between each set of actual choices and the corresponding set of 
model-optimal choices at every menu. As such, one readily observes that 
$\mathtt{JHM}(\mathcal{D})\leq \mathtt{HM}(\mathcal{D})$ 
for all $\mathcal{D}\in\mathbf{D}$, and 
$\mathtt{JHM}\equiv \mathtt{HM}$ in the subdomain of 
$\mathbf{D}$ where $|C(A)|=1$ for all $\big(A,C(A)\bigr)\in\mathcal{D}$.
Applied to the example given above, $\mathtt{JHM}$ assigns scores  
$\frac{1}{2}$, 1 and 0 to the first, second and third subject, respectively, 
whereas the ``binary'' or ``pass/fail'' extension of $\mathtt{HM}$ 
assigns scores 1, 1 and 0 instead. 
Our analysis in Sections 4 and 5 builds on this index. 
Appendix B reports the main results of Section 5 when the ``binary'' 
$\mathtt{HM}$ extension is used instead.

\section{Experimental Design}

\subsection{General Remarks}

The experiment was conducted between September 2019 and 2021 at 
the St Andrews Experimental Economics Lab with a total of 282
subjects (more details in Table \ref{tab:ExpSum}).
In the main part subjects were presented sequentially 
with a series of 50 menus that consisted of 3, 4 or 5 pairs 
of gift cards, each drawn from a grand choice set of 6 such pairs. 
Each gift card was worth \pounds 10,
thereby making each choice alternative in each menu valued at \pounds 20.
All gift cards came from popular UK and/or international brands: 
two supermarkets, two coffee shops, one bookshop, and a gift card 
that enabled dining at one of nine restaurants. 
All 6 cards in these 6 pairs could be redeemed 
in at least one venue in the local town centre 
(subjects were explicitly informed about this), with the restaurant 
gift card being redeemable in 3 such venues. 
Each of the 6 gift cards appeared in exactly 2 of the 6 pairs, 
once showing up as the first item in the pair 
and once as the second (see Appendix A for more details).

There are several reasons why the experimental choice alternatives 
were selected to be gift-card pairs, and those ones in particular. 
First, gift cards can be thought of as restricted forms of cash that 
can be used for consumption, informally traded or, indeed, gifted. 
As such, they are intrinsically valuable. 
Second, these particular gift cards were issued by some of the most popular 
leisure and grocery destinations 
for the local student population, and were all within a short walking 
distance from each other. Hence, all 
were expected to be desirable to everyone in this experiment's subject pool. 
Finally, presenting the 
choice alternatives as gift card \textit{pairs} proxies a realistic situation 
(many online retailers invite consumers to choose 
between multi-store gift card bundles rather than 
individual-store gift cards) and could 
potentially lead to some relatively hard decisions.

Out of the 63 non-empty menus that are derivable 
from this set of 6 alternatives, 
subjects were presented with the 15 binary, 20 ternary 
and 15 quaternary menus. 
Each of the 50 distinct menus was presented once and 
the order of menu presentation 
was randomized and differed between subjects. 
This choice domain is therefore rich, 
heterogeneous and symmetric in the sense that all alternatives 
are feasible in exactly the 
same number of menus (see also Appendix B for more details on this). 

The experiment's computer interface was programmed in Qualtrics 
and executed in full-screen mode on a web browser 
that prevented subjects from exiting the interface without 
the experimenter's intervention. Subject recruitment 
was done with ORSEE \citep*{ORSEE15}. 
All menus appeared as unnumbered vertical choice lists. 
The \textit{``I'm not choosing now''} option in the Free-Choice 
treatment was always the last item. 
Because subjects often had to scroll down to find and 
select that option, this positioning means that 
it was often physically harder for them to avoid/delay choice. 
There were a total of 141 participants 
in each of the two treatments. 
The 9 subjects who always deferred or always chose everything were 
excluded from the analysis because their 
choice behaviour is completely uninformative. 
Every subject received both their gift-card and cash rewards, 
as explained below.
A nine-question understanding quiz preceded the main 
part of the experiment in each of the two treatments. 
Subjects could not proceed to the main part of the 
experiment until they answered all questions correctly.

\subsection{Forced-Choice Treatment}

At the beginning of the experiment subjects in both treatments 
were allocated a monetary endowment of $I=\pounds 2.40$. 
When a menu was shown to subjects in this treatment, 
they were asked to choose one or more items from that menu. 
Subjects knew that one menu would be picked 
at random for them at the end of the experiment. 
They also knew that they would be rewarded with an element 
of their randomly selected menu, 
and that the decision they made at that menu during the 
main part of the experiment would be 
reminded to them before they are asked to make their final, 
payoff-relevant decision there. 
Once subjects were past a menu during the main part of the 
experiment they never saw it again 
unless that menu later turned out to be their randomly selected one. 
No additional information about the 
choice alternatives was provided at any point.

If a subject in this treatment chose one or more --but not all-- 
gift-card pairs from their randomly selected menu during 
the main part of the experiment, and also chose something from 
those previously selected options if that menu was determined 
to be their payoff-relevant one, then they received 
that item as an in-kind reward and $I$ as a cash reward. 
If, instead, at that point they chose something that was not among 
their previously chosen options at that menu, 
then they received that item and $I_{rev}=\pounds 1.20 <I$. 
Finally, if they had chosen everything at that menu originally, 
then they received their original endowment 
$I$ and a randomly selected element of that menu. 

\subsection{Free-Choice Treatment}

Free-/Non-Forced-Choice subjects 
were asked to either choose one or more of the alternatives 
at each menu or to delay/avoid 
making such an active choice by selecting \textit{``I'm not choosing now''}. 
What happens here if a subject makes one or more active choices from their 
randomly selected menu during 
the main part of the experiment coincides with what happens in the Forced-Choice 
treatment in the respective cases. 
But if a subject here had delayed choice at that menu originally, 
they are asked to choose an item now. 
In this case they receive this item together with an amount
$I_{def}=\pounds 2.10$ that lies strictly between the full initial endowment
$I$ and the cash reward associated with a choice reversal, $I_{rev}$. 

\begin{table}[!htbp]
\caption{Summary information on the two experimental treatments.}
\centering
\footnotesize
\begingroup
\setlength{\tabcolsep}{5pt} 
\renewcommand{\arraystretch}{1.3} 
\makebox[\textwidth][c]{
\begin{tabular}{|l| c| c| }
\hline
& \textbf{Free-Choice}		& \textbf{Forced-Choice}    \\
& \textbf{treatment}				& \textbf{treatment}		\\
\hline
\textit{Original sample}									& 141								& 141 \\
\hline
\multirow{1.5}{*}{\textit{Subjects excluded because}}		& \multirow{2}{*}{1}				& \multirow{2}{*}{3} \\
\multirow{1}{*}{\textit{they always chose everything}}		& 									& \\
\hline
\multirow{1.5}{*}{\textit{Subjects excluded because}} 		& \multirow{2}{*}{5}				&  \multirow{2}{*}{N/A} \\
\multirow{1}{*}{\textit{they always deferred}}				&									& \\
\hline
\textit{Subjects excluded - total}							& 6									&  3 \\
\hline
\textit{Subjects after all exclusions}						& 135								& 138 \\
\hline
\textit{Choice objects} 									& \multicolumn{2}{c|}{6 pairs of \pounds 10 gift cards (\pounds 20 total value)} \\
\hline
\textit{Number of menus \& decisions}						& \multicolumn{2}{c|}{50} \\
\hline
\textit{Reward frequency}									& \multicolumn{2}{c|}{Every subject}\\
\hline	
\textit{Location}											& \multicolumn{2}{c|}{St Andrews lab}\\
\hline
\multirow{1.5}{*}{\textit{Dates when the experimental}}  	& \multirow{2}{*}{17--20 Sept 2019} & \multirow{1.2}{*}{29 Jan 2020 (62)}\\
\multirow{1}{*}{\textit{sessions were conducted}}			&					   				& \multirow{1}{*}{22 Sep 2021 (79)} \\
\hline
\end{tabular}
\label{tab:ExpSum}
}
\endgroup
\end{table}

\subsection{Payment}

As soon as subjects finished all tasks, their randomly selected menu 
showed up on their screens, together with the reminder 
of the decision they had made at this menu. 
As an additional incentive for subjects to make 
deliberated and non-rushed decisions, 
they were told from the beginning that no participant would 
be able to receive their rewards and leave the lab in the first 
50 minutes of the session. 
The experimenter (this author) went to each subject's desk once they were finished  
(and after this threshold was exceeded), asked them about 
their final choice at this menu, and later gave them their 
cash and gift card rewards accordingly. 
Subjects who had chosen everything at their randomly selected menu were in turn 
invited to the experimenter's desk and, after the 
appropriate numerical range was specified on the random-number generating 
website \url{https://random.org}, and the way in which 
the numbers in this range were mapped to the relevant gift card pairs 
was agreed, a random number was generated to determine 
the pair they would be rewarded with.
The total payment per subject was approximately \pounds 22.2.

\subsection{Discussion}

Denoting by $A$ a subject's payoff-relevant randomly selected menu, 
and by $C^1(A)$, $C^2(A)$ the 
choices they made at that menu during the main and final parts, respectively, 
Table \ref{tab:design}(a) 
summarizes the incentive structure in the Forced-Choice treatment. 
Under the assumption of a utility-maximizing 
decision maker with possibly weak preferences, 
$C^1(\cdot)$ denotes their nonempty- 
but possibly multi-valued choice correspondence. 
That is, for a Forced-Choice subject $C^1(\cdot)$ 
satisfies $\emptyset\neq C^1(M)\subseteq M$ for every menu $M$. 
By contrast, $C^2(A)$ assigns a single chosen alternative to the 
subject's randomly selected menu $A$. 

\begin{table}[!htbp]
\centering
\footnotesize
\caption{\centering Incentives in the Forced- and Free-Choice 
treatments are structured around the subjects' 
1st and 2nd decisions at their randomly selected menu (denoted here by $A$).}
\setlength{\tabcolsep}{6pt} 
\renewcommand{\arraystretch}{1.3} 
\makebox[\textwidth][c]{
\begin{tabular}{|lcr|lr|c|}
\hline
\multirow{1.5}{*}{\textbf{1st decision}}	
&	
& \multirow{1.5}{*}{\textbf{2nd decision}}		
& \multirow{1.5}{*}{\textbf{Choice}}	
& \multirow{1.5}{*}{\textbf{Cash}} 	
& \multirow{1.5}{*}{\textbf{Possible}}\\
\multirow{0.9}{*}{(main stage)}			
&  
& \multirow{0.9}{*}{(payoff stage)} 
& \multirow{0.9}{*}{\textbf{reward}} 
& \multirow{0.9}{*}{\textbf{reward}} 
& \multirow{0.9}{*}{\textbf{interpretation}}\\
\hline
\multicolumn{6}{|c|}{\textbf{(a) Forced-Choice Treatment}}\\
\hline
\multirow{2}{*}{$C^1(A)=A$}		
&			
& \multirow{1.2}{*}{none}						
& \multirow{1.2}{*}{random}			
& \multirow{1.2}{*}{initial}			
& \multirow{1.2}{*}{Revealed total indifference}\\
& 
& \multirow{1}{*}{possible} 
& \multirow{1}{*}{$a\in A$}  
& \multirow{1}{*}{endowment $I$} 
& \multirow{1}{*}{is costless \& responded to literally} \\
\hline
\multirow{2}{*}{$C^1(A)\subset A$} 
& 
& \multirow{2}{*}{$C^2(A)\in C^1(A)$}		
& \multirow{2}{*}{$C^2(A)$}					
& \multirow{2}{*}{$I$}	
& \multirow{1.2}{*}{Stable revealed preference}\\
& 
& 
& 
& 
& \multirow{1}{*}{is costless} \\
\hline
\multirow{2}{*}{$C^1(A)\subset A$} 
& 
& \multirow{2}{*}{$C^2(A)\not\in C^1(A)$}		
& \multirow{2}{*}{$C^2(A)$}					
& \multirow{2}{*}{$I_{r}<I$}		
& \multirow{1.2}{*}{Unstable revealed preference} \\
&	
& 	
& 	
& 	
& \multirow{1.2}{*}{is costly}\\
\hline
\multicolumn{6}{|c|}{\textbf{(b) Free-Choice Treatment}}\\
\hline
\multirow{2}{*}{$C^1(A)=A$}		
&			
& \multirow{1.2}{*}{none}						
& \multirow{1.2}{*}{random}			
& \multirow{1.2}{*}{initial}			
& \multirow{1.2}{*}{Revealed total indifference}\\
& 
& \multirow{1}{*}{possible} 
& \multirow{1}{*}{$a\in A$}  
& \multirow{1}{*}{endowment $I$} 
& \multirow{1}{*}{is costless \& responded to literally} \\
\hline
\multirow{2}{*}{$\emptyset\neq C^1(A)\neq A$} 
& 
& \multirow{2}{*}{$C^2(A)\in C^1(A)$}		
& \multirow{2}{*}{$C^2(A)$}					
& \multirow{2}{*}{$I$}	
& \multirow{1.2}{*}{Stable revealed preference}\\
& 
& 
& 
& 
& \multirow{1}{*}{is costless} \\
\hline
\multirow{1.5}{*}{$\emptyset\neq C^1(A)\neq A$} 
& 
& \multirow{1.5}{*}{$C^2(A)\not\in C^1(A)$}		
& \multirow{1.5}{*}{$C^2(A)$}					
& \multirow{1.5}{*}{$I_{r}<I$}		
& \multirow{1.2}{*}{Unstable revealed preference} \\
&	
& 	
& 	
& 	
& \multirow{1.2}{*}{is costly, and more so than}\\
\multirow{0.5}{*}{$C^1(A)=\emptyset$}		
&	
& \multirow{0.5}{*}{$C^2(A)\in A$}				
& \multirow{0.5}{*}{$C^2(A)$}					
& \multirow{0.5}{*}{$I_{d}$ :\, $I_{r}<I_{d}< I$}	
& \multirow{1.2}{*}{revealed indecisiveness}\\
\hline
\end{tabular}
}
\label{tab:design}
\end{table}

By the very definition of indifference, a utility-maximizing subject 
who operates under the Forced-Choice design and is indifferent 
between all feasible alternatives 
at a menu is also indifferent between choosing \textit{all} 
these alternatives and \textit{any} sub-collection thereof. 
Hence, such a subject has no \textit{strict} incentive to misreport 
their indifference. 
Conversely, however, they will choose everything in a menu 
\textit{only if} they are indifferent between all alternatives 
at that menu because, by definition 
of indifference, this individual does not care which alternative 
they will get or whether this is decided by someone else or randomly 
(see also \cite*{danan10} for a formal elaboration of this point). 
Thus, it is \textit{weakly dominant} 
for this subject to choose everything feasible in a menu whenever 
they are indifferent between all items contained 
in it. 

At the same time, if a utility-maximizing subject has one 
or more optimal alternative(s) at the menu, 
and these are strictly better than something else feasible, 
then choosing everything is dis-incentivized by 
the design because doing so comes with the risk of potentially 
receiving an inferior alternative as a reward. 
The design instead incentivizes such a subject to select those 
and only those options they prefer the most, 
as this enables them to choose one of these superior options 
at the end and to also receive their full cash endowment.
For a utility-maximizing agent, therefore, the design in this treatment 
combines incentive-compatibility for truthful preference revelation in the sense of 
\cite*{azrieli-chambers-healy18} with a standard interpretation 
of multi-valued choice that is articulated in 
Kreps (\citeyear{kreps12}, p.2) as follows:
\textit{``The story is that the consumer 
chooses \textit{one} element of $A$. 
Nonetheless, we think of $c(A)$ as a subset of $A$, 
not a member or element of $A$. 
This allows for the possibility that the 
consumer is happy with any one of the several elements of $A$, 
in which case $c(A)$ lists all those elements. 
When she makes a definite choice of a single element, say $x$, 
out of $A$--when she says in effect, 
`I want $x$ and nothing else'--we write $c(A)=\{x\}$, 
or the singleton set consisting of the single element $\{x\}$. 
But if she says, `I would be happy with either $x$ or $y$', 
then $c(A)=\{x,y\}$.''}

The choice correspondence $C^1(\cdot)$ for a 
Free-Choice subject on the other hand is possibly empty- and multi-valued; 
hence, it satisfies $\emptyset\subseteq C^1(A)\subseteq A$ for every menu $A$. 
As in the Forced-Choice treatment, however, $C^2(\cdot)$ assigns a 
single chosen alternative to the relevant subject's randomly selected menu.
In this treatment too, moreover, a utility-maximizing Free-Choice subject 
is (weakly) incentivized to choose their most preferred option(s) at every menu, 
with the opportunity to defer at a cost being irrelevant to them.
Now suppose again that the subject is not a utility maximizer but has incomplete 
preferences and cannot compare any two of the feasible options at a menu. Then, 
depending on the individual's subjective perception of the decision's importance, 
they may either opt to choose everything and end up with something at random or 
opt instead to incur the relatively small cost $I-I_{def}$ 
to delay making an active 
choice at that menu themselves. The former kind of behaviour could be seen as 
analogous to recent 
findings suggesting that decision makers prefer 
to randomize when they are faced with a 
difficult decision repeatedly \citep*{agranov&ortoleva,
dwenger-kubler-weizsacker18}; we discuss this possibility 
in Appendix C.2. 
Importantly though, it could also be compatible with 
the model of Undominated Choice 
with Incomplete Preferences whenever every feasible option 
is incomparable to all others. By contrast, deferring 
at a small cost could be seen 
as a manifestation of indecisiveness-driven deferral \citep*{tversky_shafir92,
danan&ziegelmeyer06,CCGT22} and may potentially 
be compatible with the model 
of Dominant Choice with Incomplete Preferences.

\subsection{Remarks on the Asymmetric Incentives in the Design}

It is worth commenting on the discrepancy in the way 
that the design deals 
with situations where $C(A)=A$ or $\emptyset\neq C(A) \subset A$ 
at the payoff-relevant menu $A$. Recall that
in the former case subjects receive their reward from $C(A)$ at random and 
a full cash reward alongside it, whereas in the latter case 
they can themselves choose anything from $A$, possibly at the cost of 
receiving the reduced cash reward if their two choices at $A$ exhibit a reversal.
An alternative design choice here would be to break this dichotomy by 
applying the uniform randomization rule at $C(A)$ in both situations. 
This would have been a perfectly legitimate way to proceed in our 
environment of multi-valued choice. 
Introducing this dichotomy here was driven by the motivation to 
have a design that remained comparable to the one 
in \cite*{CCGT22}. The one proposed here achieves this goal because it 
does indeed reduce to the one proposed by these authors 
in the special case where $C(A)$ 
is required to be single-valued. With the alternative 
design mentioned above there would be no role for the reduced cash 
payment $I_{rev}$ that is associated with a choice reversal.
The presence of this additional payment parameter, however, 
and its lower value compared to that of the full-payment 
parameter $I$, is potentially useful in instilling a mindset towards
preference-guided choices by subjects.
On the other hand, the unified choice-reward rule 
where randomization is applied on any non-empty $C(A)$
is immune to the possibility of subjects'
choosing dominated alternatives, which in theory is left open 
by the dichotomous rule in the current design. While this 
theoretical possibility can indeed not be ruled out, it is worth 
keeping in mind that a subject who includes dominated alternatives in
their first-round choice $C(A)\subset A$ not only does 
not benefit from doing so but, in fact, spends additional time and 
effort --thereby incurring a cost-- in the process.

\section{Individual-Level Analysis}

\subsection{Non-Parametric Model Fitting}

We now turn to the tests of our possibly multi- 
and/or empty-valued experimental choice data 
for compatibility with each of the three models introduced in Section 2. 
These tests were done by applying the JHM
method that was introduced in Section 2.3.
We refer to the JHM score of a model on a given subject's dataset 
as the model's \textit{distance score} on that dataset.\footnote{We 
use the term \textit{``distance score''} rather than 
\textit{``distance''} partly because the underlying 
function is not a proper metric and 
partly to highlight the ranking aspect of this 
JHM-based goodness-of-fit approach.} 
The model-based extension of HM was for single-valued (forced or free) 
choices was originally introduced
in \citet*{CCGT16,CCGT22}\footnote{The 2016 working-paper 
version of that study indirectly 
constructed choice correspondences by augmenting 
the subjects' actual choices with 
their survey responses in indifference-vs-preference questions that 
followed binary menus only. Before constructing 
the subjects' indifference classes 
from these data, the survey responses were analysed for consistency, 
and responses that eventually led to intransitive indifferences 
were discarded. Whenever this was the case, the subjects' actual choices were 
interpreted as revealing strict preferences, regardless of the survey response.} 
Since \cite*{prest}, moreover, this has been 
extended considerably and made freely available online 
in the open-source desktop 
application \Prests\!\!.\footnote{With retainment of its original 
focus on the model of Rational 
Choice/Utility Maximization with strict preferences, 
the HM approach has been extended in other intuitive directions by 
\citet*{apesteguia&ballester15} 
and \citet*{dean&martin}. However, although the ``Swaps'' 
measure of \citet*{apesteguia&ballester15} 
that pertains to general choice environments is also readily computable 
in \PrestsFootnote\!\!, 
it is currently unclear how it should be computed 
when subjects choose multiple items 
from the same menu, or how it should be extended 
in models that generalise utility maximization.}

More specifically, a brute-force algorithm was implemented for these 
combinatorial-optimi-zation computations. 
This involved the production of all choice datasets that are generated by all 
possible instances of every model, 
and comparing each such dataset against every subject's own dataset in order 
to find the model(s) and instance(s) 
that are closest to it in the minimum distance-score sense. 
This method therefore detects perfect as well as 
\textit{approximate} model fits, and in the latter case 
it quantifies the approximation in an intuitive way that 
may be interpreted as the number of ``mistakes'' 
made by an agent who might be portrayed as 
following a particular decision rule.\footnote{For other approximations that 
have been proposed or applied recently for/at 
distinct classes of models and choice environments we refer the reader 
to \cite*{choi&etal07,echenique&etal,choi&etal,apesteguia&ballester15,dean&martin,
bouacida-martin21,dembo-kariv-polisson-quah,fudenberg-lian-gao21,
cherchye-demuynck-lanier-derock23,echenique-imai-saito23} and references therein.}

Despite the large numbers of weak orders (4,683), 
partial orders (130,023) and, especially, 
incomplete preorders (209,527) that are defined on a set of 6 alternatives 
(OEIS, \citeyear{OEIS21}), the above tool makes such an \textit{exact} 
computation possible very quickly --in less than 12 seconds-- 
for all subjects in each treatment, for each of the three models. 
Furthermore, although the brute-force algorithm is linear in the number of subjects 
but exponential in the number of alternatives, the model-rich distance-score 
method itself is scalable and can be extended to analyse datasets that are 
derived from much larger sets of alternatives by employing powerful open-source 
constraint solvers.

\begin{figure}[hbtp!]
\centering
\caption{\centering Distributions of all subjects' Jaccard-Houtman-Maks
distance scores for each of the three models, 
and the corresponding scores from simulations.\vspace{10pt}}
\begin{subfigure}[b]{0.47\textwidth}
\centering
\caption{Forced-Choice treatment}
\includegraphics[width=0.99\textwidth]{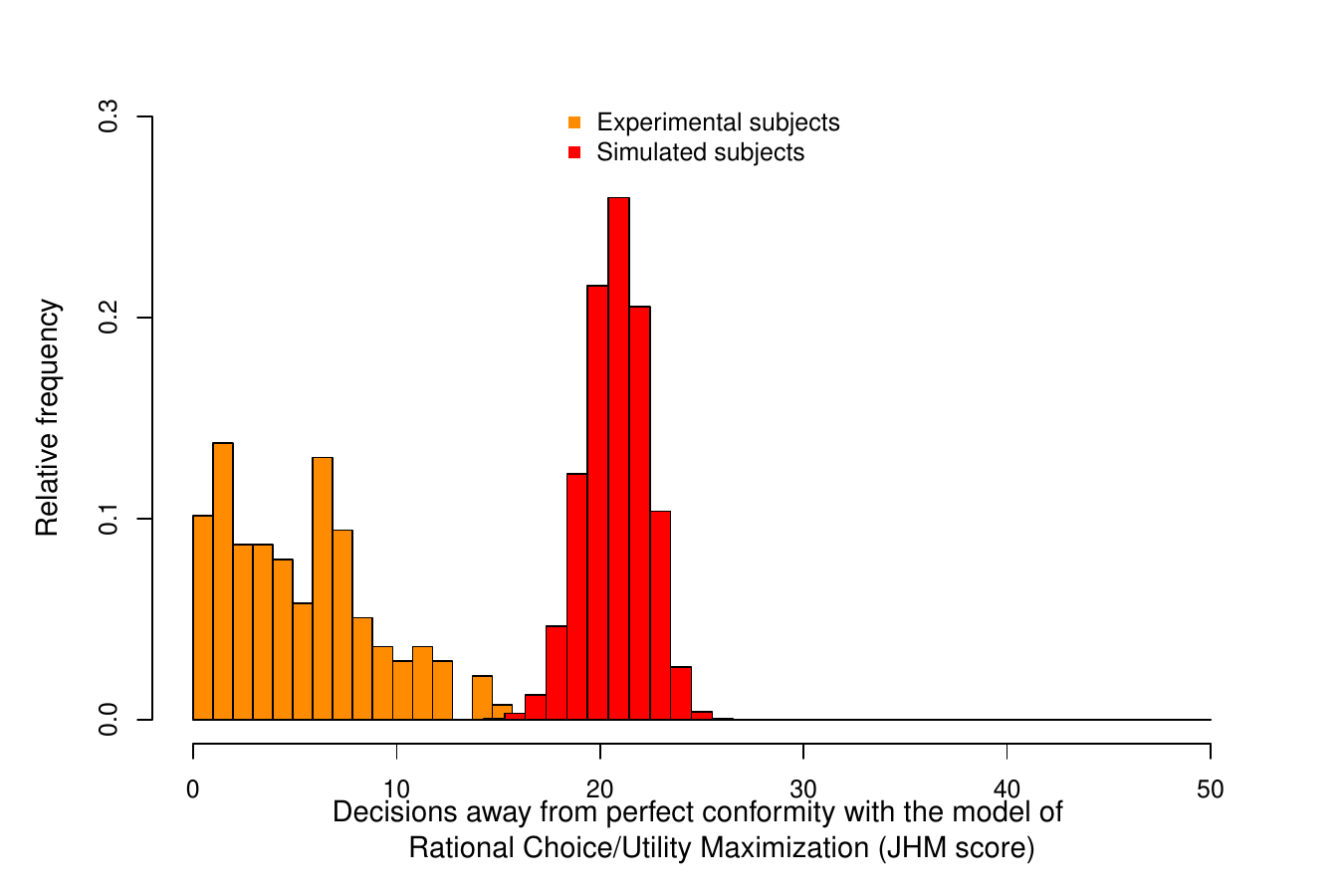}
\label{fig:fc_um}
\end{subfigure}\hspace{15pt}
\begin{subfigure}[b]{0.47\textwidth}
\centering
\caption{Free-Choice treatment}
\includegraphics[width=0.99\textwidth]{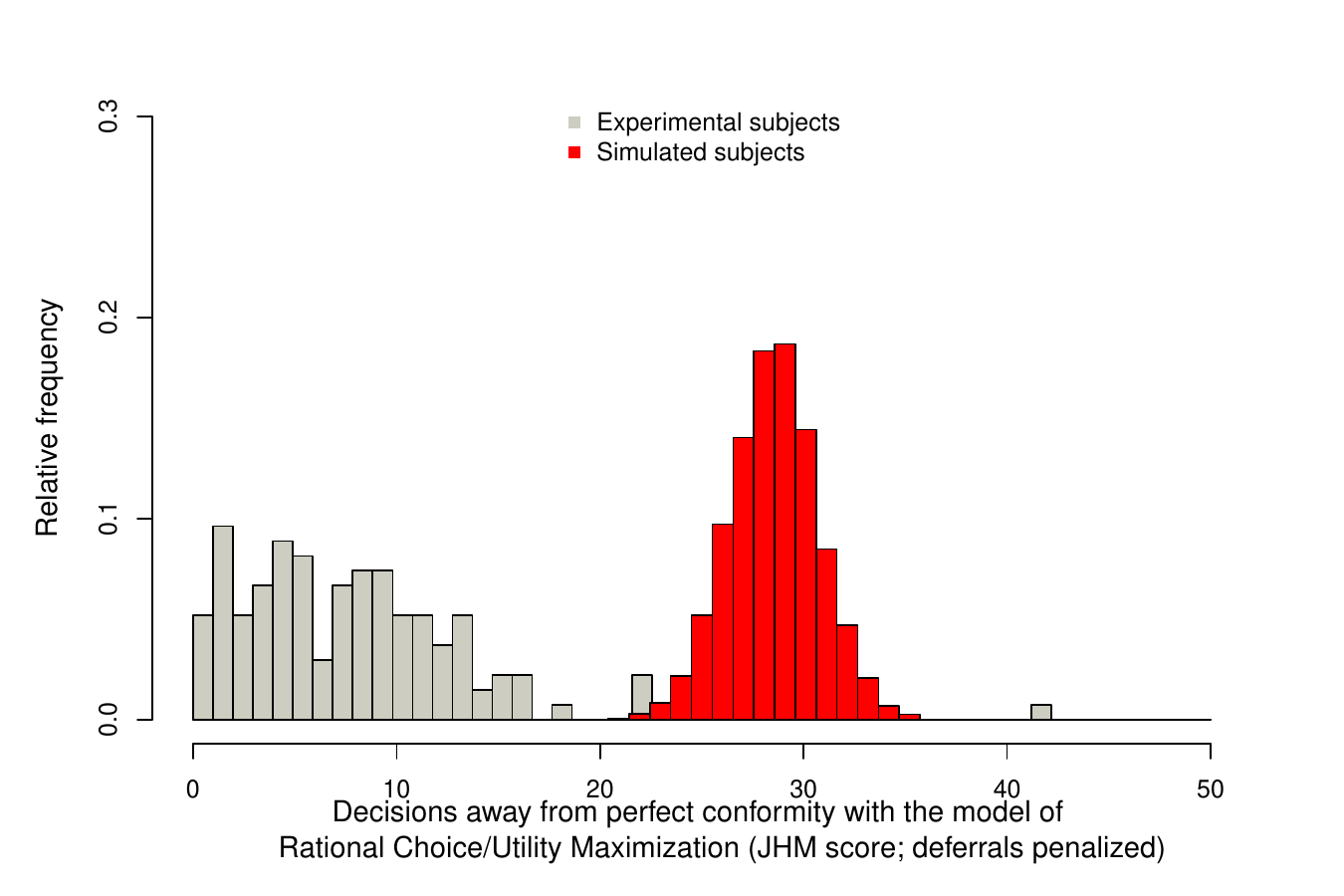}
\label{fig:nfc_um}
\end{subfigure}
\begin{subfigure}[b]{0.47\textwidth}
\centering
\includegraphics[width=0.99\textwidth]{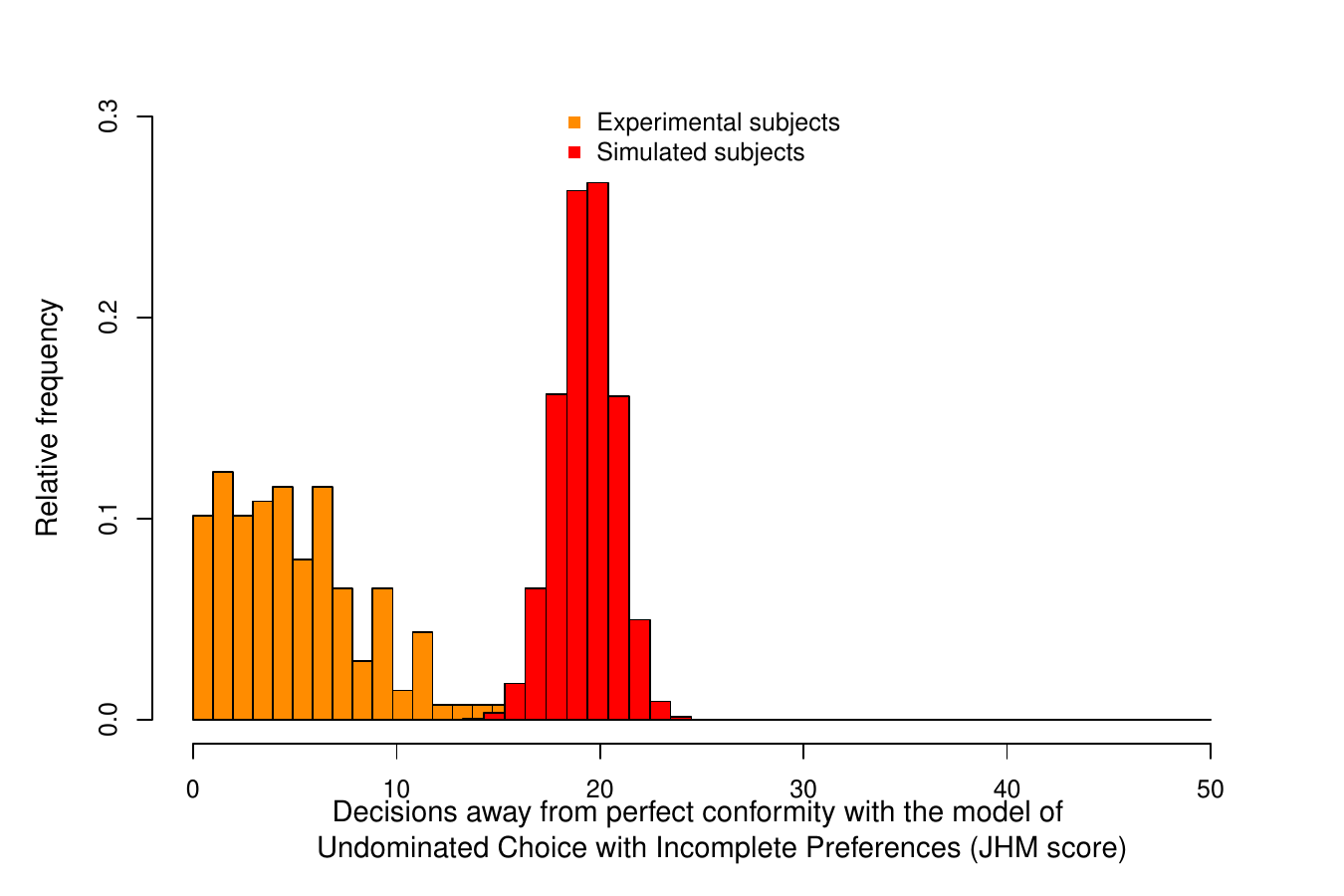}
\label{fig:fc_uc}
\end{subfigure}\hspace{15pt}
\begin{subfigure}[b]{0.47\textwidth}
\centering
\includegraphics[width=0.99\textwidth]{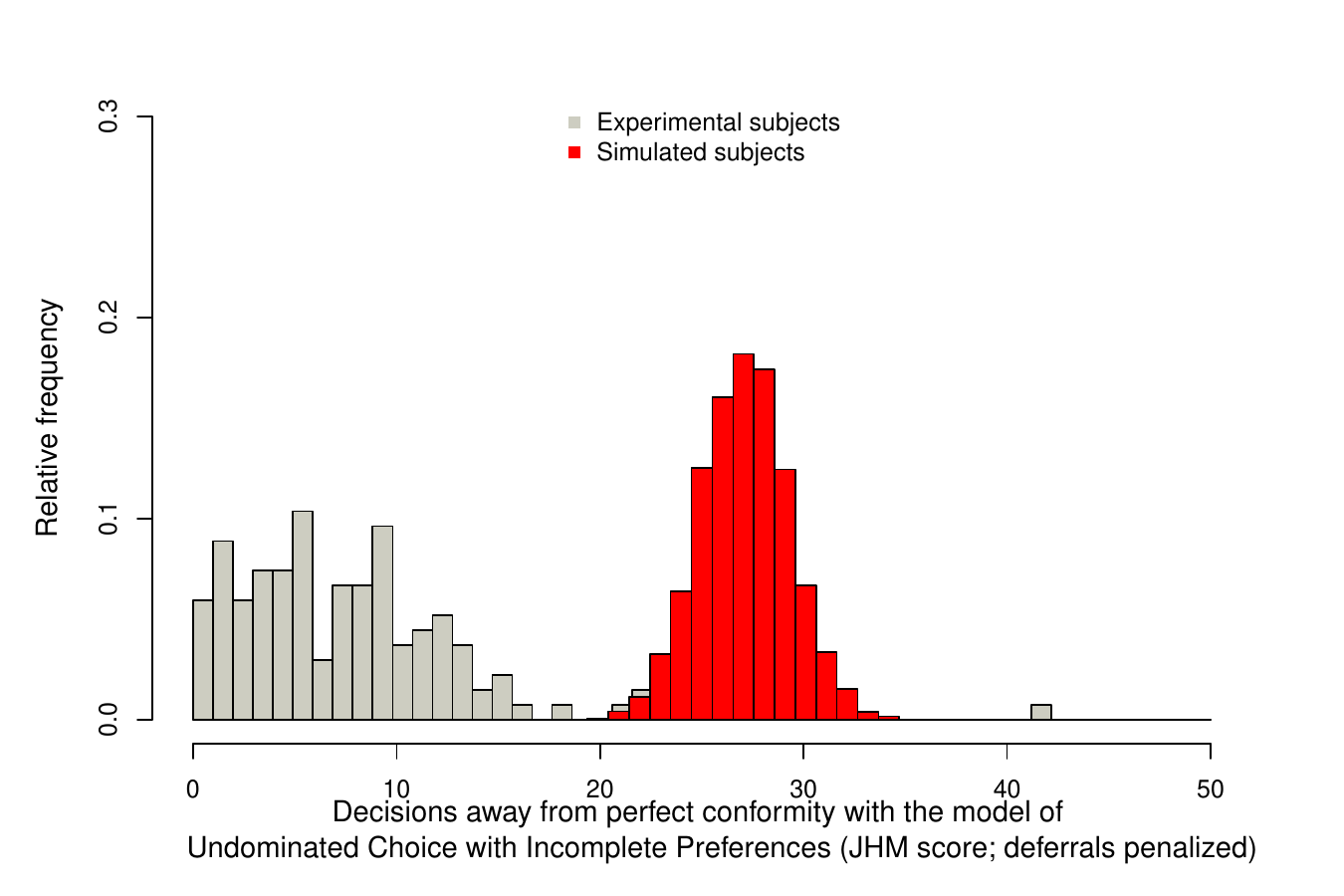}
\label{fig:nfc_uc}
\end{subfigure}		
\begin{subfigure}[b]{0.47\textwidth}
\centering
\caption*{}
\end{subfigure}\hspace{15pt}
\begin{subfigure}[b]{0.47\textwidth}
\centering
\includegraphics[width=0.99\textwidth]{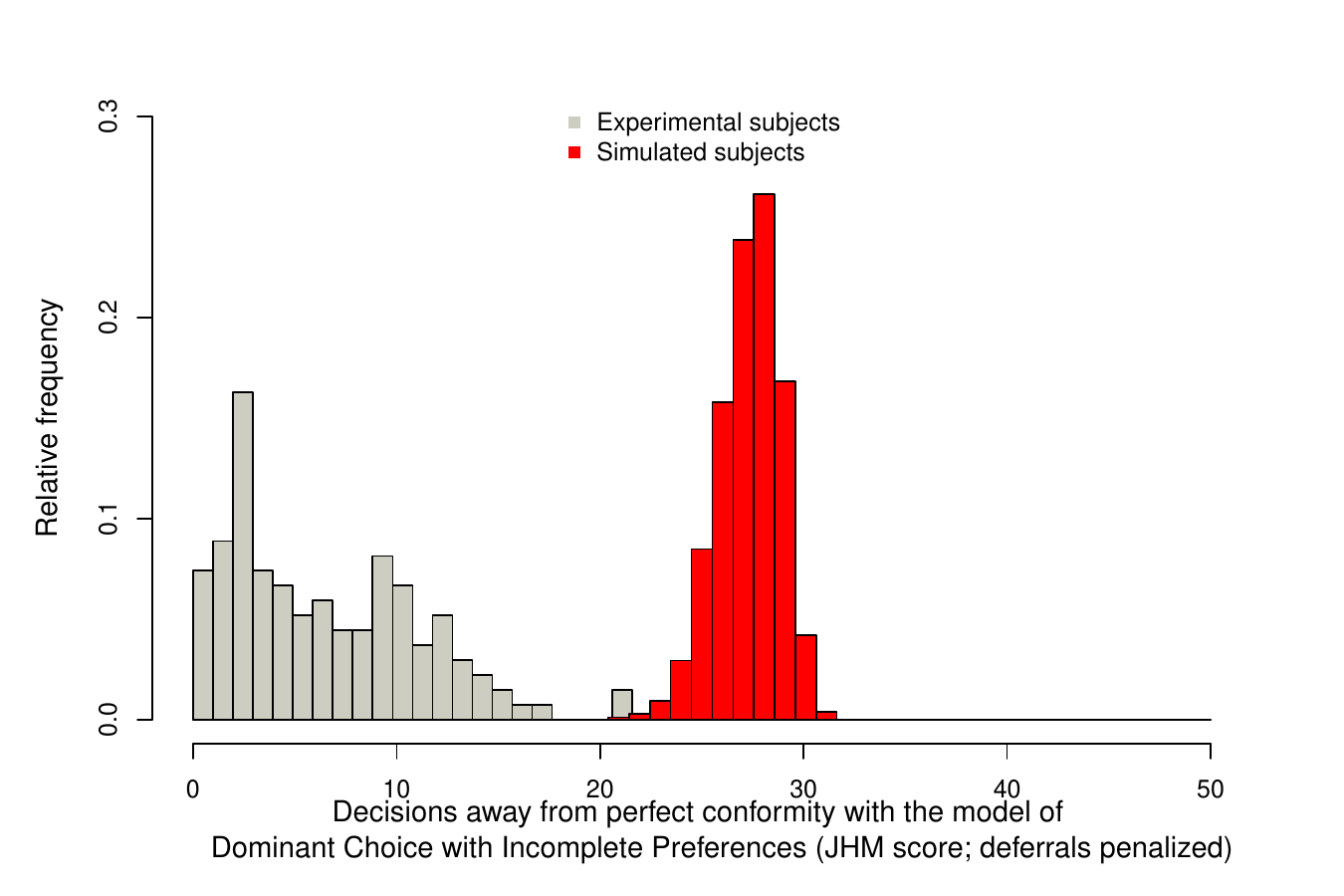}
\label{fig:nfc_mdc}
\end{subfigure}	
\label{fig: models}
\end{figure}

Table \ref{tab:models} and Figure \ref{fig: models} present summaries of this 
goodness-of-fit analysis for both Forced- and Free-Choice subjects. 
In line with standard practices in empirical revealed preference analysis 
whereby one also wishes to understand the extent to which a certain behaviour or 
perfect/approximate model fit could have been generated 
randomly \citep*{bronars87}, 
we also performed our model analysis on 10,000 simulated datasets 
of artificial uniform-random behaving subjects under both 
a Forced- and a Free-Choice configuration 
(details on the simulations' structure are available in Appendix C). 
The JHM distance-score distributions of simulated subjects 
are juxtaposed in the relevant graphs of Figure \ref{fig: models} 
to those of the experimental subjects for each of the three models, 
while the minimum values from the simulated-subject JHM distributions 
are also presented in Table \ref{tab:models}.

\begin{table}[!htbp]
\centering
\footnotesize
\caption{\centering 
Classification of subjects who are perfectly/approximately 
explainable by one of the three models under the 
Jaccard-Houtman-Maks goodness-of-fit 
method.\vspace{10pt}}
\begin{subtable}{\textwidth}\centering
\setlength{\tabcolsep}{5pt} 
\renewcommand{\arraystretch}{1.3} 
\makebox[\textwidth][c]{
\begin{tabular}{|l|rl|rl|rl|rl|}
	\hline
	&&& 
	\multicolumn{2}{c}{\textbf{Undominated}}& 
	\multicolumn{2}{|c|}{\textbf{Dominant}}	&& \\
	&\multicolumn{2}{c|}{\textbf{Utility}} 	&
	\multicolumn{2}{c}{\textbf{Choice with}}& 
	\multicolumn{2}{|c|}{\textbf{Choice with}}&& \\
	\multicolumn{1}{|c|}{}	&
	\multicolumn{2}{c|}{\textbf{Maximization}}	&
	\multicolumn{2}{c}{\textbf{Incomplete}} 	&
	\multicolumn{2}{|c|}{\textbf{Incomplete}}	& 
	\multicolumn{2}{c|}{\multirow{1}{*}{\textbf{All}}}\\
	&&& \multicolumn{2}{c}{\textbf{Preferences}}& 
	\multicolumn{2}{|c|}{\textbf{Preferences}}	&&  \\
	\hline 
	& \multicolumn{8}{c|}{\textbf{Forced-Choice treatment} ($N=138$)}\\
	\hline 					
	Subjects with JHM = 0	& 
	\multicolumn{2}{c|}{5 (4\%)}		& 
	\multicolumn{2}{c|}{0}		&
	\multicolumn{2}{c|}{--}		& 
	\multicolumn{2}{c|}{5 (4\%)}\\
	\hline
	Subjects with JHM $\leq$ 5 	& 
	\multicolumn{2}{c|}{61 (44\%)}		& 
	\multicolumn{2}{c|}{16 (11.5\%)} 		&
	\multicolumn{2}{c|}{--} 			& 
	\multicolumn{2}{c|}{77 (56\%)}\\
	\hline 
	Mean/median best score (JHM $\leq$ 5)	& 
	\multicolumn{2}{c|}{2.01/1.83}			& 
	\multicolumn{2}{c|}{3.56/4}  		& 
	\multicolumn{2}{c|}{--}				& 
	\multicolumn{2}{c|}{2.34/2}\\	
	\hline			
	\multirow{1.5}{*}{Mean / median best-model} 		& 
	\multicolumn{2}{c|}{\multirow{2}{*}{1.23/1}} 	& 
	\multicolumn{2}{c|}{\multirow{2}{*}{1.19/1}}  	& 
	\multicolumn{2}{c|}{\multirow{2}{*}{--}}		& 
	\multicolumn{2}{c|}{\multirow{2}{*}{1.22/1}} \\
	\multirow{1}{*}{preference orderings (JHM $\leq$ 5)} 
	& & & & & & & & \\
	\hline 
	Minimum score in simulations	& 
	\multicolumn{2}{c|}{14.87} 		& 
	\multicolumn{2}{c|}{13.92} 		& 
	\multicolumn{2}{c|}{--}			& 
	\multicolumn{2}{c|}{--}\\
	\hline
	& \multicolumn{8}{c|}{\textbf{Free-Choice treatment} ($N=135$)}\\ 
	\hline 
	Subjects with JHM = 0	& 
	\multicolumn{2}{c|}{3 (2\%)} 		& 
	\multicolumn{2}{c|}{0}		& 
	\multicolumn{2}{c|}{7 (5\%)}		& 
	\multicolumn{2}{c|}{10 (7\%)}\\
	\hline
	Subjects with JHM $\leq$ 5 	& 
	\multicolumn{2}{c|}{30 (22\%)}		& 
	\multicolumn{2}{c|}{7 (5\%)} 		&
	\multicolumn{2}{c|}{37 (27\%)} 		& 
	\multicolumn{2}{c|}{74 (55\%)}\\
	\hline 
	Mean/median best score (JHM $\leq$ 5)	& 
	\multicolumn{2}{c|}{2.12/1.75} 		& 
	\multicolumn{2}{c|}{2.37/2.08}			& 
	\multicolumn{2}{c|}{1.93/2}			&  
	\multicolumn{2}{c|}{2.05/2}\\ 
	\hline 
	\multirow{1.5}{*}{Mean/median best-model} 	& 
	\multicolumn{2}{c|}{\multirow{2}{*}{1/1}}&  
	\multicolumn{2}{c|}{\multirow{2}{*}{1/1}}& 
	\multicolumn{2}{c|}{\multirow{2}{*}{1.16/1}}& 
	\multicolumn{2}{c|}{\multirow{2}{*}{1.08/1}}\\
	\multirow{1}{*}{preference orderings (JHM $\leq$ 5)} 
	& & & & & & & & \\
	\hline			
	Minimum score in simulations & 
	\multicolumn{2}{c|}{21.03} &  
	\multicolumn{2}{c|}{19.67} &  
	\multicolumn{2}{c|}{20.08} & 
	\multicolumn{2}{c|}{--} \\
	\hline 
\end{tabular}
}
\end{subtable}
\label{tab:models}
\caption*{\scriptsize \centering 
Note: model-score ties were always broken in 
favour of Rational Choice/Utility Maximization 
(no other ties emerged). 
}
\end{table}

For human subjects, Rational Choice often tied with one,  
but never both, of the other two models. 
The classification presented in Table \ref{tab:models}
broke ties in favour of that model whenever a subject's JHM distance 
score was produced by Rational Choice and another model 
(see the notes of Table \ref{tab:models} for more details). 
Under this tie-breaking assumption, and in line also 
with model-approximation methods used in the above-cited studies, 
the two exhibits show the number, proportion 
and relative-frequency distributions of distinct subjects that 
are on average within 10\% (equivalently, within 5 decisions) 
away from being explainable perfectly by an instance of some model, 
separately for each of the two experimental treatments. 
We chose this conservative approximation range 
for two reasons: (i) simulations suggest that a JHM distance score of 5 for 
any of the three models is extremely unlikely to occur randomly 
in this decision environment (see Table \ref{tab:models} 
and Figure \ref{fig: models}); 
(ii) for subjects with a JHM score not greater than 5 there is typically 
only one best-matching preference ordering that explains their behaviour 
under the respective subject-optimal model.

The proportion of Forced-Choice subjects with a perfect 
or 10\%-approximate model fit was 4\%  (all under Rational Choice) 
and 56\%, respectively, with 44\% and 11.5\% of 
those in the latter group classified under Rational Choice 
and Undominated Choice with Incomplete Preferences. 
Furthermore, the proportion of Free-Choice subjects with a 
perfect or 10\%-approximate fit was 7\% and 55\%, respectively. 
22\%, 5\% and 27\% of all subjects in this treatment 
were categorized under Rational Choice, Undominated and Dominant Choice 
with Incomplete Preferences, respectively. 
In addition, 3 and 7 of these subjects were \textit{perfectly} 
compatible with the first and third model, respectively, with 
one subject among the former three and five among the latter 
seven revealing indifferences 
in their weakly ordered and weakly preordered preferences. 

The model fit is slightly better on average in the 
Free-Choice treatment than in the Forced-Choice one 
(mean/median best JHM scores: 2.05/2 vs 2.34/2), 
although the distributions are not significantly different. 
Similarly, and in line with the results from 
the consistency analysis that is presented in Section 5, 
the proportion of subjects who were best-matched by one of the 
two models of consistent active choices 
(i.e. Rational Choice and Dominant Choice with Incomplete Preferences)
is higher in the Free-Choice treatment (49\% vs 44\%), 
while the proportion of subjects best-matched 
by the third model (Undominated Choice with Incomplete Preferences) 
that predicts generally inconsistent active choices is lower 
in that treatment (5\% vs 11.5\%). 
Moreover, when Rational Choice was not the best-matching model, 
its distance score was on average 1.4 and 5.7 units 
higher in the Forced- and Free-Choice treatments, respectively. Finally, 
the highest JHM distance score of 5 for experimental subjects in this 
approximate model-fitting analysis is far below the corresponding 
minimum scores of simulated random-behaving subjects, 
which ranged between 14 and 21.

\begin{figure}[!htbp]
\centering
\caption{\centering 
All 37 subjects that were best-matched by Dominant 
Choice with Incomplete Preferences 
were unlikely to have achieved this classification at random conditional on 
the menus where they deferred.\vspace{-30pt}}
\includegraphics[width=0.70\textwidth]{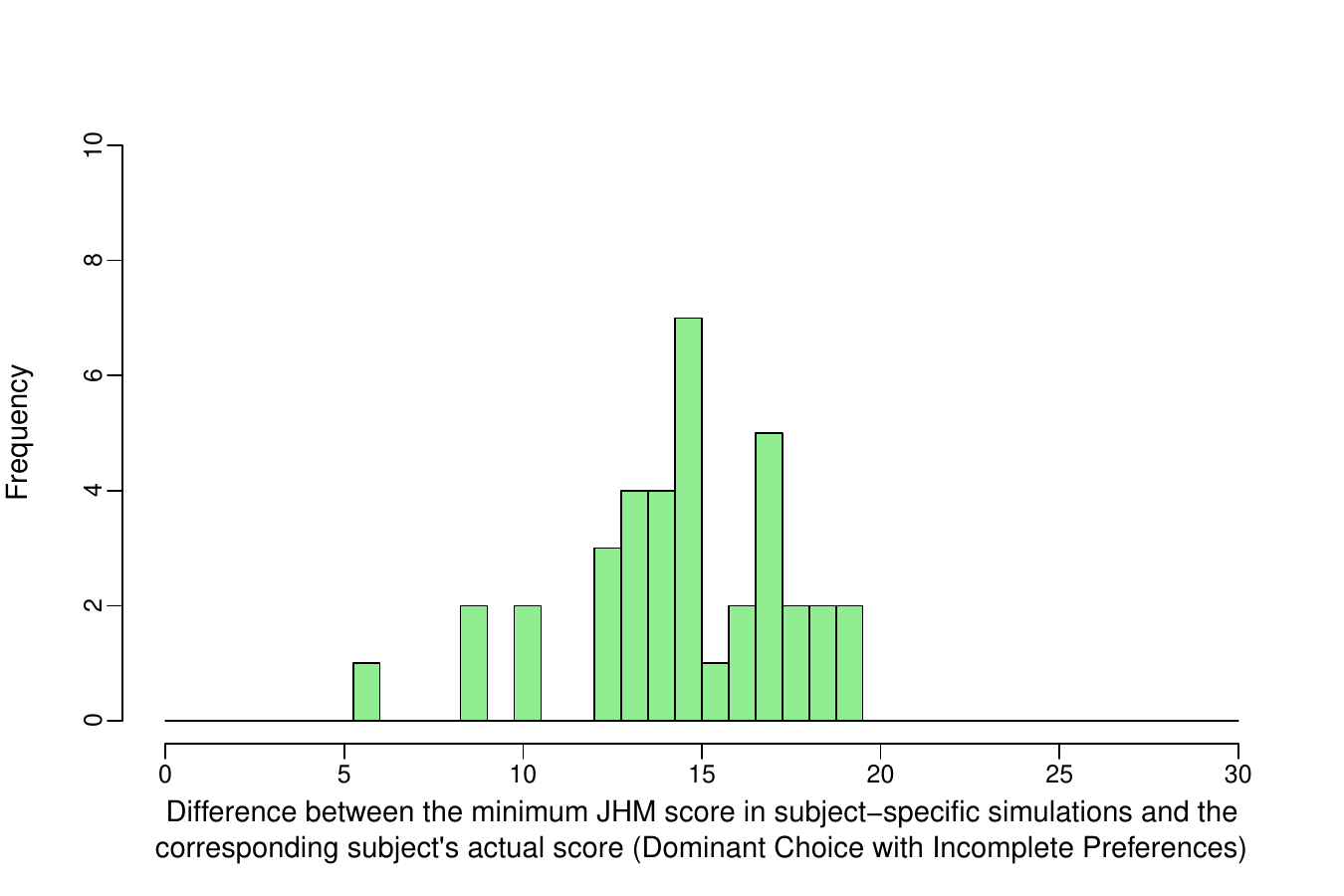}
\label{fig:mdc_human_simul_diffs}
\end{figure}

For the 37 subjects who are best-explained 
by Dominant Choice with Incomplete Preferences, moreover, we conducted 
an additional robustness check. 
By analogy to the way in which the ``predictive success'' \citep{selten91}
score of  the Rational-Choice model on Free-Choice subjects' active-choice consistency 
is evaluated in Section 5.4 below, we generated a set of 10,000 
random-behaving simulated subjects for each of the 37 
human experimental participants who were best-matched by that model, 
holding fixed in each such subject's simulations block 
the menus at which the relevant human participant deferred. 
This allows for evaluating the predictive success of the 
Dominant-Choice model's 10\%-approximation range in this decision environment. 
Figure \ref{fig:mdc_human_simul_diffs} shows the 
histogram with JHM distance-score differences for this model 
between the 2.5th percentile value in each subject-specific 10k-simulations 
block and the relevant subject's actual score. 
For all 37 subjects the difference is positive and typically 
much greater than 10 score units (i.e. active-choice or deferral decisions). 
This analysis suggests that Dominant Choice with Incomplete Preferences 
is not excessively permissive in this environment, and that its good fit is 
useful towards explaining behaviour and eliciting preferences.

\subsection{Recovery of Strict Preferences, Indifference and Indecisiveness}

As was highlighted earlier, the collected choice data and non-parametric 
combinatorial-optimization method presented earlier 
jointly allow us to separate the strict part from the potential indifference 
and/or incomparability parts of each subject's revealed preferences, 
conditional on the model that best explains the subject's overall behaviour.
For the vast majority of all subjects that are perfectly or approximately 
matched by some model this analysis reveals a unique weak/strict 
and complete/incomplete preference ordering that generates their 
minimum JHM distance score, with the mean (median) number of such orderings being 
1.22 (1) and 1.08 (1) in the Forced- and Free-Choice treatments.
Thus, despite the relatively large number of 50 distinct menus and decisions in the 
experiment and the fact that this large number nevertheless leads to an incomplete
collection of menus, this model-rich analysis generates sharp 
subject-specific preference estimates.

\begin{figure}[!htbp]
\centering
\caption{\centering A subject's complete weak preferences, 
optimally recovered by Utility Maximization.}
\includegraphics[width=0.1\textwidth]{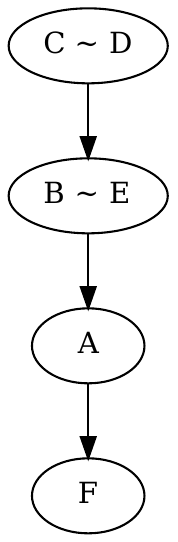}
\label{fig:um}
\end{figure}

\begin{figure}[!htbp]
\centering
\caption{\centering A subject's incomplete preferences, optimally recovered 
by Undominated Choice with Incomplete Preferences, 
with (left) or without (right) indifferences.}
\includegraphics[width=0.25\textwidth]{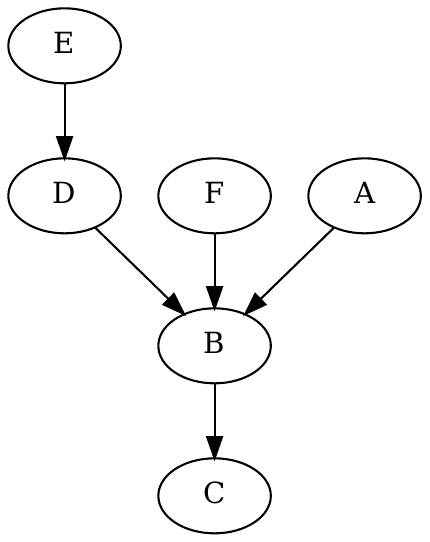}\hspace{40pt}
\includegraphics[width=0.185\textwidth]{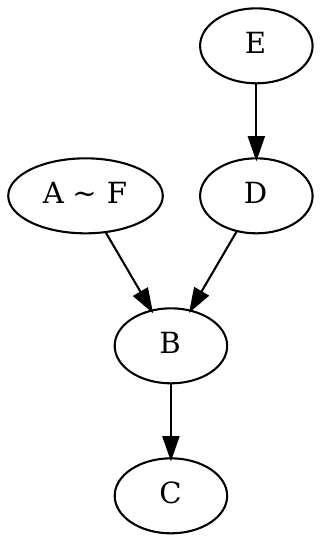}
\label{fig:ind-ind-1}
\end{figure}

\begin{figure}[!htbp]
\centering
\caption{\centering A subject's incomplete weak preferences,
optimally recovered by Dominant Choice with Incomplete Preferences.}
\includegraphics[width=0.28\textwidth]{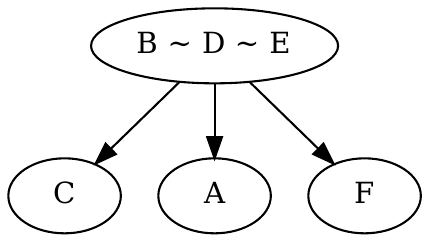}
\label{fig:ind-ind-2}
\end{figure}

We now turn to the important task of distinguishing 
between indifference and incomparability. 
To this end, it is perhaps worth recalling first that indifferences 
are generically non-existent in a large class of the benchmark \cite{bewley} class 
of incomplete preference relations under uncertainty that are defined 
on a space of bundles of continuous commodities or uncertain acts \citep*{gerasimou18EL}. 
By contrast, however, this is very much not the case when such preferences 
are over finitely many discrete options, as in the choice environment analysed 
in this paper. Indeed, our analysis uncovers that out of the 60 subjects (22\% of the total) 
that the two models of incomplete-preference maximization explain optimally 
across the two treatments, for 46 of them (77\%) it is possible to separate 
the strict, indifference and incomparability/indecisiveness components 
of their preferences (robustly so for the model of Dominant Choice with 
Incomplete Preferences). The empirical documentation of this theory-based separation 
of the three  binary relations that are possible when preferences are 
incomplete is a novel contribution of this paper. Figures \ref{fig:ind-ind-1} 
and \ref{fig:ind-ind-2} show the directed graphs of two subjects' incomplete 
weak/strict preferences that were optimally recovered by these models; 
Figure \ref{fig:um} does so for a utility-maximizing subject.

\begin{table}[!htbp]
\centering
\footnotesize
\caption{\centering Revealed indifferences 
conditional on each subject's best-fitting model.}
\setlength{\tabcolsep}{5pt} 
\renewcommand{\arraystretch}{1.3} 
\begin{tabular}{|l|cc|cc|}
\hline
& \multicolumn{2}{|c|}{\multirow{1.5}{*}{\textbf{Indifference-revealing}}}
& \multicolumn{2}{|c|}{\textbf{Mean (median)}}\\
& \multicolumn{2}{|c|}{\multirow{1}{*}{\textbf{subjects}}}
& \multicolumn{2}{|c|}{\textbf{indifferences}}\\
\hline
Utility Maximization 
& \qquad 71 & (79\%)
& \quad 2.89 & (2)\\
\hline
Dominant Choice with Incomplete Preferences 
& \qquad 28 & (76\%) 
& \quad 2.26& (2)\\
\hline
Undominated Choice with Incomplete Preferences
& \qquad 16 & (69.5\%)
& \quad 0.69 & (0.5)\\
\hline
Overall
& \qquad 115 & (77\%) 
& \quad 2.4& (1.08)\\
\hline
\end{tabular}
\label{tab:indiff-summary}
\end{table}

Looking across the three models and two treatments, 
Table \ref{tab:indiff-summary} summarizes the key revealed-indifference information
that is derived from each subject's best-fitting model and demonstrates 
the descriptive relevance of indifference in these data. 
In particular, 79\%, 76\% and 69.5\% of all subjects who are 
best-explained by Utility Maximization, Dominant Choice and Undominated Choice
with Incomplete Preferences, respectively, have at least one model-compatible 
preference relation that reveals at least one indifference between distinct 
gift-card pairs. Across all models, subjects and their 
respective model-compatible preference relations, finally, the average 
(median) number of indifference comparisons between distinct gift-card pairs 
is 2.4 (1.08). Considering that there are 15 possible binary comparisons on a set 
of 6 elements, this means that the indifference relation accounts for 16\% of subjects'
preference comparisons, on average. 
This proportion is highest for Utility Maximization (19\%), followed by Dominant (15\%)
and Undominated Choice with Incomplete Preferences (5\%).

\subsection{Cross-Validation Test of Model-Fitting Analysis}

We finally performed a cross-validation exercise to assess the out-of-sample 
robustness of the model-fit estimates that were reported earlier.
More specifically, we split each subject's data into the first and second
half (recall that each subject had a randomly different menu-presentation order)
and conducted the same goodness-of-fit analysis separately for each half.
We then counted the subjects for whom the same model was found to be optimal 
in each of the first and second 25 decisions, as well as is all 50.

In the Forced-Choice treatment, 50 of the 77 subjects (65\%) were best-matched 
by the same model --possibly in different modes-- in both halves and overall. 
From the remaining 17 subjects, moreover, 14 were best-matched by Undominated 
Choice with Incomplete Preferences in the first half and by Rational Choice in
the second. This suggests the possibility that some subjects may have learned 
their preferences and behaved increasingly more rationally during the experiment.

In the Free-Choice treatment on the other hand, where all three models have 
meaningful explanatory power and therefore more possibilities arise, 
things are less straightforward. More specifically, 15 of the 74 subjects 
in this treatment (20\%) were best-matched by the same model both in the first
and second half and overall. The learning hypothesis that was suggested above 
is relevant here too, with 17 subjects (23\%) that were best-matched by one of 
the two models of incomplete-preference maximization in the first half being
optimally described by utility maximization in the second half. 
Another 8 subjects (11\%) went in the opposite direction, while 7 (9\%) 
were as if they alternated between the two models of choice with 
incomplete preferences across the two halves.

In summary, this cross-validation exercise shows that a sizeable proportion 
of subjects who are within 5 decisions (10\%) away from being perfectly 
explainable by one of the three models are best-matched by the same model 
throughout, particularly in the Forced-Choice treatment, while for another  
considerable proportion it is possible that preference learning may have 
contributed to a better fit of Rational Choice in the second half.

\section{Aggregate-Level Analysis}

\subsection{Choice Sizes}

\begin{figure}
\centering
\caption{Relative frequencies of the different choice sizes, 
adjusted for feasibility.\vspace{-10pt}}
\includegraphics[width=0.85\textwidth]{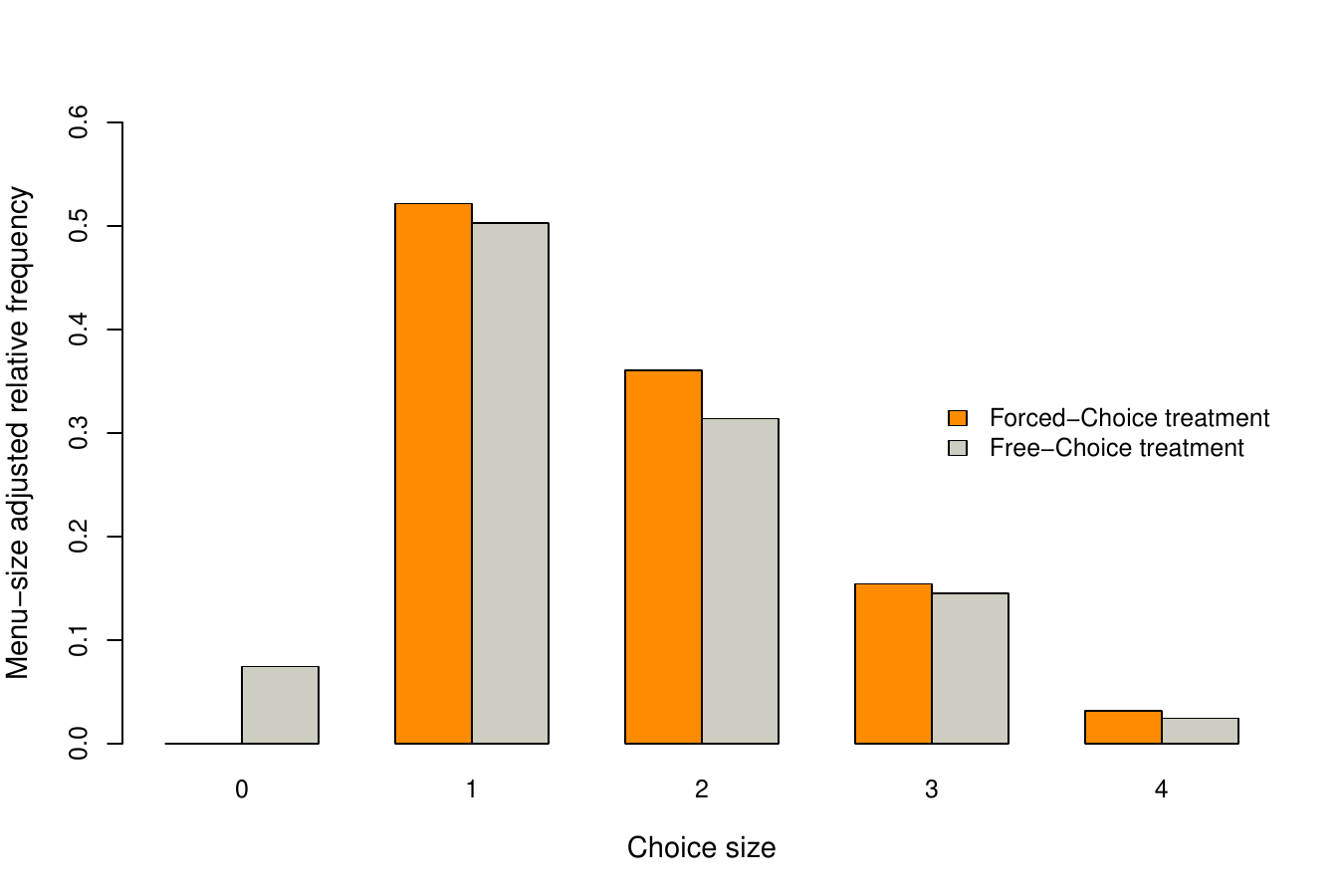}
\label{fig:adjusted_choice_frequencies}
\end{figure}

Turning to patterns in the aggregate data, 
we start with a key new variable of interest, 
the subjects' \textit{choice size}, 
which is defined simply as the number of gift-card pairs 
that were chosen at each menu. 
This variable ranges between 0 and 4 in the Free-Choice treatment, 
and between 1 and 4 in the 
Forced-Choice treatment. However, because choice sizes 
1 and 2 (as well as 0 in the former treatment) 
are always feasible, whereas 3 and 4 are not always so, we adjust 
their relative frequencies accordingly. 
Specifically, Figure \ref{fig:adjusted_choice_frequencies} 
presents the distributions of menu-size 
adjusted choice sizes in the two treatments, which are derived 
once their absolute frequencies are 
divided by the total number of menus where these choice 
sizes might be observed. For example, 
the denominators here are $50\times N$ and $15\times N$ 
for choice sizes 1 and 4, respectively, 
where $N$ is the total number of subjects in the relevant treatment. 

This comparison shows that the modal adjusted choice 
size was 1 in both treatments, 
at a rate of just over 50\%, and was followed by sizes 2 and 3. 
The \textit{deferral rate} 
in the Free-Choice treatment, defined as the relative 
frequency of a zero choice size, 
is 7.4\%. This is slightly higher than the 6.9\% \textit{choose-everything} rate, 
which is obtained by weighted-averaging 
the relative frequencies where $n$ out of $n$ 
items were chosen, for $n=2,3,4$ (Table \ref{tab:choice_size_rates_times}). 
The latter rate is also lower than the corresponding one in the Forced-Choice 
treatment (8.5\%), and the difference is statistically significant ($p<0.001$; 
two-sided Fisher's exact test). 
This difference is consistent with the intuition that, 
in the absence of the possibility to delay choice 
when faced with a potentially difficult 
decision, subjects are more likely to choose 
everything that is available in the menu, 
either as a result of following a more general decision rule or, 
in the context of our 
experimental design, possibly due to a preference for randomization. 
Finally, the rate 
at which subjects chose all 4 alternatives 
was very low in both treatments, 
at 3.2\% and 2.5\% in the Forced- 
and Free-Choice treatments, respectively.

\begin{table}[!htbp]
\centering
\footnotesize
\caption{\centering Choice sizes and the corresponding relative 
frequencies and average response times \newline at menus of different sizes.}
\setlength{\tabcolsep}{5pt} 
\renewcommand{\arraystretch}{1.3} 
\makebox[\textwidth][c]{
\begin{tabular}{|l|c|c|c|c|c|}
\hline
& \multicolumn{5}{|c|}{\textbf{Alternatives chosen in the Forced-Choice treatment}}\\
& \multicolumn{5}{|c|}{(average response times, in seconds, in parenthesis)}\\
\hline
& \textbf{0}& \textbf{1} 	 		& \textbf{2}			& \textbf{3} 			& \textbf{4}\\
\hline 
\textbf{Binary menus}	& -- 		& 83.77\%  \hfill (5.82)& 16.23\% \hfill (7.56)	&	--					& -- \\
\hline 
\textbf{Ternary menus}	& --		& 46.19\% \hfill (7.31) & 47.25\% \hfill (8.44)	&  6.56\% \hfill (8.88)	& --\\
\hline 
\textbf{Quaternary menus}& --		& 28.50\% \hfill (8.27)& 41.01\% \hfill (10.06)& 27.29\% \hfill (11.25)	& 3.18\% (13.13) \\
\hline 
\hline 
& \multicolumn{5}{|c|}{\textbf{Alternatives chosen in the 
Free-Choice treatment}}\\
& \multicolumn{5}{|c|}{(average response times, 
in seconds, in parenthesis)}\\
\hline
& \textbf{0} 			& \textbf{1} 	 		& \textbf{2}			& \textbf{3} 			& \textbf{4}  \\
\hline 
\textbf{Binary menus}	& 11.95\% \hfill (6.19)	& 74.52\%  \hfill (6.01)& 13.53\% \hfill (7.93)	&	--					& -- \\
\hline 
\textbf{Ternary menus}	& 6.85\% \hfill (9.62)	& 44.67\% \hfill (7.25) & 43.33\% \hfill (8.93)	&  5.15\% \hfill (8.29)	& --\\ 
\hline 
\textbf{Quaternary menus}& 3.70\% \hfill (12.71)& 33.53\% \hfill (8.98)	& 33.28\% \hfill (9.93)	& 27.02\% \hfill (11.24)&2.47\% (12.92)  \\
\hline
\end{tabular}
}
\label{tab:choice_size_rates_times}
\end{table}

Table \ref{tab:choice_size_rates_times} elaborates 
further on this theme by also showing how the relative frequencies of 
different choice sizes vary with the number of alternatives 
at different menus, and by also presenting the corresponding 
average response times. These conditional relative frequencies 
are uniformly higher in the Forced-Choice treatment for 
choice sizes 1 and 2 in binary and ternary menus. 
In addition, proportionally more subjects in that treatment chose 
2, 3 or 4 gift-card pairs in quaternary menus too. 
As far as deferrals in the Free-Choice treatment are concerned, 
these were more likely in binary menus ($\approx$ 12\%) 
than in ternary ($\approx$ 7\%) or quaternary menus ($\approx$ 4\%). 

\subsection{Response Times}

\begin{figure}[hbtp!]
\centering
\caption{\centering Response times are positively 
correlated with menu and choice size in both treatments.\vspace{10pt}}
\begin{subfigure}[b]{1\textwidth}
\centering
\caption{Forced-Choice treatment}
\includegraphics[width=0.49\textwidth]{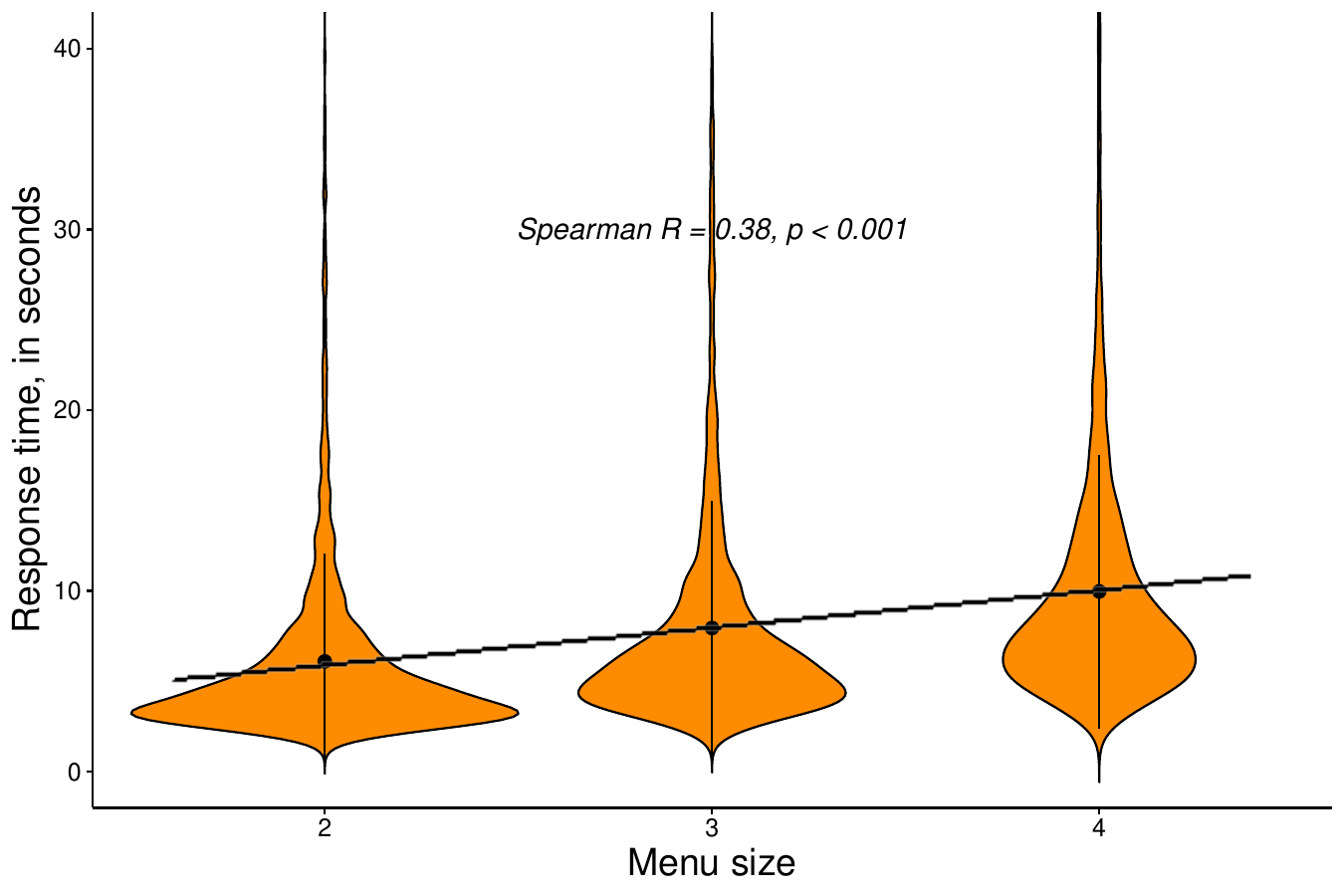}\hspace{5pt}
\includegraphics[width=0.49\textwidth]{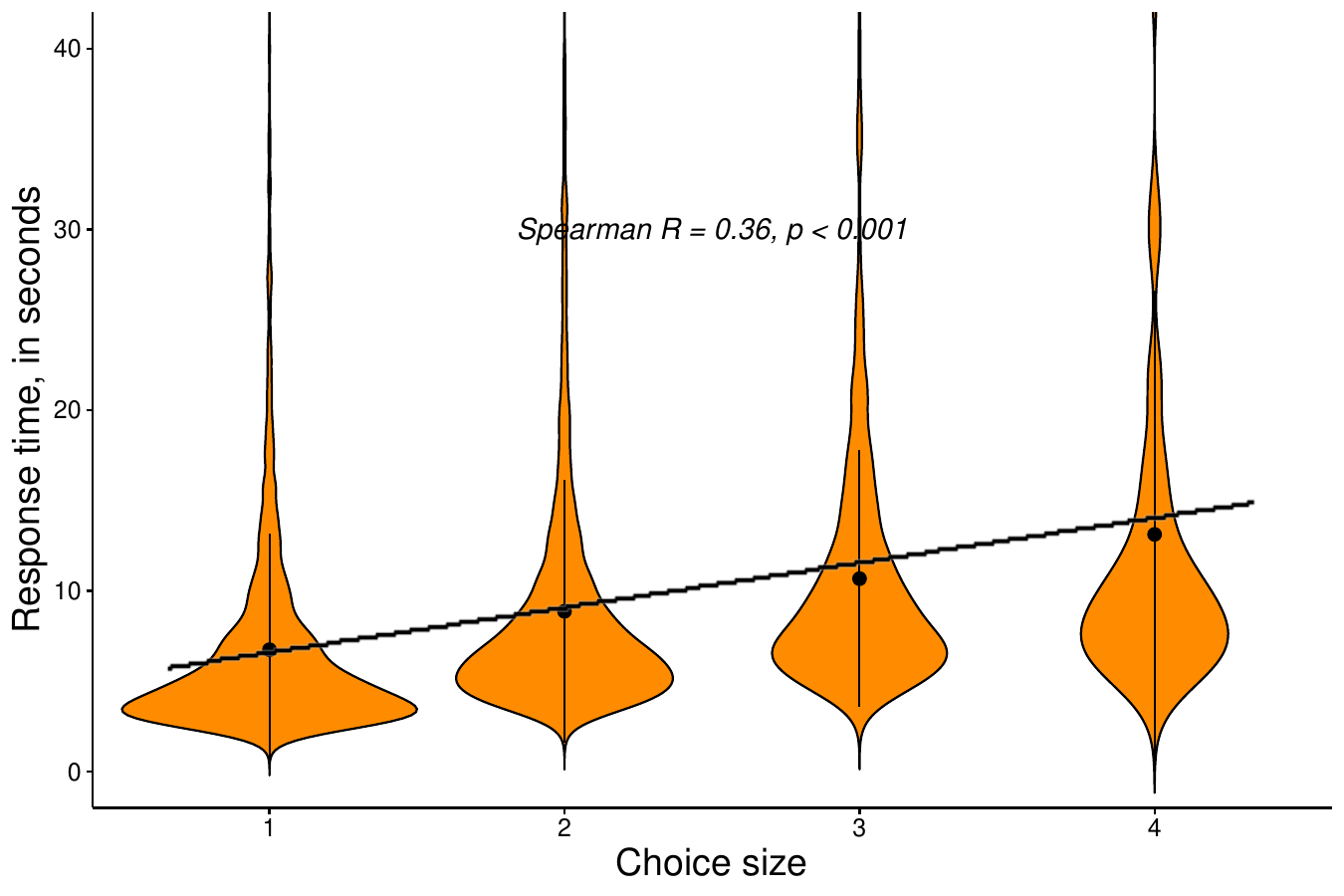}
\end{subfigure}
\begin{subfigure}[b]{1\textwidth}
\centering
\caption{Free-Choice treatment}
\includegraphics[width=0.49\textwidth]{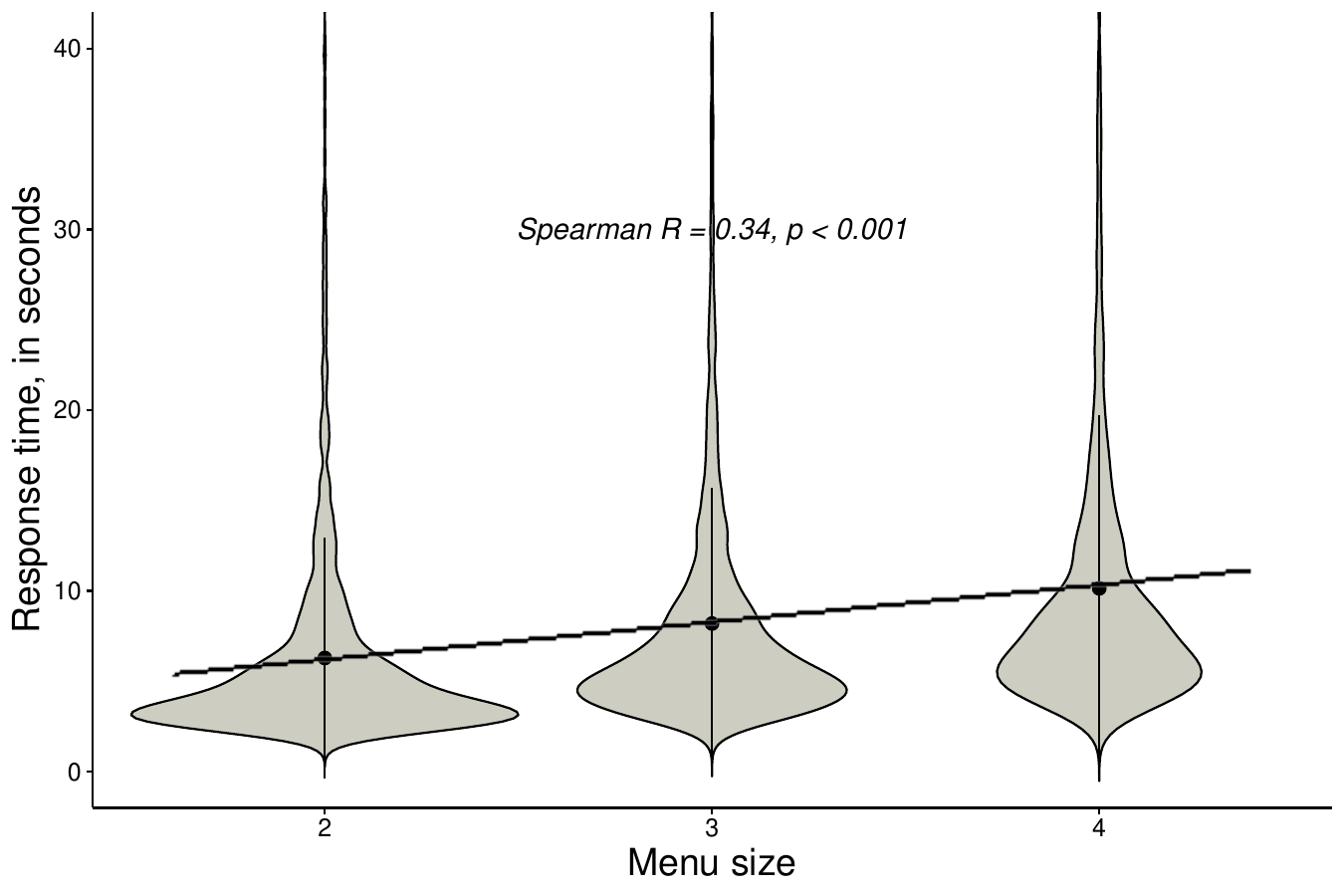}\hspace{5pt}	\includegraphics[width=0.49\textwidth]{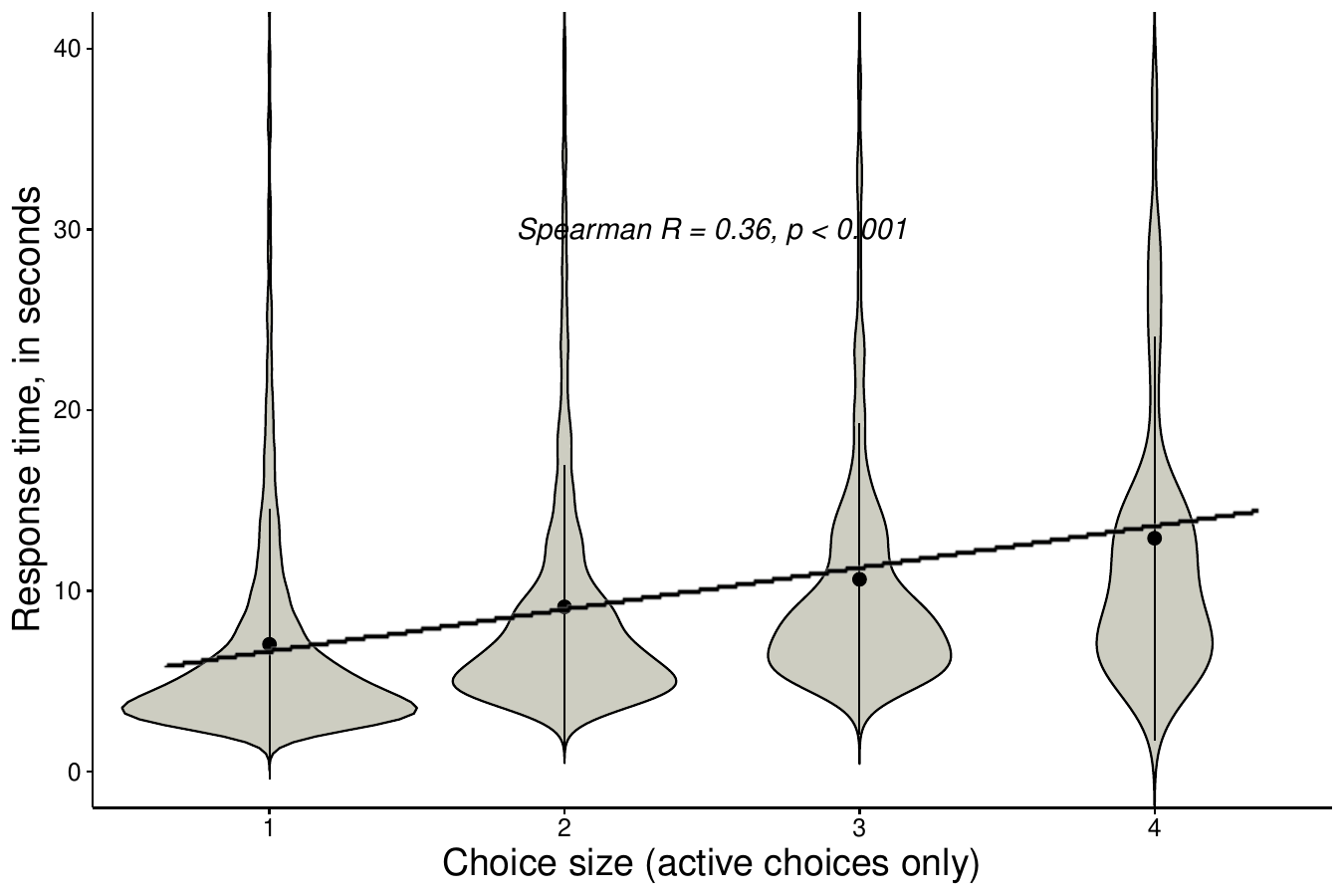}
\includegraphics[width=0.49\textwidth]{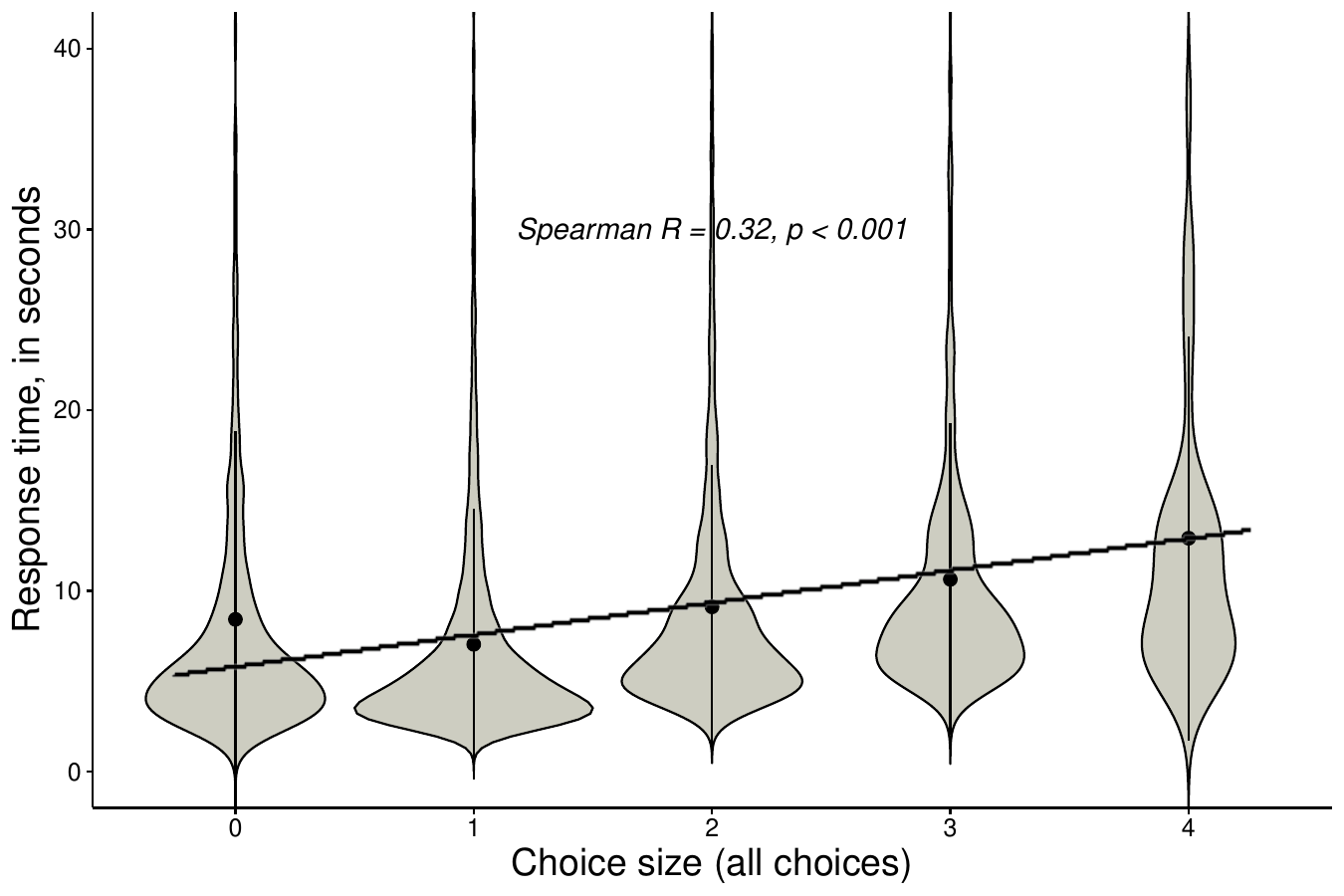}
\end{subfigure}
\caption*{\raggedright 
\scriptsize
\centering Note: The violin plots are thicker (thinner) 
in regions with more (fewer) observations, 
and also show the conditional average response times and 95\% 
confidence intervals per menu/choice size.}
\label{fig:spearman_correlations_times}
\end{figure}

Notably, the average response time in both treatments 
was always lowest ($\approx$ 6 secs) 
when subjects chose a \textit{single} 
pair of gift cards, while it increased monotonically 
(and statistically significantly) 
in similar ways as the size of 
choices and menus increased (Figure \ref{fig:spearman_correlations_times}). 
These novel empirical facts might be seen as intuitive 
evidence suggesting that the subjects' decision was easier 
at menus where they chose a single gift-card pair, 
perhaps because that was their clearly preferred one. 
The latter explanation would be in line with the one often 
mentioned in the literature of \textit{two-alternative forced-choice 
decisions} via sequential-sampling processes such as the \textit{drift-diffusion 
model} and extensions \citep{ratcliff&mckoon08,alos-ferrer-fehr-netzer},
although the novelty of identifying a possibly structure for the 
evolution of this process in \textit{free- and non-binary} choices 
would remain. 

However, the findings are also consistent with an alternative 
interpretation whereby the time was shorter in those 
cases because subjects only had to move the mouse cursor 
on their computer to check a single choice box on their screen, 
thereby mechanically resulting in a shorter 
response time than when multiple items were selected 
from relatively larger menus. This argument is less 
relevant in the case of deferral decisions, however, 
where the average response time increases monotonically 
from (approximately) 6 to 10 to 13 secs when the menu 
includes 2, 3 and 4 alternatives, respectively. 
This suggests that the decision to defer was not 
generally based on some menu-irrelevant strategy 
(e.g. to reach the end of the experiment quickly), 
but was influenced instead by the relevant menu's composition 
and, presumably, the subjects' preferences at that menu. 
Finally, there is no significant difference in the 
distributions of response times in the subjects' active choices 
across treatments ($p=0.612$; two-sided Mann-Whitney $U$ test).

\begin{figure}[!htbp]
\centering
\caption{\centering Distributions of the average choice 
proportions and menus where subjects 
chose everything or deferred.\vspace{10pt}}
\label{fig:basics}
\begin{subfigure}[b]{0.47\textwidth}
\centering
\caption{Forced-Choice treatment}
\includegraphics[width=0.99\textwidth]{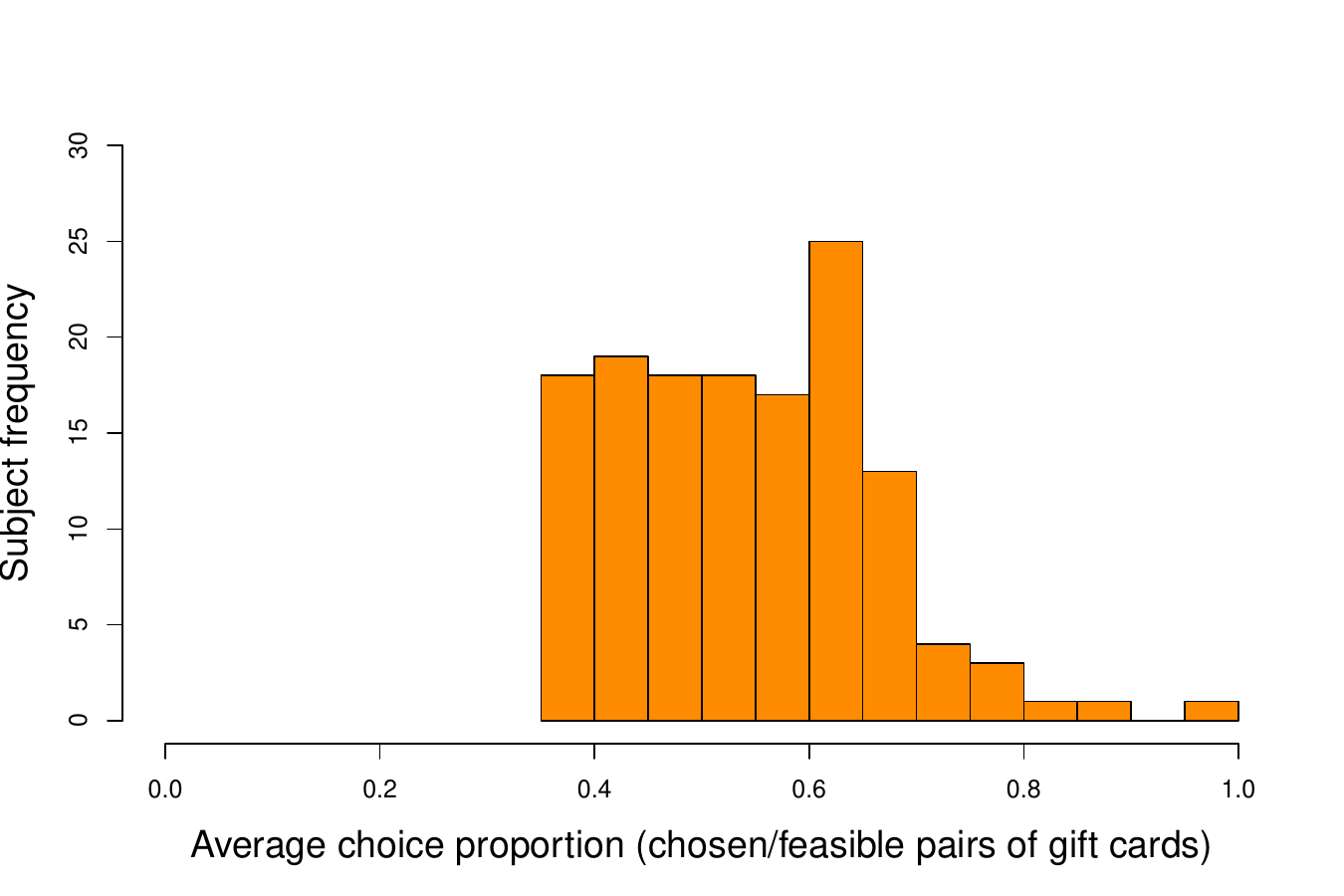}
\label{fig:choice_proportion_distribution_fc}
\end{subfigure}\hspace{15pt}
\begin{subfigure}[b]{0.47\textwidth}
\centering
\caption{Free-Choice treatment}
\includegraphics[width=0.99\textwidth]{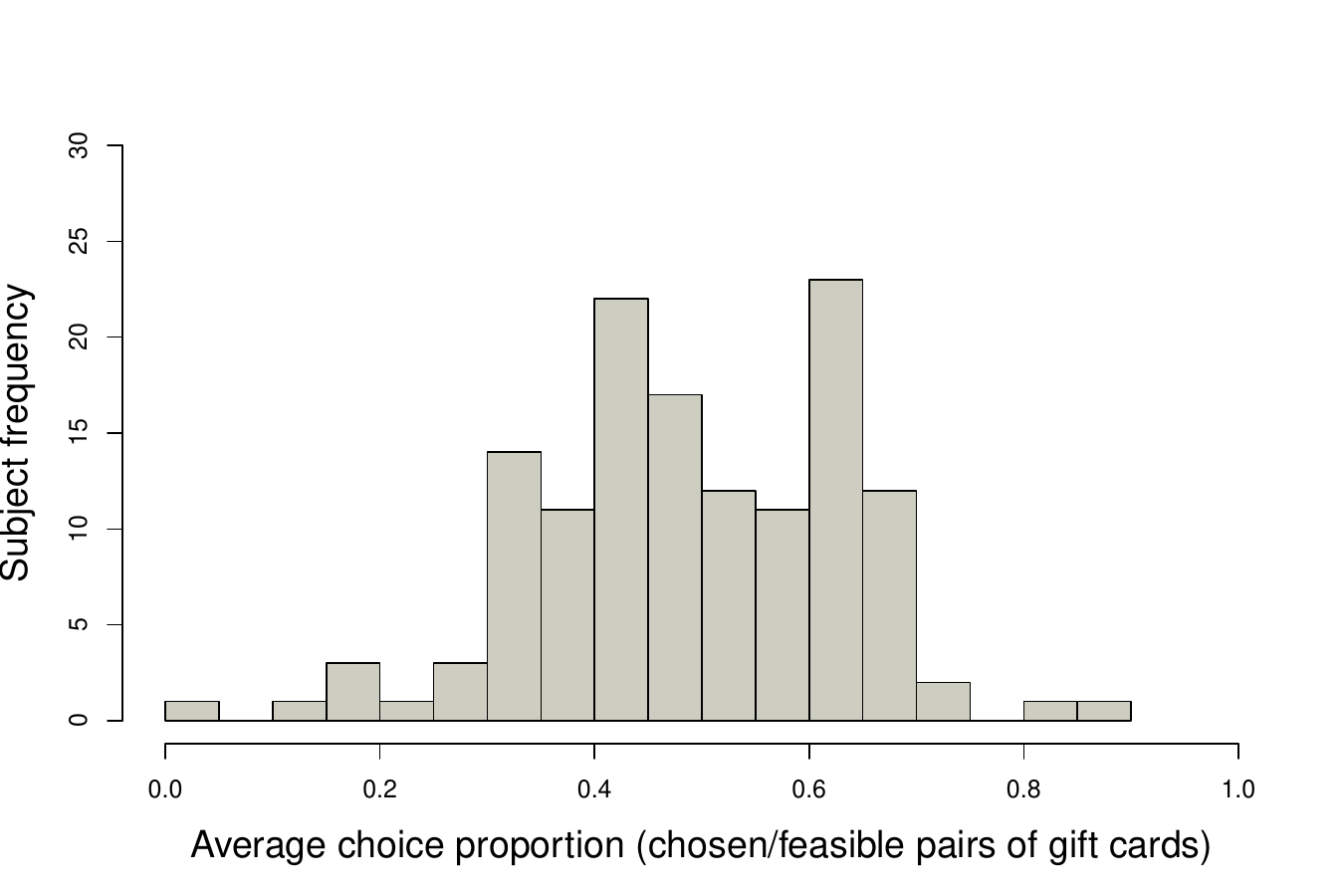}
\label{fig:choice_proportion_distribution_nfc}
\end{subfigure}
\begin{subfigure}[b]{0.47\textwidth}
\centering
\includegraphics[width=0.99\textwidth]{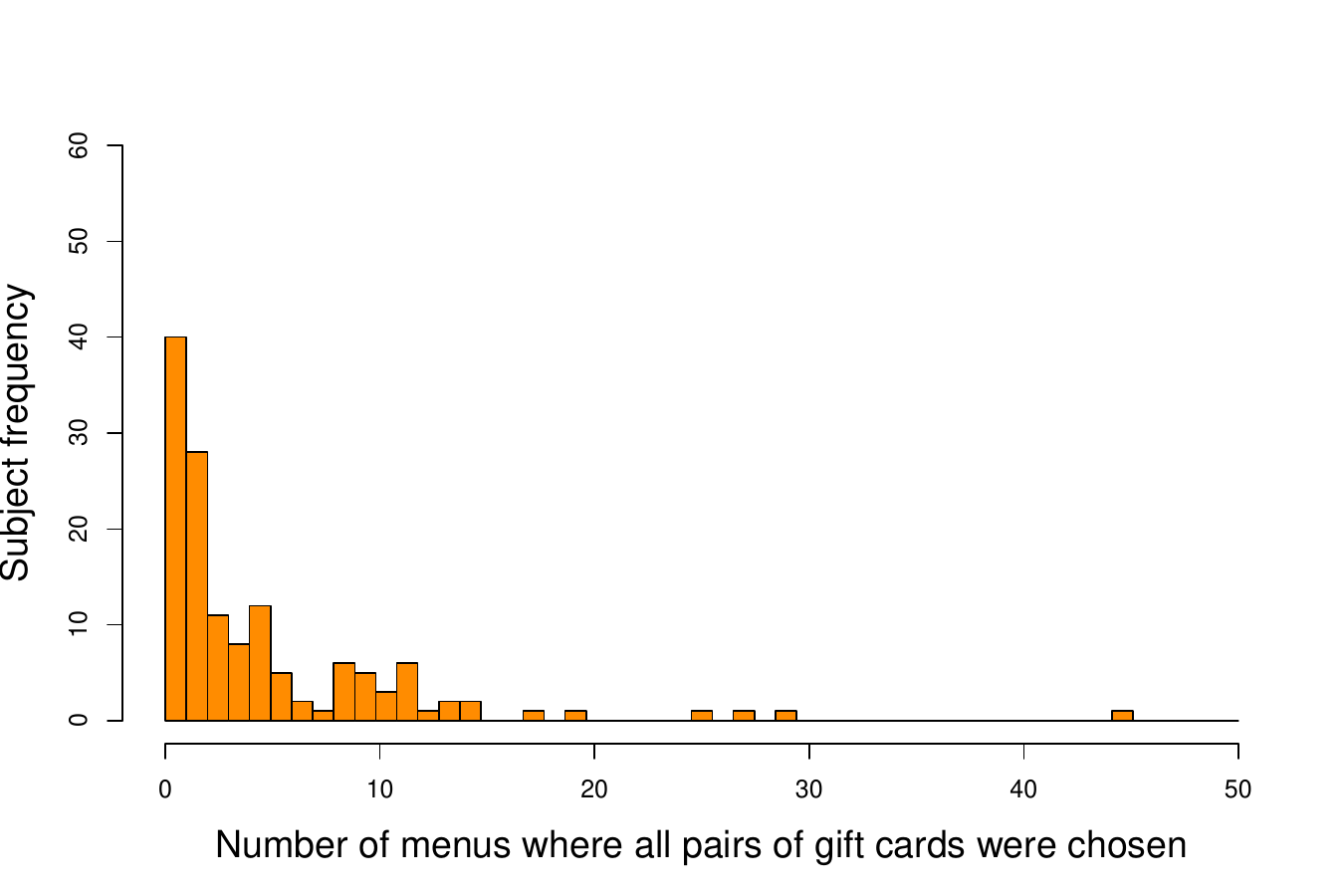}
\label{fig:choice_proportion_one_fc}
\end{subfigure}\hspace{15pt}
\begin{subfigure}[b]{0.47\textwidth}
\centering
\includegraphics[width=0.99\textwidth]{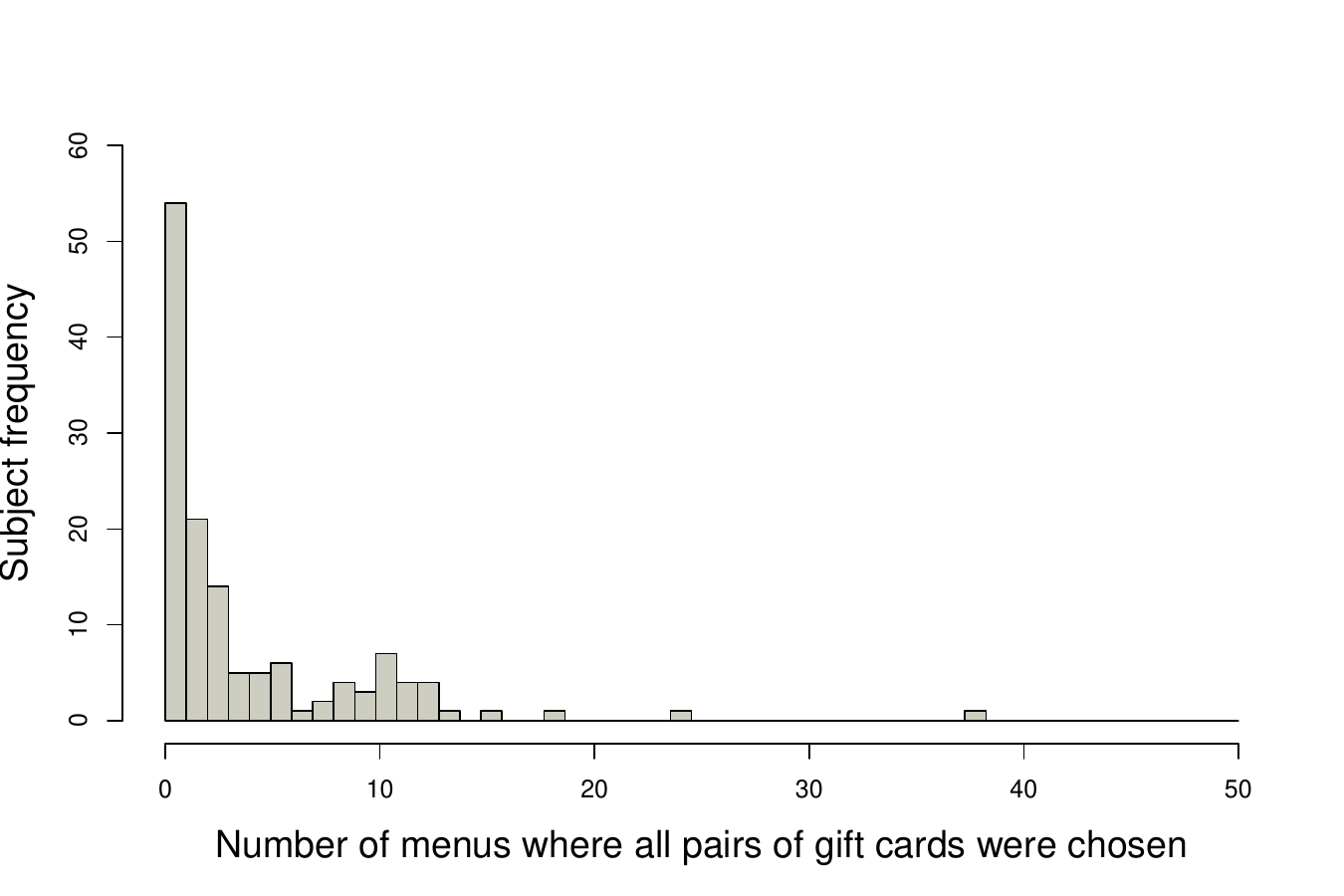}
\label{fig:choice_proportion_one_nfc}
\end{subfigure}
\begin{subfigure}[b]{0.47\textwidth}
\centering
\caption*{}
\end{subfigure}\hspace{15pt}
\begin{subfigure}[b]{0.47\textwidth}
\centering
\includegraphics[width=0.99\textwidth]{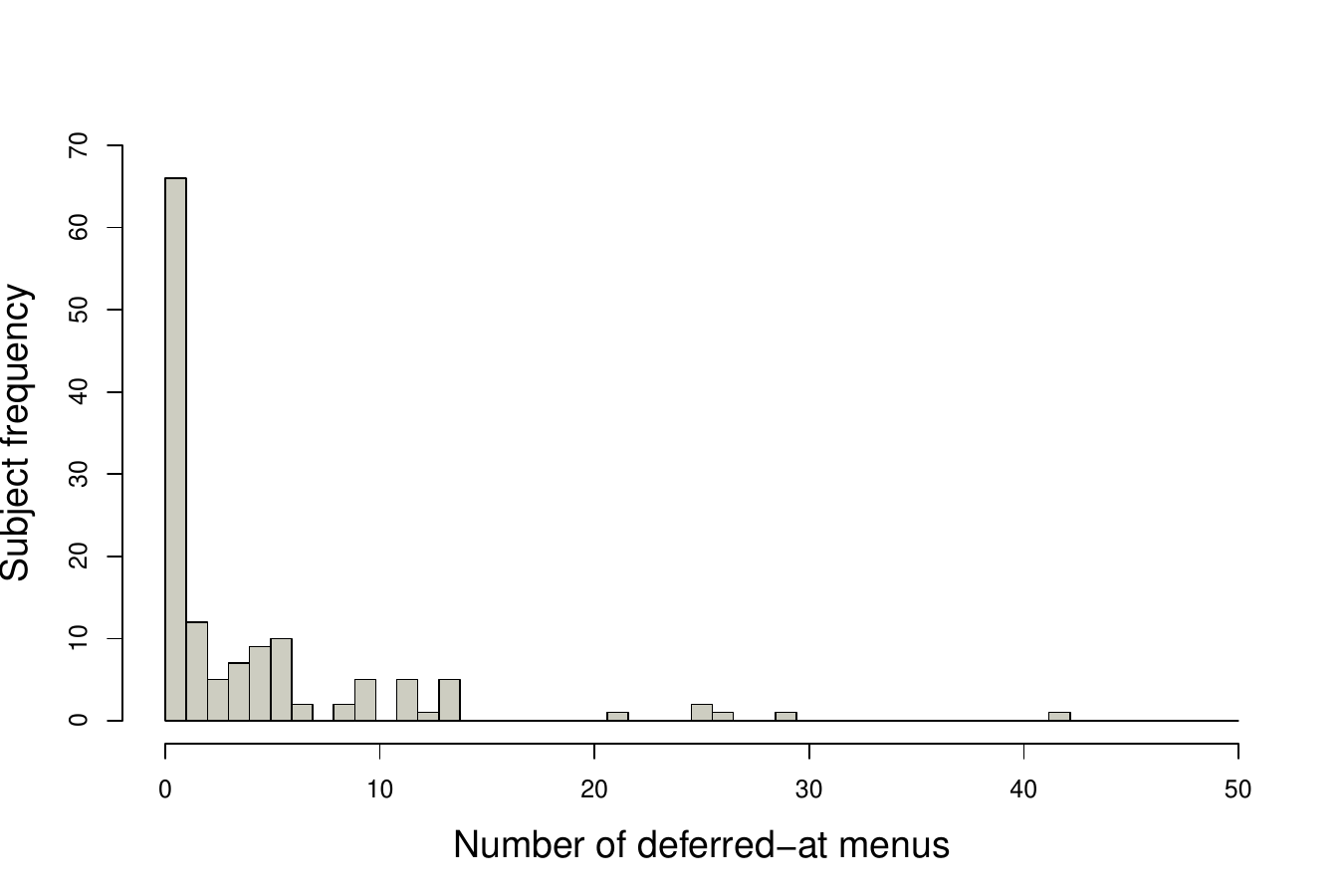}
\label{fig:deferrals_distribution_nfc}
\end{subfigure}
\caption*{\raggedright \scriptsize Notes: Unless otherwise noted, 
all $p$-values below are from 
two-sided Mann-Whitney $U$ tests. The distribution of average 
choice proportions (top panel) 
is significantly different between treatments, 
both when deferrals are included ($p=0.006$) and 
when they are not ($p<0.001$; figure not shown). 
The per subject distribution of menus where 
everything was chosen (middle panel) is not 
significantly different between treatments ($p=0.127$), 
although the rate at which subjects exhibited 
such behaviour is higher in the Forced-Choice 
treatment (8.4\% vs 6.9\%; $p<0.001$; 
two-sided Fisher's exact test), and so is the 
proportion of subjects who did so at least once (71\% vs 60\%; $p=0.057$; 
two-sided Fisher's exact test). 
See also main text. In the Non-Forced-Choice 
treatment the distribution of the number 
of menus where everything was chosen 
is not significantly different from the corresponding distribution where 
subjects deferred (bottom right panel; $p=0.498$).}
\end{figure}

\subsection{Choice Proportions}

Next, we define a subject's \textit{choice proportion} 
at a menu as the number of chosen alternatives divided by the 
number of feasible alternatives at that menu. 
The distributions of average choice proportions across the two 
treatments are shown in Figure \ref{fig:basics} 
(top panel) and are significantly different (higher in the Forced-Choice 
treatment) both when deferrals are included 
and when they are not ($p=0.006$ and $p<0.001$, respectively; 
two-sided Mann-Whitney $U$ tests). 
The mean/median choice proportions across all subjects in the Forced- and 
Non-Forced-Choice treatments were 0.54/0.5 and 0.49/0.5, respectively. 
That is, subjects in both treatments 
tended to choose around half of the feasible alternatives, on average.

Additionally, 69 subjects (51.1\%) in the Free-Choice 
treatment deferred at least once, with deferrals 
per subject ranging between 1 and 42 
(see absolute subject frequencies in 
Figure \ref{fig:basics}-(b); bottom panel), and with a mean/median 
occurrence of 7.28/5 menus. Moreover, 
81 subjects in this treatment (60\%) chose all feasible gift-card pairs 
at least once, with such occurrences 
per subject ranging between 1 and 38 menus 
(Figure \ref{fig:basics}-(b); middle panel), 
and with a mean/median occurrence of 3.42/1 menus. 
By contrast, in the Forced-Choice treatment there were 98 subjects (71\%) 
who chose everything at least once, 
with occurrences per subject ranging 
between 1 and 45 (Figure \ref{fig:basics}-(a); middle panel), 
and with a mean/median of 4.22/2 menus. 
The difference in the two proportions is borderline 
(in)significant at the 5\% level ($p=0.058$; 
two-sided Fisher's exact test). 
These findings lend further support to the possibility that, 
unable to delay their choice 
when faced with a hard decision, 
subjects are more likely to choose everything.

\subsection{Choice Consistency}

For this analysis --which is summarized in Table \ref{tab:consistency}-- 
we compute and compare the JHM scores 
of subjects' active choices under Utility Maximization 
while accounting for the possibility of non-trivial indifferences.
That is, we compare subjects' single- and/or multi-valued choices 
--ignoring any deferral decisions-- that are compatible with 
indifference-permitting utility maximization. 
First, we focus on the distributions of the JHM scores across the 
two treatments. Their mean (5.38 v 4.43), median (4.96 v 3.50) 
and standard deviation (4.02 v 3.75) are uniformly 
and significantly ($p=0.045$) lower in the Free-Choice
than in the Forced-Choice treatment. 
Thus, the number of partially or fully ``mistaken'' --from 
the point of view of utility maximization-- active choices 
is significantly smaller for subjects who were not asked to always
choose some gift-card pair(s). 
This is also reflected in our second test for treatment effects
in active-choice consistency, 
where we compare the proportions of subjects 
who made perfectly (JHM score of 0) or approximately 
(JHM score less than or equal to 1, 2, 3, 4 or 5) 
consistent active choices in the two treatments. 
There were approximately 4\% and 13\% perfectly consistent 
subjects in the Forced- and 
Free-Choice treatments, respectively ($p=0.007$). 
The pattern is similar for approximately consistent subjects,
with 14.5\% v 26\% making up to one fully mistaken choice ($p=0.023$),
27.5\% v 37\% up to two ($p=0.092$), 36\% v 50\% up to three ($p=0.028$),
43\% v 54\% up to four ($p=0.069$) and 51\% v 59\% up to five ($p=0.181$).

\begin{table}[!htbp]
\centering
\footnotesize
\caption{Active-choice consistency in the two treatments.\vspace{5pt}}
\setlength{\tabcolsep}{6pt} 
\renewcommand{\arraystretch}{1.3} 
\makebox[\textwidth][c]{
\begin{tabular}{|l|c|c|c|}
\hline
& \textbf{Forced Choice} & \textbf{Free Choice} & $p$-value\\
\hline
Mean JHM & 5.38 & 4.43 & \multirow{3}{*}{0.045}\\
\cline{1-3}
Median JHM & 4.96 & 3.50 & \\
\cline{1-3}
St.Dev JHM & 4.02 & 3.75 & \\
\hline
JHM $=0$ & 5 & 17 & 0.007\\
\hline
JHM $\leq 1$ & 20 & 35 & 0.023\\
\hline
JHM $\leq 2$ & 38 & 51 & 0.092\\
\hline
JHM $\leq 3$ & 50 & 67 & 0.028\\
\hline
JHM $\leq 4$ & 59 & 73 & 0.069\\
\hline
JHM $\leq 5$ & 70 & 80 & 0.181\\
\hline
$N$ & 138 & 135 & \\
\hline
\end{tabular}
}
\caption*{\centering 
\scriptsize Note: All $p$-values are from two-sided Mann-Whitney (1st)\linebreak
and Fisher's exact (2nd to 7th) tests.}
\label{tab:consistency}
\end{table}

To account for the fact that deferring at a menu 
can never decrease a decision maker's active-choice 
consistency we also carry out subject-specific simulations 
to find how likely it would be for a possibly 
deferring subject to attain their Jaccard-Houtman-Maks score given 
that they made their active choices only at 
those particular menus where they did so. More specifically, 
in line with \cite*{selten91}, 
\cite*{beatty-crawford11} and \cite*{CCGT22}, 
we compute in each treatment the Selten measure of predictive 
success of utility maximization on subjects' active choices. 
We do so as follows:
\begin{enumerate}
\item[(i)] for every experimental subject, create 10,000 
datasets from as many artificial subjects who 
were restricted to make uniform-random active choices 
only at those menus where the experimental subject did so;
\item[(ii)] find the proportion of such subject-specific 
simulated datasets within this block that have a zero 
Jaccard-Houtman-Maks score when utility maximization 
under both strict and weak preferences are accounted for. 
Then, from the proportion, $p_i$, of perfectly consistent 
human subjects in treatment $i$, subtract the 
average proportion, $a_i$, of artificial 
subjects who are also perfectly consistent.
\end{enumerate} 
The closer the difference $m_i:=p_i-a_i$ is to 1, 
the higher the proportion of subjects whose active choices 
were consistent with utility maximization, and the more likely 
it is that this could not have happened randomly. 
Low but positive values on the other hand could arise 
either because (i) relatively many experimental 
subjects were consistent but this could probably 
be due to chance; or (ii) because relatively few subjects 
were consistent and consistency in this environment 
was unlikely to occur by chance. 

It turns out that the latter case applies to the data 
from both our treatments, where we find 
$$
\begin{array}{llllllll}
m_{FC} & = & p_{FC} - a_{FC} & \approx & 0.0362 - 0 & = & 0.0362\\
m_{NFC} & = & p_{NFC} - a_{NFC} & \approx & 0.126 - 0 & = & 0.126
\end{array} 
$$
In line with the comparisons presented in Table \ref{tab:consistency},
however, one may extend this analysis further to the case of 
\textit{approximate} active-choice compliance with utility maximization.
Taking as our approximation threshold the JHM distance score of 5 
that corresponds to 10\% of the maximum value that this score can take,
we find that a similar gap remains in the predictive ``approximate'' success 
of utility maximization between treatments:
$$
\begin{array}{lllllllll}
m^{10\%}_{FC} & = & p^{10\%}_{FC} - a^{10\%}_{FC} & \approx 
& 0.507 - 0 & = & 0.507 \\
m_{NFC}^{10\%} & = & p^{10\%}_{NFC} - a^{10\%}_{NFC} & \approx 
& 0.592 - 0.074 & = & 0.585
\end{array} 
$$

The results from all tests that are presented in this subsection  
document a clear negative effect that the act of 
forcing choice exerts on subjects' choice consistency.
This complements in some important ways the respective finding that 
was originally documented in \cite*{CCGT22}. 
In particular, the results here show that forced choices 
are less consistent even when subjects have to decide from nearly 
twice as many menus and even when the experimental design allows 
them to make multi-valued choices that in turn enables the analyst 
to test for the possibility that their behaviour is well-approximated 
by utility maximization with or without indifferences.

\begin{figure}[!htbp]
\centering
\caption{\centering Choice consistency is negatively 
correlated with average response times in both treatments.\vspace{10pt}}
\label{fig:corr_times_hm}
\begin{subfigure}[b]{0.47\textwidth}
\centering
\caption{\footnotesize Forced-Choice treatment}
\includegraphics[width=0.99\textwidth]{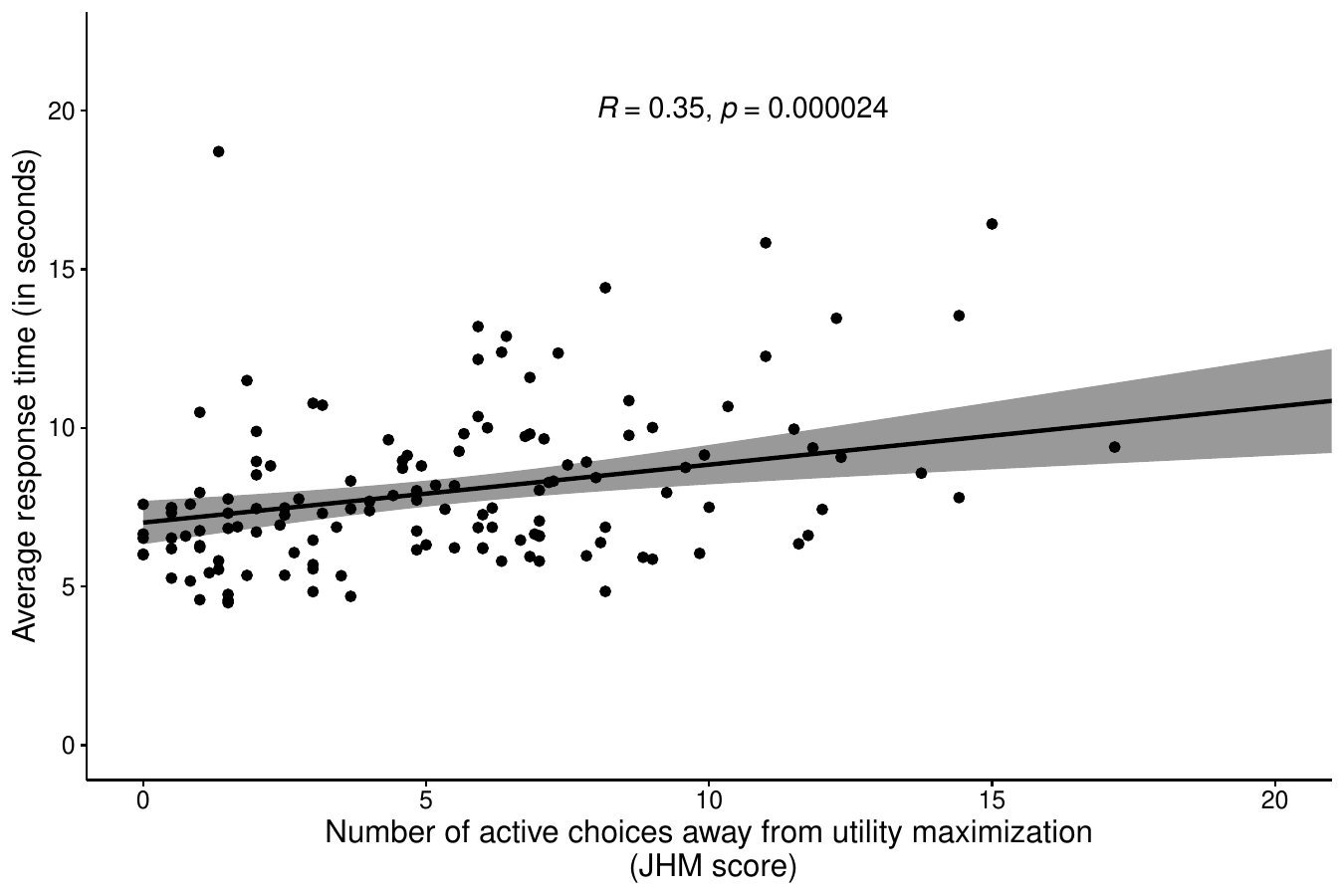}
\label{fig:corr_times_hm_fc}
\end{subfigure}\hspace{15pt}
\begin{subfigure}[b]{0.47\textwidth}
\centering
\caption{\footnotesize Free-Choice treatment}
\includegraphics[width=0.99\textwidth]{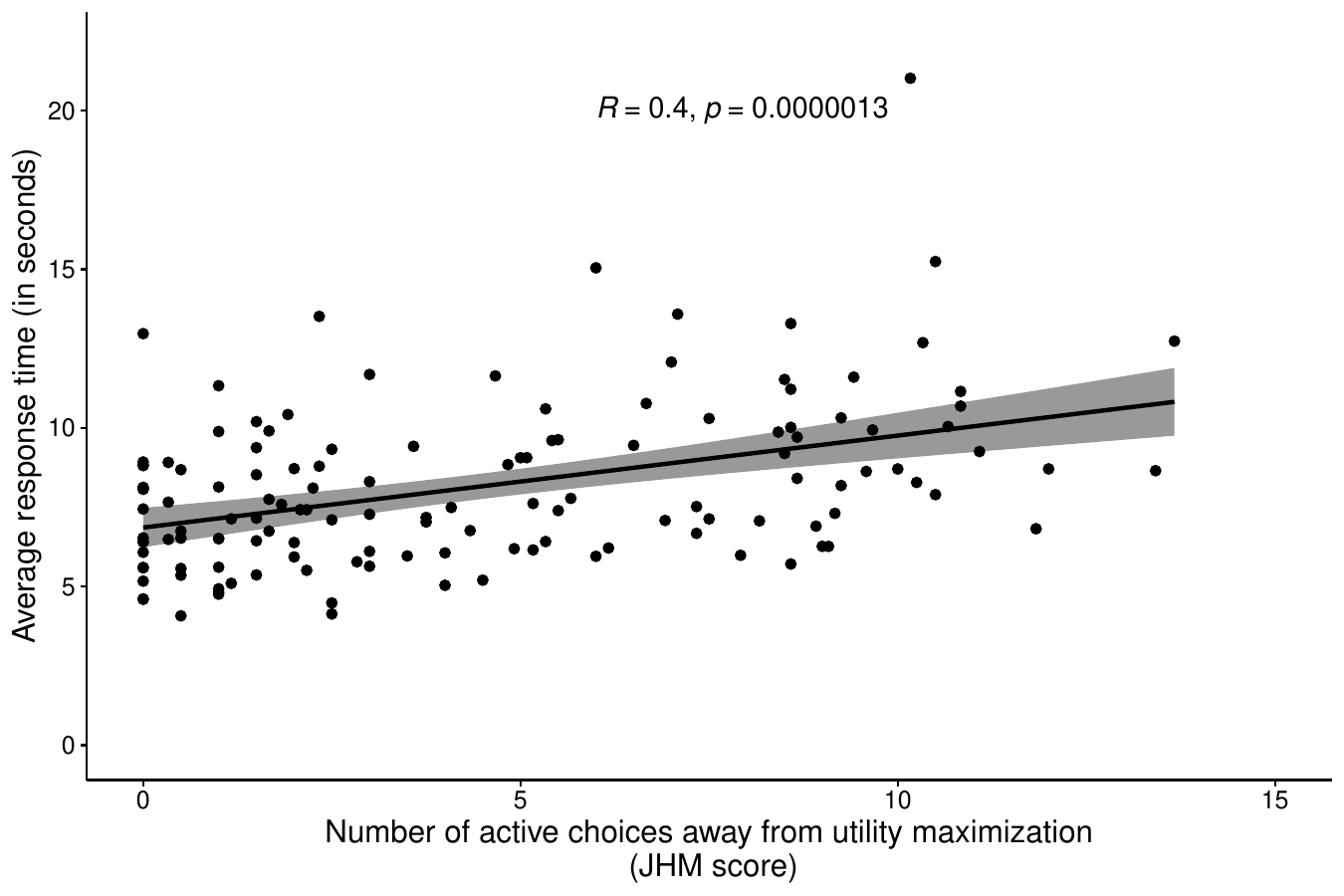}
\label{fig:corr_times_hm_nfc}
\end{subfigure}
\begin{subfigure}[b]{0.47\textwidth}
\centering
\includegraphics[width=0.99\textwidth]{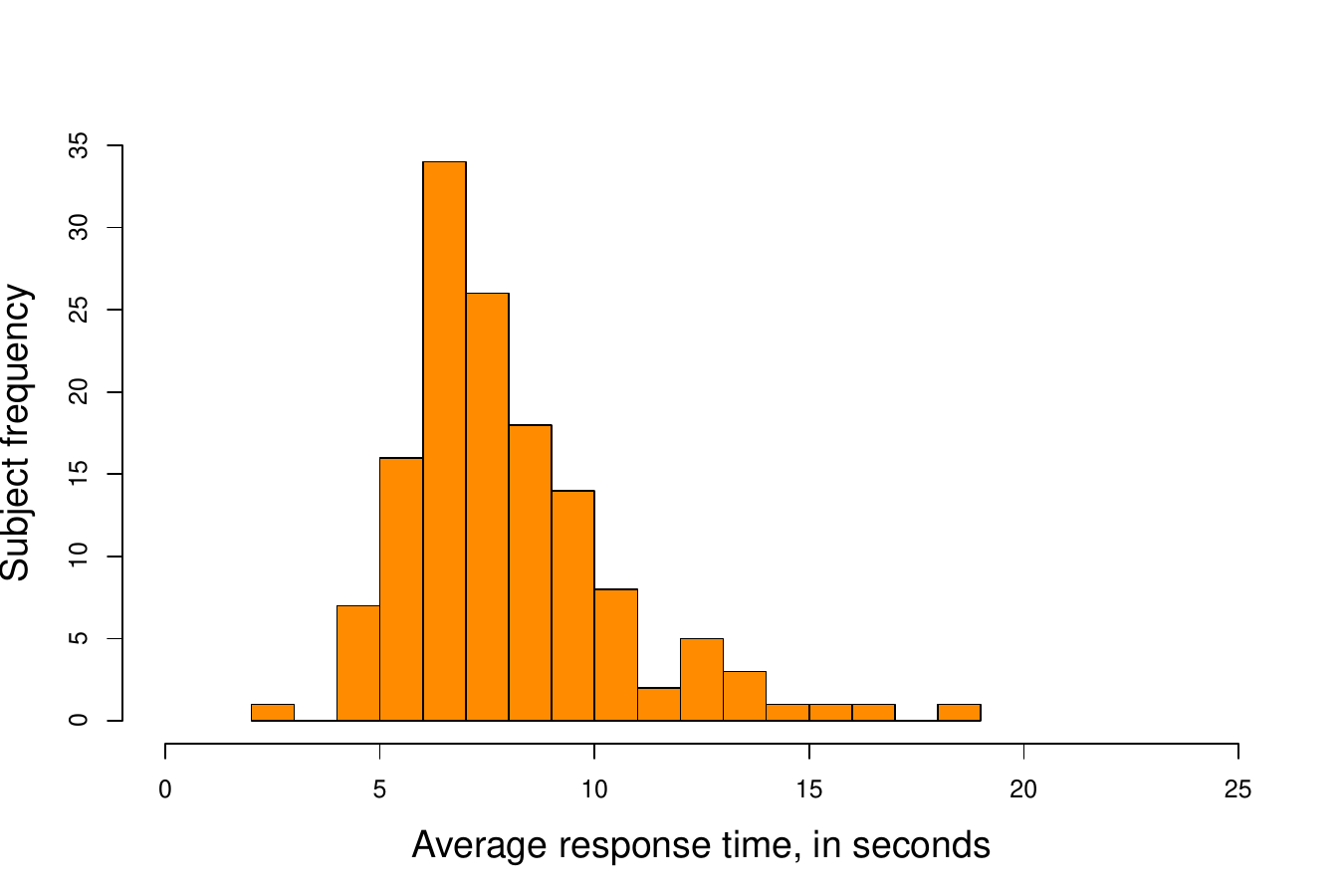}
\label{fig:times_distribution_fc}
\end{subfigure}\hspace{15pt}
\begin{subfigure}[b]{0.47\textwidth}
\centering
\includegraphics[width=0.99\textwidth]{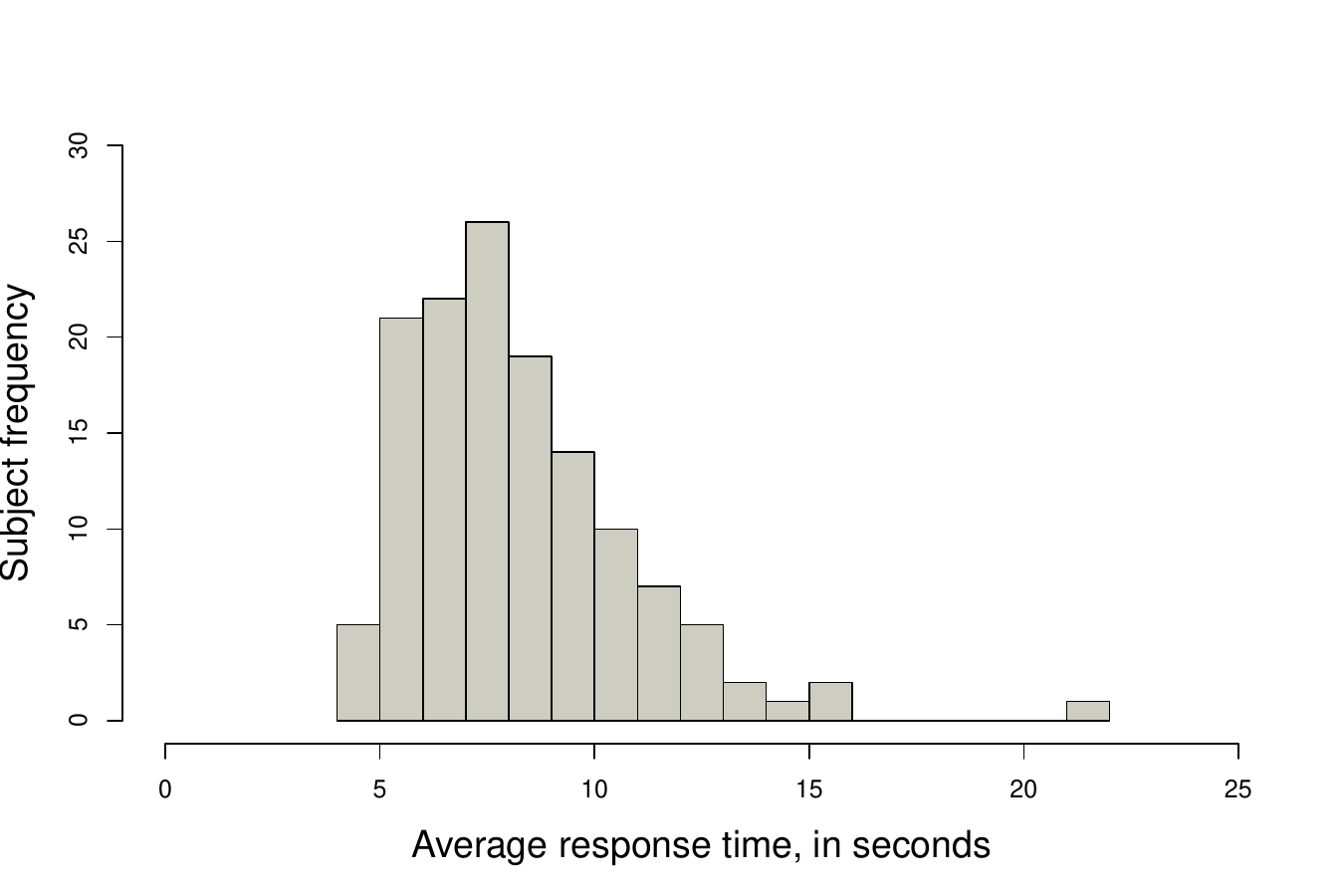}
\label{fig:times_distribution_nfc}
\end{subfigure}
\caption*{\raggedright \scriptsize  \centering
Notes: $R$ is the Spearman 
coefficient and $p$ is the $p$-value. 
Shaded areas indicate 95\% confidence intervals. 
There is no significant difference in the average 
active-choice response times between treatments 
($p=0.550$; 2-sided Mann-Whitney test).}
\end{figure}

We also ask how a subject's choice consistency, 
as captured by the above Jaccard-Houtman-Maks index, 
is related to their response times. 
Figure \ref{fig:corr_times_hm} (top panel) shows 
that there is a significant 
--and very similar in size, with a Spearman coefficient
of 0.35 and 0.45-- 
\textit{positive} correlation in both the forced- and 
free-choice treatments between subjects' 
active-choice Jaccard-Houtman-Maks indices and their 
average response times. 
That is, choice consistency is \textit{negatively} 
correlated with the time it takes for subjects to make their 
active choices. 
This interesting finding is broadly 
consistent with the important prediction 
of the drift-diffusion model mentioned earlier,  
which predicts that shorter response times are more likely 
when a clearly preferred alternative exists, 
which would imply in turn more consistent active-choice behaviour.
The novelty here in this respect is that such a finding is 
documented in a rich dataset that comprises both binary and 
non-binary menus.
Another possible explanation for this finding is 
that subjects with a higher cognitive ability 
(an unmeasured variable in this study) are both 
more consistent and faster than subjects with 
a lower cognitive ability. 
A distinct possible explanation is that spending 
more time before deciding is more characteristic of people who 
are prone to second thoughts and therefore more 
likely to be involved in choice reversals when presented 
with a series of decision problems. Conversely, 
it is also possible that subjects who had neither a stable 
preference relation over the choice alternatives 
nor a clear decision rule ended up spending more time 
on average at each menu, but without this extra 
time ultimately alleviating the effects of their ambivalence.

\begin{figure}[!htbp]
\centering
\caption{Choice consistency is negatively 
correlated with the average choice size.\vspace{10pt}}
\label{fig:corr_choice_sizes_hm}
\begin{subfigure}[b]{0.47\textwidth}
\centering
\caption{\footnotesize Forced-Choice treatment}
\includegraphics[width=0.99\textwidth]{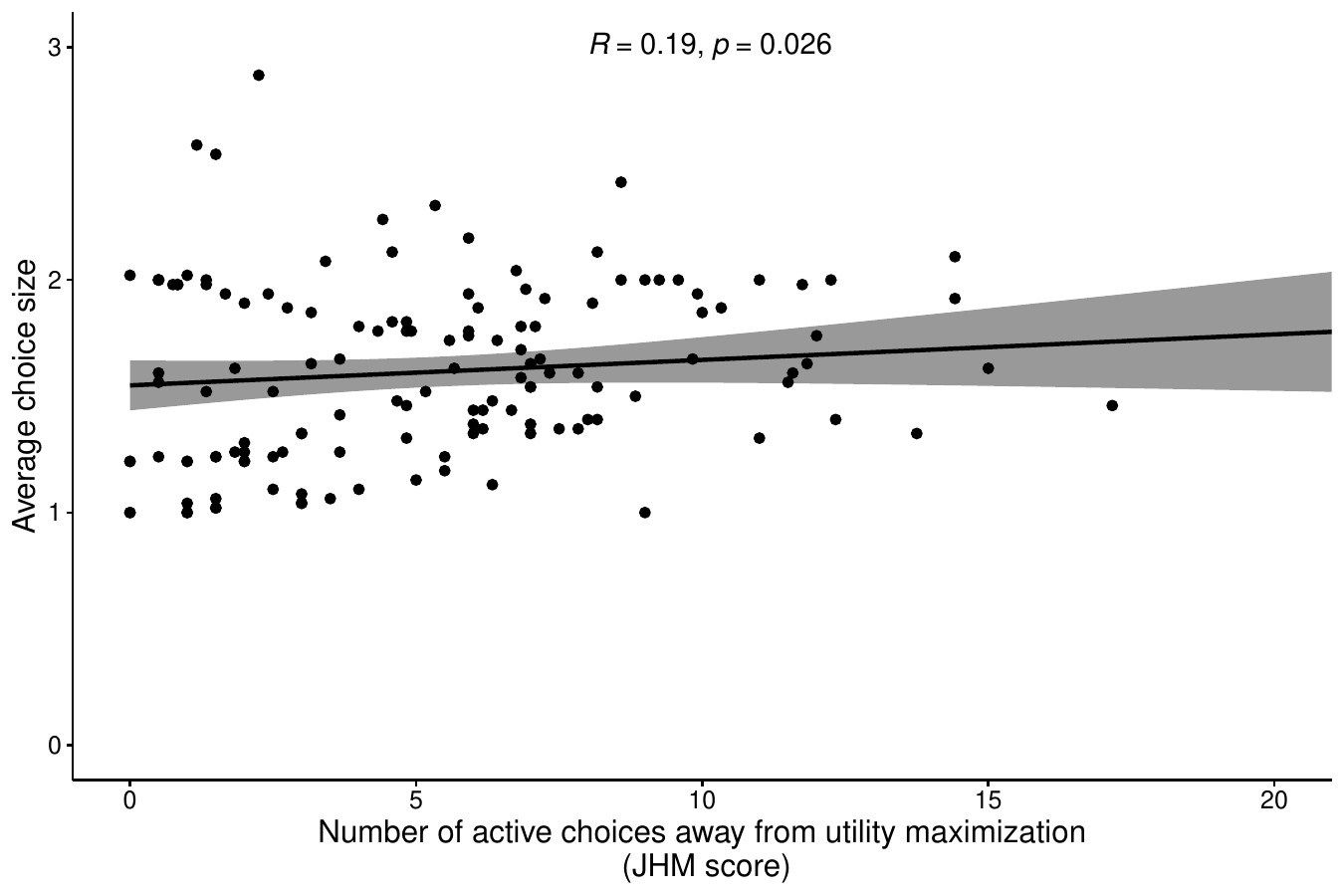}
\label{fig:corr_choice_sizes_hm_fc}
\end{subfigure}\hspace{15pt}
\begin{subfigure}[b]{0.47\textwidth}
\centering
\caption{\footnotesize Free-Choice treatment}
\includegraphics[width=0.99\textwidth]{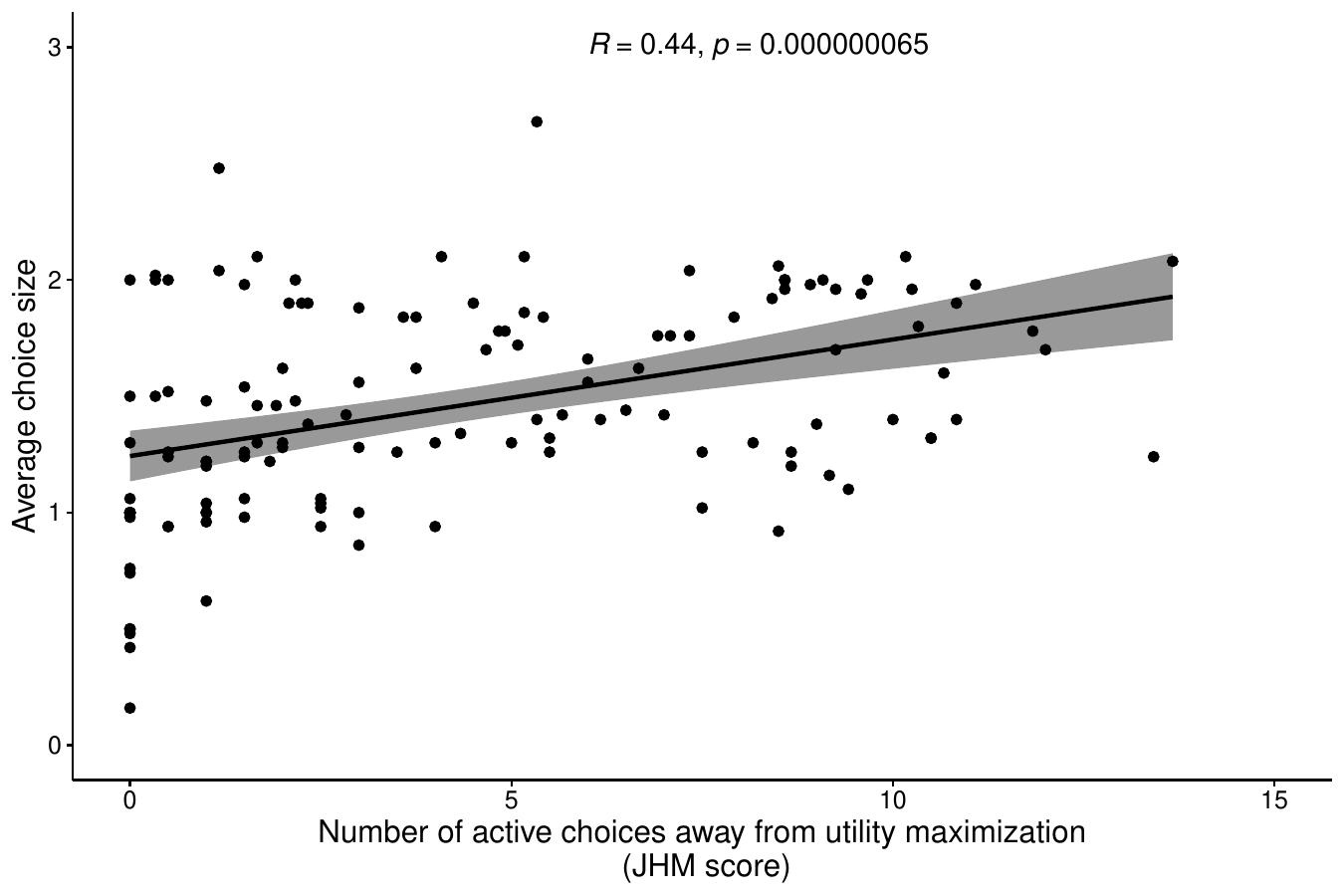}
\label{fig:corr_choice_sizes_hm_nfc}
\end{subfigure}
\caption*{\raggedright\scriptsize Note: $R$ is the Spearman 
correlation coefficient and $p$ is the $p$-value. 
Shaded areas indicate 95\% confidence intervals.}
\end{figure}

Figure \ref{fig:corr_choice_sizes_hm} further shows that there is also a 
significant negative correlation 
between subjects' active-choice consistency and their average choice sizes in 
each of the two treatments, 
with this relationship being more than twice as pronounced for 
Free-Choice subjects ($R=0.44$ vs $R=0.19$ on 
Jaccard-Houtman-Maks index). 
In line with our preceding findings and discussion, 
an intuitive explanation 
for this difference is that, unlike Free-Choice subjects, 
their Forced-Choice counterparts 
could not defer at menus where they might perhaps have wished to do so, 
opting instead for more alternatives per menu. 
But while delaying choice when confronted with a difficult problem safeguards 
the consistency of one's behaviour, choosing more (possibly all) alternatives 
could do the opposite 
because it opens up more possibilities 
for choice reversals/cycles to emerge. 

\begin{figure}[!htbp]
\centering
\caption{\centering
Deferring subjects in the Free-Choice treatment 
are significantly more consistent:\linebreak 
their JHM scores first-order stochastically dominate 
those of non-deferring subjects.\vspace{-25pt}}
\includegraphics[width=0.85\textwidth]{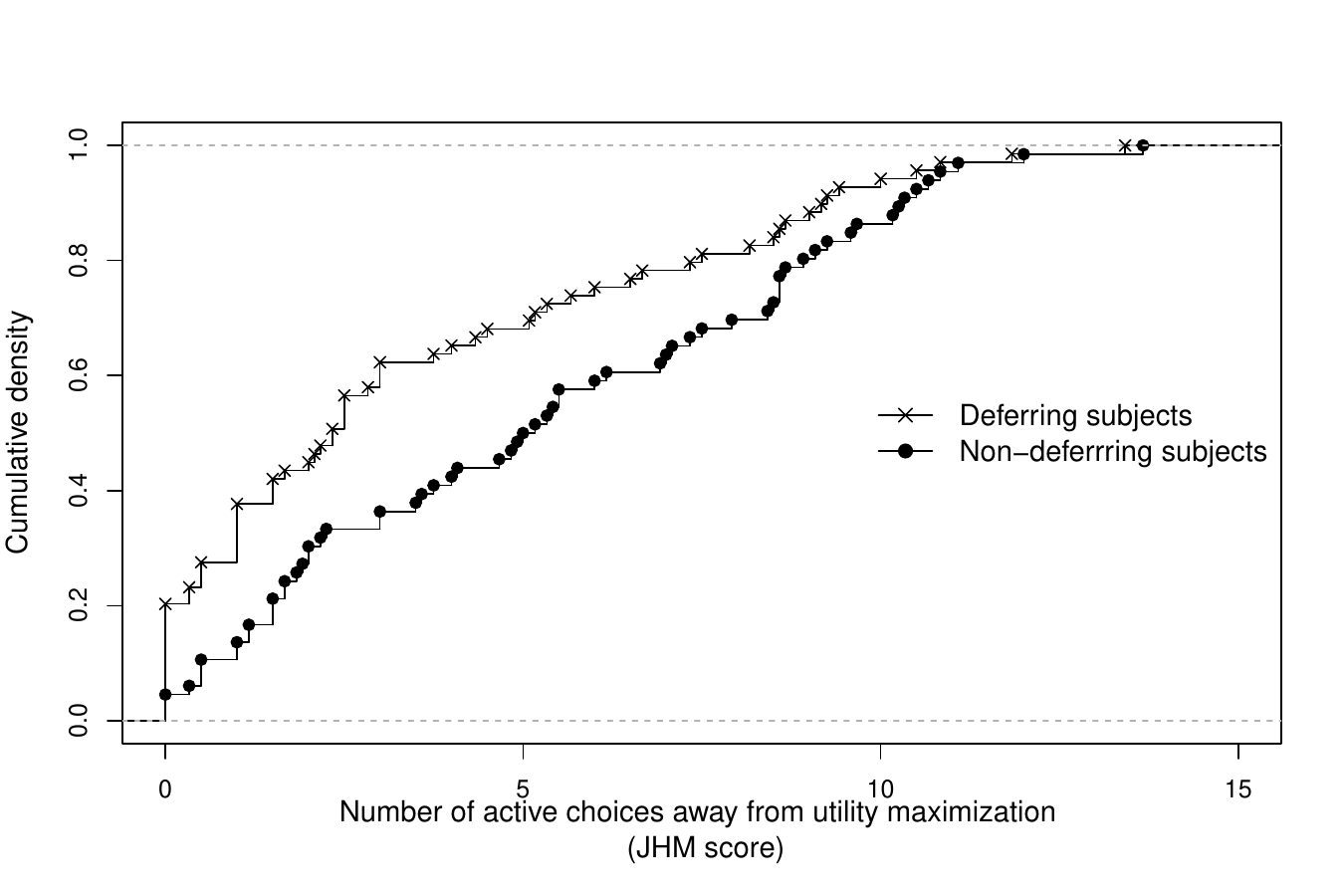}
\label{fig:fosd}
\end{figure}

Some additional support to this explanation, finally, 
is obtained by also comparing 
the active-choice consistency 
of Free-Choice subjects who deferred at least once to the consistency of 
those who did not. 
More specifically, 14 of the 69 (20\%) deferring subjects were perfectly 
consistent in this sense, while 3 of the 66 (4.5\%) 
non-deferring ones were so (equivalently, 14 out of the 17 perfectly 
consistent in this treatment deferred at least once). 
The difference between these two proportions is significant 
($p=0.008$; two-sided Fisher's exact test). 
Moreover, the two groups had an average/median Jaccard-Houtman-Maks score of 
3.58/2.33 and 5.31/5.08, respectively, 
and the difference in the two distributions is significant 
too ($p=0.003$; two-sided Mann-Whitney 
$U$ test). In fact, as shown in Figure \ref{fig:fosd}, 
deferring subjects are uniformly more consistent than non-deferring ones 
in the sense that the distribution of their 
JHM scores first-order stochastically dominates 
the corresponding distribution of the latter subjects. 
That is, for every number less than or equal to $n$, and for every $n$, 
non-deferring subjects were more likely to 
be strictly more than $n$ active choices away from utility maximization. 
This within-treatment effect further highlights the important mediating 
role of deferrals for consistency.

\section{Concluding Remarks}

This paper proposes and implements theory-guided methods of 
data collection and analysis that aim to 
contribute towards eliciting a decision maker's possibly 
weak or incomplete preferences and, where relevant, 
towards distinguishing between their indifference and 
indecisiveness parts. On the data-collection side, 
the paper contributes an incentivized experimental design 
with \textit{multi-valued} forced- and 
free-choice treatments. On the data-analytic side, 
it deploys a model-rich combinatorial-optimization 
method that allows for recovering an individual's 
possibly weak or incomplete preferences from their 
multi-valued choices in a model-optimal way. 
This method is a novel extension of the celebrated \cite*{houtman&maks} 
approach that is fine-tuned to the generally multi-valued nature 
of the choice data by accounting for the (Jaccard) similarity 
between the observed and model-optimal choices, and penalizing 
a subject's proximity score relative to a model accordingly. 
We showed how this method can be fruitfully applied on the general 
model of rational choice with potential indifferences, 
as well as to two models of incomplete-preference maximization 
that allow for --and distinguish between-- indifference and/or incompleteness.
The method is obviously general, however, and in the future can be 
applied on other models of bounded rationality that predict 
multi-valued choices.

Despite the relatively large number of decisions, 
the behaviour of 56\% of all subjects in 
our sample is either perfectly or 
approximately matched by some simple but richly 
structured deterministic model of complete or incomplete 
preference maximization, with uniquely 
recovered and indifference-exhibiting preferences 
in the vast majority of cases. Rational choice 
accounts for the behaviour of over 60\% of all subjects 
in this group (34\% of the total), 
while the two models of incomplete-preference 
maximization together account for the remaining subjects. 
Importantly, the optimal preference relation 
that is recovered conditional on a subject's 
best-matching model typically features a non-trivial 
indifference relation that may encompass up to 19\% 
of all possible comparisons, and therefore highlights 
the importance of accounting for indifferences 
in revealed-preference analyses. 

Our analysis points to the usefulness of multi-valued 
and indifference-/incomparability-permitting choice experiments. 
It also highlights the importance of methodologically 
pluralistic methods that allow for recovering preferences 
and/or choice rules by comparing observed behavioural with the 
predictions made by utility maximization 
as well as by models of bounded-rational choice.

\bibliographystyle{ecta}
\bibliography{MultiValued}

\setcounter{page}{0}

\thispagestyle{empty}

\vfill

\renewcommand*{\thepage}{\footnotesize{\arabic{page}}}

\appendix

\section{Experiment Details}

\subsection{Instructions: Non-Forced-Choice Treament (verbatim)}

\vspace{-30pt}

\hspace{-40pt}\includegraphics[scale=0.85]{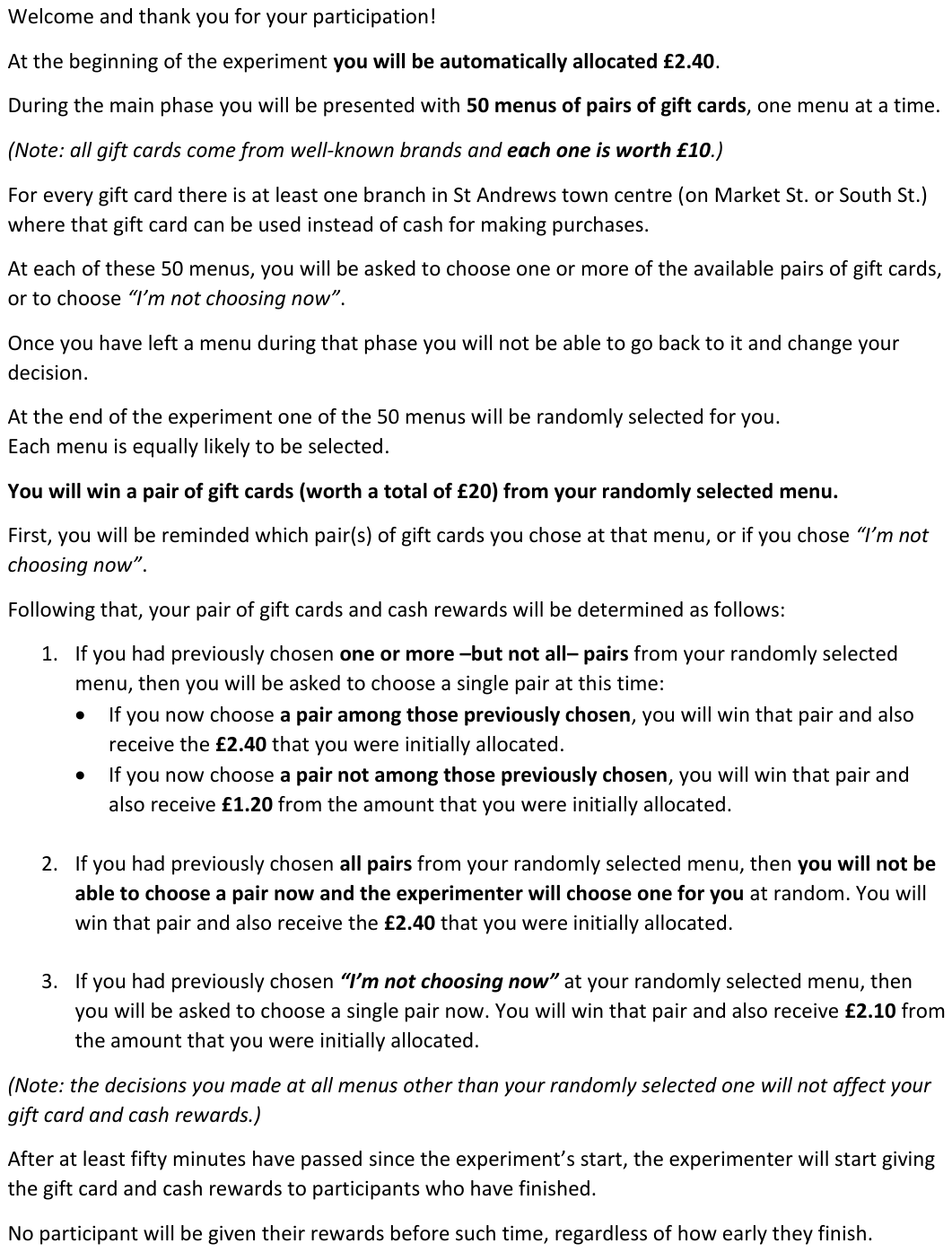}

\subsection{Instructions: Forced-Choice Treament (verbatim)}

\includegraphics[scale=0.85]{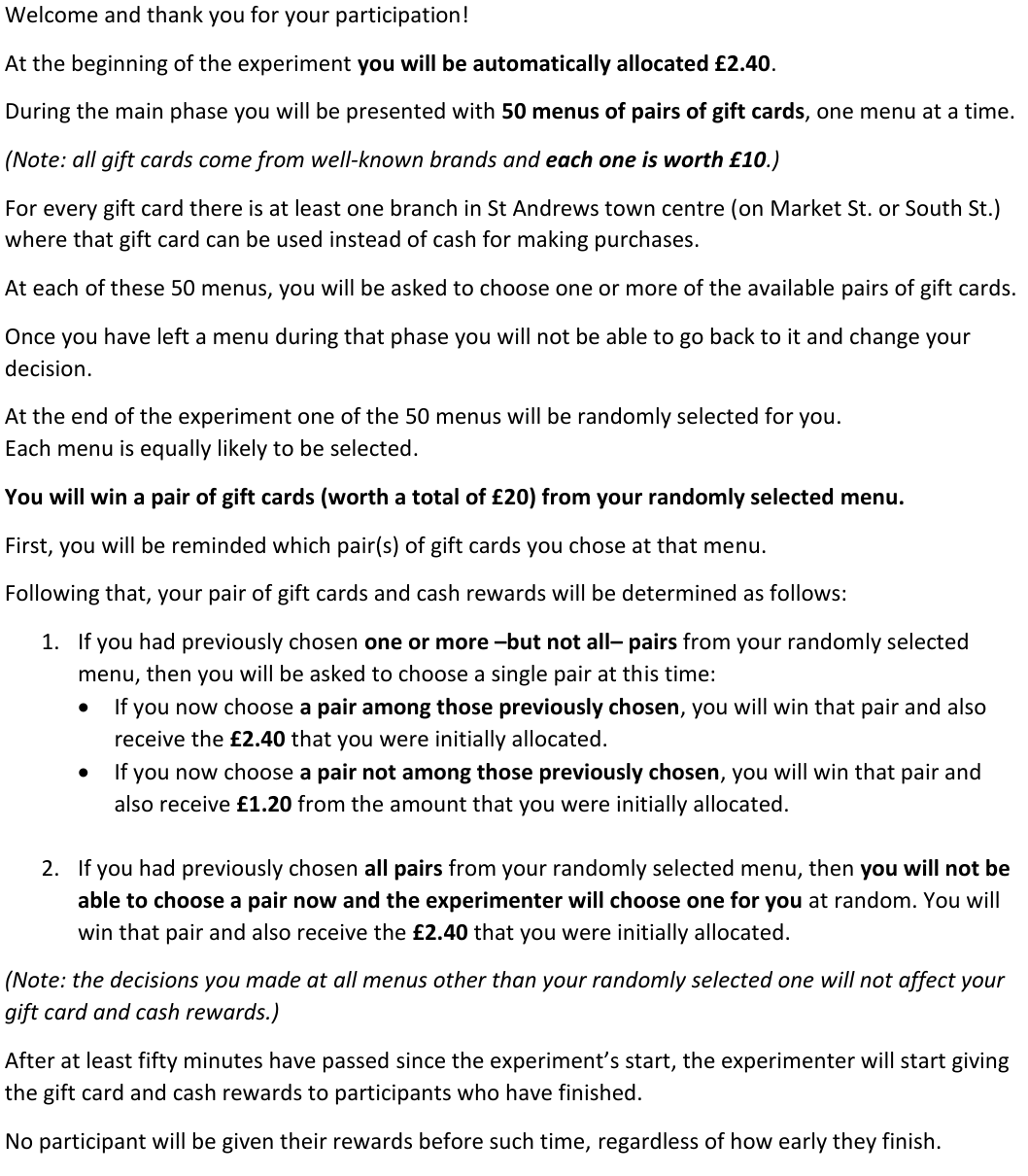}

\pagebreak

\subsection{The six pairs of \pounds 10 gift cards used in the experiment}

\begin{figure*}[!htbp]
	\centering
	\begin{subfigure}[b]{0.46\textwidth}
		\centering
		\includegraphics[width=\textwidth]{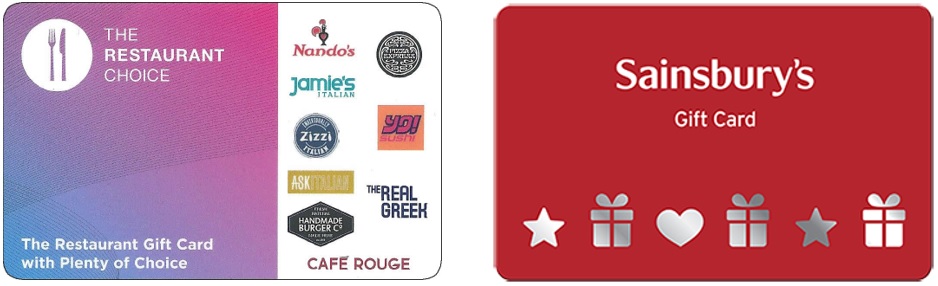}
	\end{subfigure}
	\hfill
	\begin{subfigure}[b]{0.46\textwidth}  
		\centering 
		\includegraphics[width=\textwidth]{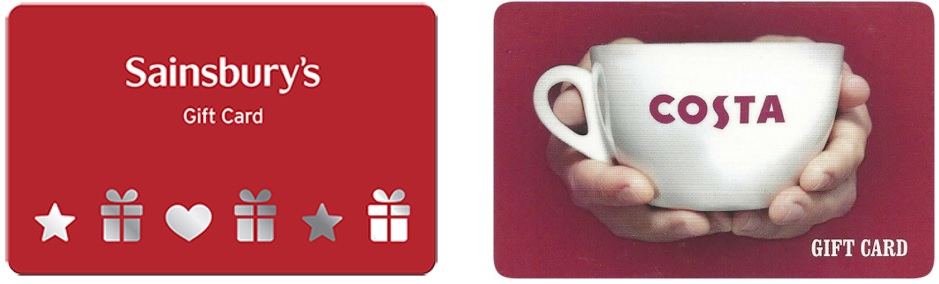}
	\end{subfigure}
	\vskip\baselineskip\vspace{5pt}
	\begin{subfigure}[b]{0.46\textwidth}   
		\centering 
		\includegraphics[width=\textwidth]{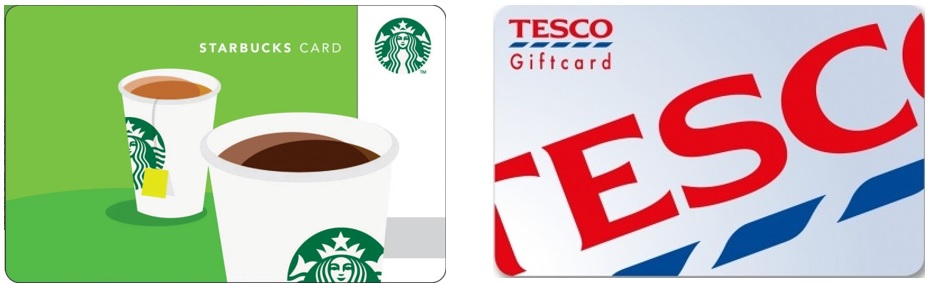}
	\end{subfigure}
	\hfill
	\begin{subfigure}[b]{0.46\textwidth}   
		\centering 
		\includegraphics[width=\textwidth]{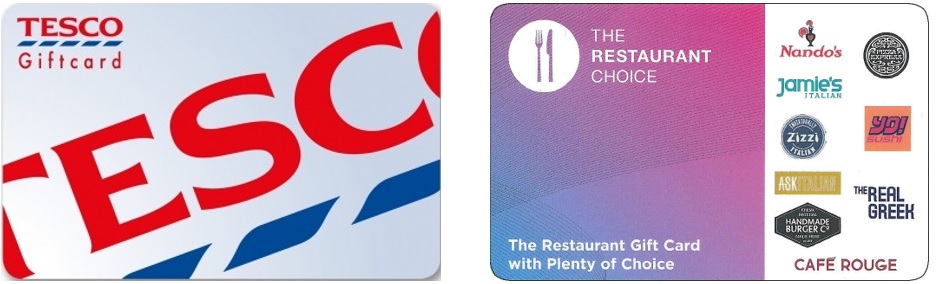}
	\end{subfigure}
	\vskip\baselineskip\vspace{5pt}
	\begin{subfigure}[b]{0.46\textwidth}   
		\centering 
		\includegraphics[width=\textwidth]{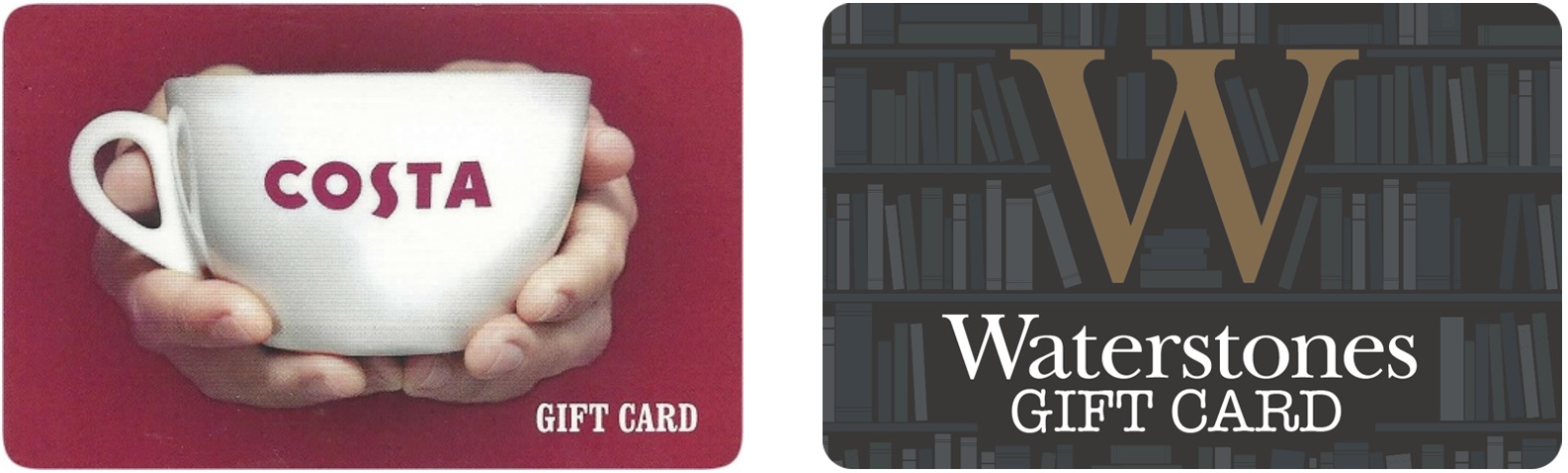}
	\end{subfigure}
	\hfill
	\begin{subfigure}[b]{0.46\textwidth}   
		\centering 
		\includegraphics[width=\textwidth]{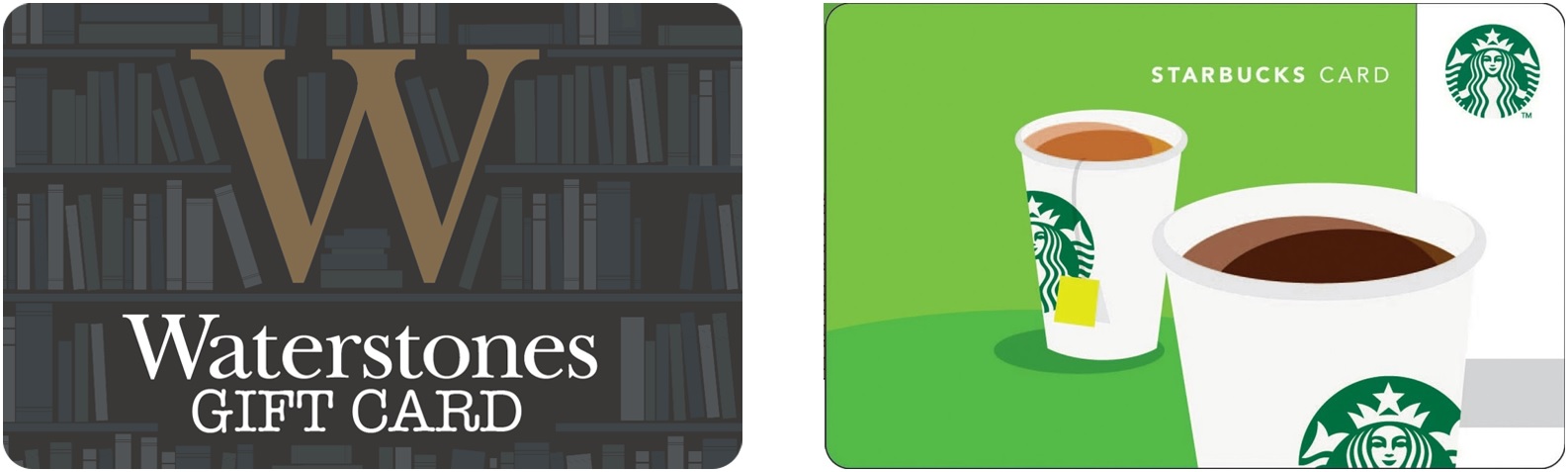}
	\end{subfigure}
\end{figure*}

\subsection{Example decision problem shown in the main part*}

\begin{center}
	\includegraphics[scale=0.75]{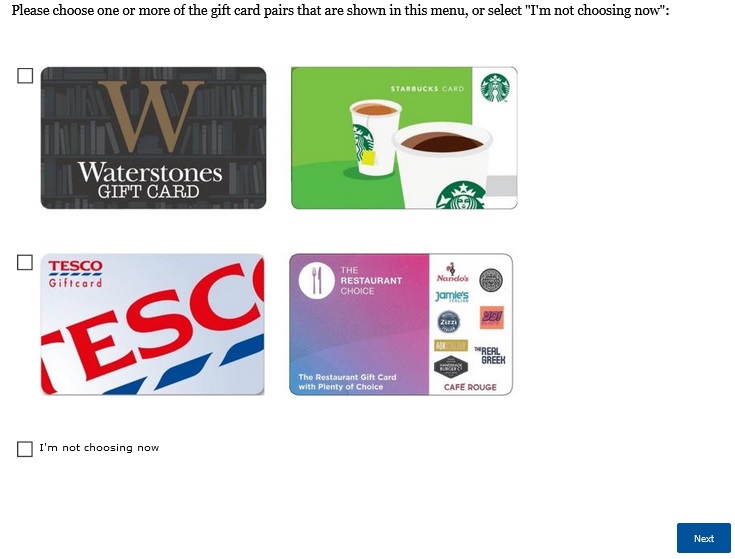}
\end{center}

\vspace{-15pt}

{
	\scriptsize \noindent * This example is taken from the Non-Forced-Choice treatment. 
	Presentation in the Forced-Choice treatment is identical except 
	that \textit{``I'm not choosing now''} is unavailable and the instruction 
	reads \textit{``Please choose one of more gift cards from this menu''}.	
}

\subsection{Example randomly selected menu shown at the end}

\begin{center}
	\includegraphics[scale=0.85]{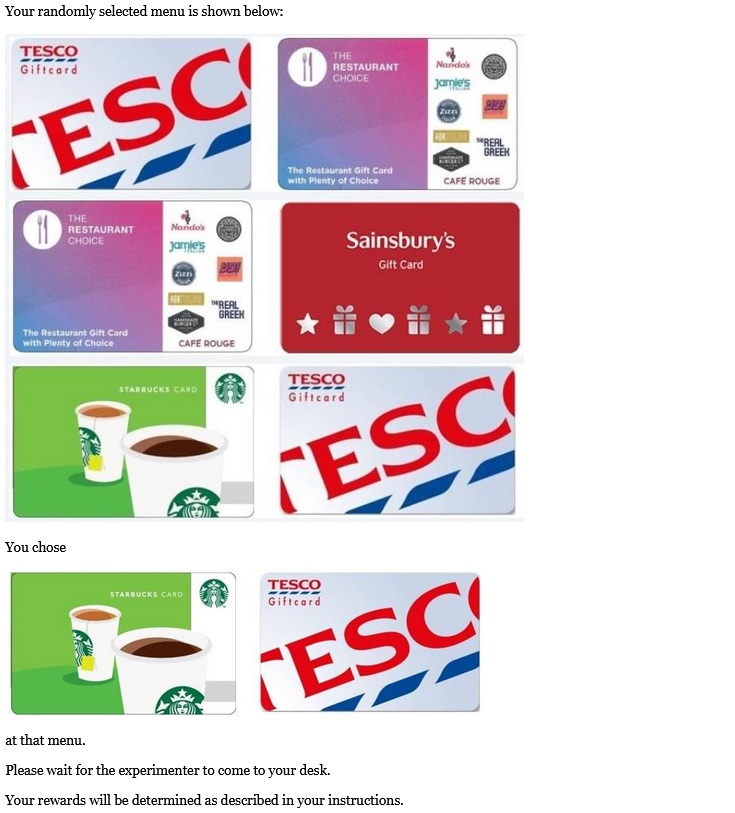}
\end{center}

\pagebreak

\section{Model Goodness-of-Fit Based on the Binary Extension of HM}

\begin{table}[!htbp]
	\centering
	\footnotesize
	\caption{\centering 
		Classification of subjects who are perfectly/approximately 
		explainable by some model under the multi-valued 
		choice extensions of HM that follow \eqref{hm} 
		with $\succsim$ and $\mathcal{R}$/$\mathcal{Q}$ replacing 
		$\succ$ and $\mathcal{P}$.\vspace{10pt}}
	\begin{subtable}{\textwidth}\centering
		\setlength{\tabcolsep}{5pt} 
		\renewcommand{\arraystretch}{1.3} 
		\makebox[\textwidth][c]{
			\begin{tabular}{|l|rl|rl|rl|rl|}
				\hline
				&& 						& \multicolumn{2}{c}{\textbf{Undominated}} & \multicolumn{2}{|c|}{\textbf{Dominant}} 	&& \\
				&\multicolumn{2}{c|}{\textbf{Utility}} 	&\multicolumn{2}{c}{\textbf{Choice with}} & \multicolumn{2}{|c|}{\textbf{Choice with}}&& \\
				\multicolumn{1}{|c|}{}	&\multicolumn{2}{c|}{\textbf{Maximization}}	    &\multicolumn{2}{c}{\textbf{Incomplete}} & \multicolumn{2}{|c|}{\textbf{Incomplete}}	& \multicolumn{2}{c|}{\multirow{1}{*}{\textbf{All}}}\\
				&&						& \multicolumn{2}{c}{\textbf{Preferences}} & \multicolumn{2}{|c|}{\textbf{Preferences}}&&  \\
				\hline 
				& \multicolumn{8}{c|}{\textbf{Forced-Choice treatment} ($N=138$)}\\
				\hline 					
				\% of subjects with score $=$ 0 	& \multicolumn{2}{c|}{3.62\%}	& \multicolumn{2}{c|}{0\%}	&\multicolumn{2}{c|}{0\%}	& \multicolumn{2}{c|}{3.62\%}\\
				\hline
				\% of subjects with score $\leq$ 5 & \multicolumn{2}{c|}{33.33\%}	& \multicolumn{2}{c|}{2.17\%} &\multicolumn{2}{c|}{0\%} & \multicolumn{2}{c|}{35.50\%}\\
				\hline
				\% of subjects with score $\leq$ 10 & \multicolumn{2}{c|}{46.38\%}	& \multicolumn{2}{c|}{8.70\%} &\multicolumn{2}{c|}{0\%} & \multicolumn{2}{c|}{55.07\%}\\
				\hline 
				Mean/median best score ($\leq 10$)	& \multicolumn{2}{c|}{4.06/4}	& \multicolumn{2}{c|}{7.25/8}  &  \multicolumn{2}{c|}{--}	& \multicolumn{2}{c|}{4.56/4}\\			
				\hline			
				Minimum score in simulations	& \multicolumn{2}{c|}{25} & \multicolumn{2}{c|}{23} & \multicolumn{2}{c|}{--}	& \multicolumn{2}{c|}{23}\\
				\hline 
				\multirow{1.5}{*}{Mean/median best-model} & \multicolumn{2}{c|}{\multirow{2}{*}{1.17/1}}  & \multicolumn{2}{c|}{\multirow{2}{*}{1.00/1}}  & \multicolumn{2}{c|}{\multirow{2}{*}{--}}	& \multicolumn{2}{c|}{\multirow{2}{*}{1.15/1}} \\
				\multirow{1}{*}{preference orderings (score $\leq 10$)} & & & & & & & & \\
				\hline
				& \multicolumn{8}{c|}{\textbf{Non-Forced-Choice treatment} ($N=135$)}\\ 
				\hline 
				\% of subjects with score $=$ 0 			& \multicolumn{2}{c|}{2.22\%}	& \multicolumn{2}{c|}{0\%}	& \multicolumn{2}{c|}{5.19\%} 			& \multicolumn{2}{c|}{7.41\%}\\
				\hline
				\% of subjects with score $\leq$ 5 & \multicolumn{2}{c|}{18.51\%}	& \multicolumn{2}{c|}{2.96\%} &\multicolumn{2}{c|}{21.48\%} & \multicolumn{2}{c|}{42.96\%}\\
				\hline
				\% of subjects with score $\leq 10$		& \multicolumn{2}{c|}{24.44\%} & \multicolumn{2}{c|}{5.19\%} & \multicolumn{2}{c|}{28.15\%} & \multicolumn{2}{c|}{57.78\%}\\
				\hline 
				Mean/median best score ($\leq 10$)	& \multicolumn{2}{c|}{4.15/3} & \multicolumn{2}{c|}{5.57/5}	& \multicolumn{2}{c|}{3.39/3}	&  \multicolumn{2}{c|}{3.91/3}\\ 
				\hline			
				Minimum score in simulations & \multicolumn{2}{c|}{27} &  \multicolumn{2}{c|}{26} &  \multicolumn{2}{c|}{18} & \multicolumn{2}{c|}{18} \\
				\hline 
				\multirow{1.5}{*}{Mean/median best-model} & \multicolumn{2}{c|}{\multirow{2}{*}{1.00/1}} &  \multicolumn{2}{c|}{\multirow{2}{*}{1.29/1}} &  \multicolumn{2}{c|}{\multirow{2}{*}{1.08/1}}	&  \multicolumn{2}{c|}{\multirow{2}{*}{1.06/1}} \\
				\multirow{1}{*}{preference orderings (score $\leq 10$)} & & & & & & & & \\
				\hline 
			\end{tabular}
		}
	\end{subtable}
	\label{tab:models-appendix}
	\caption*{\raggedright \scriptsize Notes: Model-score ties were always broken in 
		favour of Rational Choice/Utility Maximization (no other ties emerged). 
		When Rational Choice/Utility Maximization was not a subject's optimal model, 
		the difference between this and the optimal model's extended HM score 
		was on average 4.2 and 6.5 in the Forced- and 
		Non-Forced-Choice treatments, respectively.}
\end{table}

\pagebreak

\section{Analysis of Other Decision Models \& Heuristics}

Although not constructed so as to enable a thorough 
investigation of such behaviours, 
utilizing two important aspects of the experimental 
design and its implementation allows for inferring that an additional 10\% 
of the analysed subjects could be thought of as exhibiting a systematic 
\textit{satisficing} behaviour 
\citep*{simon56PR,caplin-dean-martin,reutskaja-etal11} 
or a \textit{preference for randomization} 
\citep*{agranov&ortoleva,agranov-ortoleva20,
	dwenger-kubler-weizsacker18,cettolin&riedl,agranov-healy-nielsen22}. 
Additionally, while the experiment was not designed to test for this either, 
we also investigated the potential presence of a ``preference for flexibility'' 
over menus  \citep*{kreps79,DLR01}. 
We did so by entertaining the possibility that some 
subjects might have ``metavisualised'' the actual 
experimental task of choosing potentially 
many alternatives \textit{from each menu} and responded to it by choosing 
\textit{from the collection of all menus} that are derivable from that menu. 
Reassuringly, this analysis does not 
detect any notable patterns pointing towards 
a systematic such ``metavisualised'' preference for flexibility.
Details of these additional analyses are presented in this appendix.

\subsection{Satisficing}

\cite*{simon56PR} famously coined the term \textit{satisficing} to describe 
resource-constrained agents who, 
instead of searching through all available alternatives in order to find the best, 
only search until they find one that meets an acceptability threshold. 
Two recent forced-choice studies that tested the satisficing 
hypothesis in economics are \cite*{reutskaja-etal11} and 
\cite*{caplin-dean-martin}. 
The former found weak evidence for such behaviour 
in choice from lists of food snacks under intense time pressure, 
whereas the latter found strong evidence 
in choice from lists of monetary amounts that were described 
verbally through a series of additions and 
subtractions, without (or with limited) time pressure. 
The fact that all 50 menus in both treatments 
of our experiment were presented vertically as unnumbered lists 
allows us to add to this literature by 
complementing the preceding model-based analysis 
with a test for satisficing in the present 
time-unconstrained decision environment of 
multi-valued choice over pairs of gift cards.

To this end, we first conduct a relatively narrow 
but potentially informative test 
of satisficing by focusing on the frequency with which subjects opted to 
choose only the \textit{first} alternative 
that appeared in the menus they saw, and then asking whether this frequency 
can be viewed as being above 
and beyond what might be reasonably interpretable differently. Specifically, 
we find that 5 and 4 subjects 
(3.3\%) in the Forced- and Free-Choice treatments, respectively, 
who were not classified as approximately 
explainable by one of the three deterministic models discussed earlier chose 
only the first option at 
frequencies that strictly exceeded the 97.5\% cut-off values of 0.28 and 0.29 
that are derived from the 
relevant simulations. These numbers rise to 27 and 20 (17\%), 
respectively, if the non-randomness 
criterion is retained but the model-classification requirement is dropped. 

In addition, we compute and study the average position of 
each subject's chosen item(s) in the 50 menus' 
list orderings because, intuitively, an unusually low average value 
of this metric could also be indicative 
of satisficing behaviour for the subject in question. 
We find that for an additional 7 and 5 (4.4\%) subjects 
in the Forced- and Free-Choice treatments, respectively, 
who were not classified as approximately 
explainable by one of the three deterministic models that were 
discussed above, the average positions of their 
chosen item(s) in the 50 menus' list orderings were strictly below 
the 2.5\% percentile cut-off value of 1.84 
that is derived from simulated uniform-random forced choices. 
These two distinct analyses therefore suggest 
that a total 21 subjects (7.7\%) who were not 
model-classified in the main part of the paper
might be thought of as exhibiting a 
systematic satisficing behaviour.

\begin{figure}[!htbp]
	\centering
	\caption{\centering Forced-Choice subjects are 
		more likely to select items that are 
		higher up in the menu list.}
	\begin{subfigure}[b]{0.47\textwidth}
		\centering
		\caption{Forced-Choice treatment}
		\includegraphics[width=0.99\textwidth]{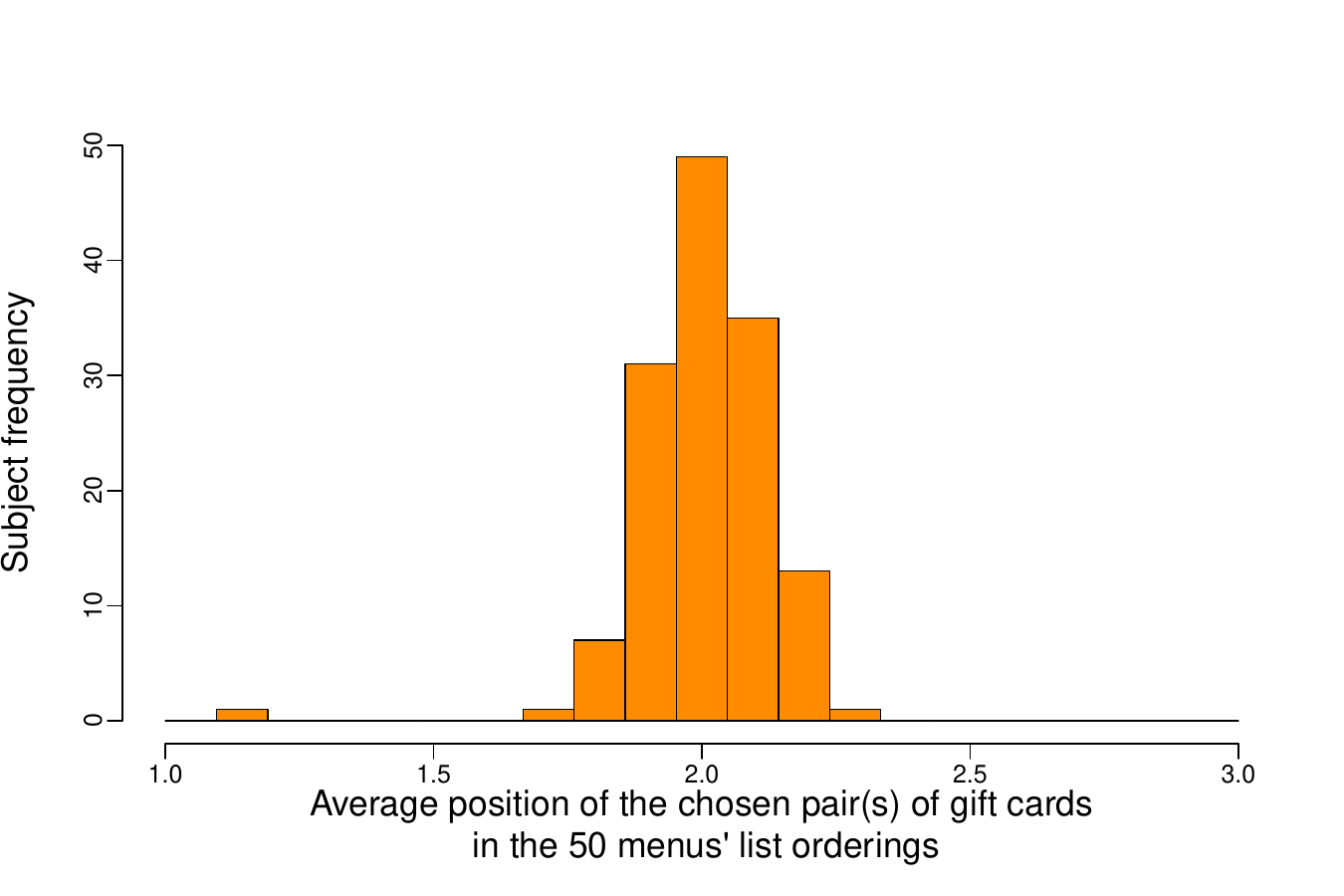}
		\caption*{\scriptsize Distribution range: $[1.15, 2.25]$}
	\label{fig:average_choice_order_fc}
\end{subfigure}\hspace{15pt}
\begin{subfigure}[b]{0.47\textwidth}
	\centering
	\caption{Free-Choice treatment}
	\includegraphics[width=0.99\textwidth]{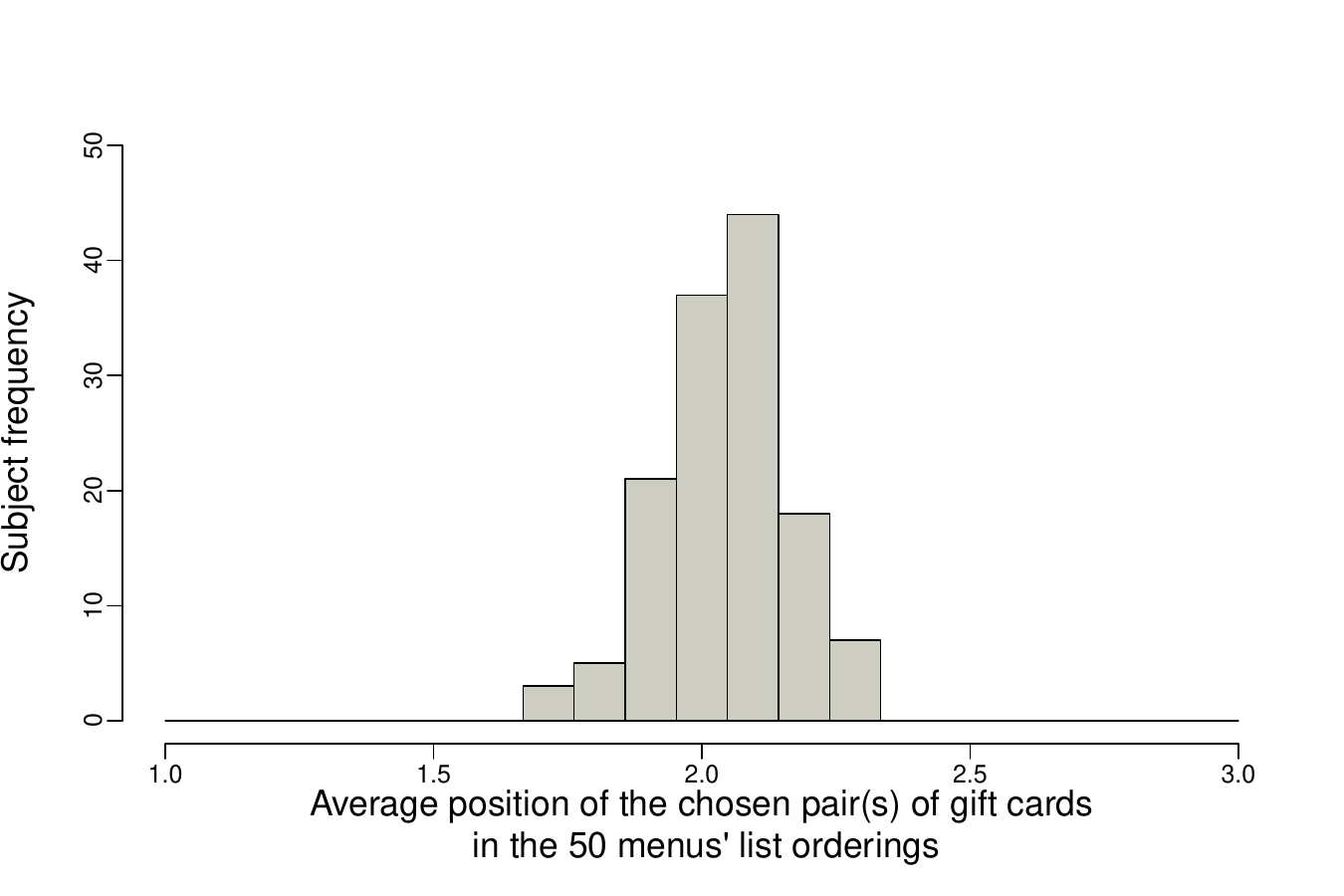}
	\caption*{\scriptsize Distribution range: $[1.74, 2.32]$}
\label{fig:average_choice_order_nfc}
\end{subfigure}
\caption*{\scriptsize Note: $p=0.009$; two-sided Mann-Whitney $U$ test.}
\label{fig:satisficing}
\end{figure}

Interestingly, a cross-treatment comparison 
of how the average positions of all 
chosen alternatives are distributed 
(see Figure \ref{fig:satisficing}) indicates 
that Forced-Choice subjects may be 
significantly more likely to choose items 
that appear higher up in the menu list 
($p=0.009$; two-sided Mann-Whitney $U$ test). 
This finding could be explained 
intuitively as follows: for individuals who are 
either insufficiently motivated to 
make 50 careful decisions over 2--4 pairs 
of gift cards in an experimental lab or find the 
task to be cognitively challenging, 
being unable to avoid/defer the 
decision at a small cost could make the use of 
a non-compensatory decision heuristic 
such as satisficing more likely.

\subsection{Preference for Randomization}

\cite*{agranov&ortoleva,agranov-ortoleva20}, 
\citet*{dwenger-kubler-weizsacker18}, \cite{cettolin&riedl} 
and \cite*{agranov-healy-nielsen22}, 
among others, have recently documented a \textit{preference for randomization} 
in repeated choices from binary menus 
of money lotteries.\footnote{\cite*{ong-qiu23} report such a preference 
in an ultimatum-bargaining setting where 
receivers are willing to incur a cost in order to randomize between 
acceptance and rejection of an offer.} 
Such a preference refers to subjects' frequent tendency to change their 
choices from one occurrence 
of a menu to the next, and their willingness to even incur a small cost in 
order to have their choice determined randomly.
In addition to the indifference-permitting goodness-of-fit analysis of the 
three models that was presented earlier, 
the aspect of our experimental design whereby subjects are rewarded with 
a random alternative at a menu if 
they chose all alternatives at that menu allows us to test for the existence 
of a similar preference for 
randomization in our data, even without any choice repetitions. 
Being another by-product of the experiment's 
features, however, this additional test is limited in its scope because 
the experimental design was not 
constructed to detect all possible manifestations of preference for randomization 
in multi-valued choice environments.

As with our analysis of satisficing, to proceed with this investigation we regard 
a subject as potentially 
exhibiting such a preference if: (i) they are not approximately explainable by 
one of the three deterministic 
models of preference maximization; and (ii) the number of menus were they chose 
everything strictly exceeds 
the 97.5\% simulations-based cut-off values of 14 and 11 menus in the Forced- 
and Free-Choice treatments, 
respectively. Both criteria are satisfied by 3 and 4 subjects (2.5\%) in 
the two treatments, and these rise to 6 and 9 (5.5\%) when the first criterion is 
dropped.\footnote{These estimates are conservative 
and may be better seen as lower bounds. The reason is that 
the experimental design only allows subjects who might have 
exhibited such a systematic preference for 
randomization to reveal it only when they were faced with 
difficulty deciding between \textit{all} 
feasible alternatives at a given menu, not from a proper subset thereof.} 
In addition, none of these 7 subjects 
belongs to the ``satisficing'' 	category that was defined above. 
This finding adds to the existing and growing 
literature on preference for randomization by showing that such a 
preference could potentially 
manifest itself in binary as well as \textit{non-binary} menus of 
\textit{riskless} alternatives, even when 
the choice-deferral option is feasible and acts as another obvious 
way for an individual to deal with 
a difficult decision, and even in non-repeated-choice environments.

\subsection{Menu Meta-Visualization and Menu Preferences for Flexibility}

Subjects in our experiment were asked to make choices \textit{from} menus.
However, because the design specifically allowed them to choose multiple 
items at each menu, an analyst might be tempted to view the resulting choices 
as choices \textit{over} menus instead.
The argument here is that subjects who are presented with menu $A$ and choose 
the elements of some $C(A)\subseteq A$ might be thought of as choosing 
the \textit{menu} $C(A)$ from the ``metavisualised'' collection of feasible menus 
$\{A':A'\subseteq A\}$.\footnote{In the free-choice 
treatment only, this collection 
also includes the empty 
subset of $A$.} Such a view is contrary to the experiment's motivation, design 
and the way in which each choice problem was presented to subjects in the 
experimental interface. Furthermore, it effectively assumes that participants 
are endowed with sufficiently rich memory and other cognitive resources 
to be able to mentally translate each of the 50 distinct menus $A$ into 
a collection of 4, 8 or 16 (when $|A|=2,3,4$, respectively) 
weak sub-menus of $A$ and to form preferences over those.
With these caveats in mind, and to make our analysis as comprehensive as possible, 
we now re-examine our data through the lens of this distinct point of view.

The leading theories of menu preferences in the literature portray 
decision makers as exhibiting either a preference for \textit{flexibility} 
\citep*{kreps79,DLR01} or a preference for \textit{commitment} 
\citep*{gul-pesendorfer01,gul-pesendorfer05}.\footnote{\cite*{toussaert18}
reports on an experimental design that was 
explicitly constructed to test for the presence of 
such preferences.} 
The former preference is manifested at a menu $A$ if 
$B$ is preferred to $A$ whenever $B\supset A$. 
By contrast, a preference for commitment in the 
Gul-Pesendorfer theory of temptation and self-control
is formalized with the Set Betweeness axiom 
whereby ``$A$ is preferred to $B$'' implies ``$A$ is preferred to $A\cup B$
is preferred to $B$''.

Under the postulated metavisualization and the associated cognitive-richness 
and mental-preference hypotheses, we proceed now to treating the subjects' 
choices over alternatives as choices over menus.
Because the choice alternatives and menus in our experiment
do not seem to invite any temptation and self-control trade-offs,
our aim here is to test for any systematic preferences for flexibility
at the individual subject level. 
We acknowledge from the outset, however, 
that this analysis, while targeting the hallmark of the respective 
decision theories cited above, do not constitute a comprehensive test thereof. 
The reason is that these models rely on several additional 
axioms on preferences and assumptions on the structure of the 
preference domain which go beyond the riskless choice 
environment of our experiment.

\begin{figure}[!htbp]
\caption{\centering Subjects' cumulative densities of 
flexibility violations at the 50 menus\linebreak
under the menu meta-visualisation hypothesis.}
\vspace{-15pt}
\centering
\includegraphics[width=0.75\textwidth]{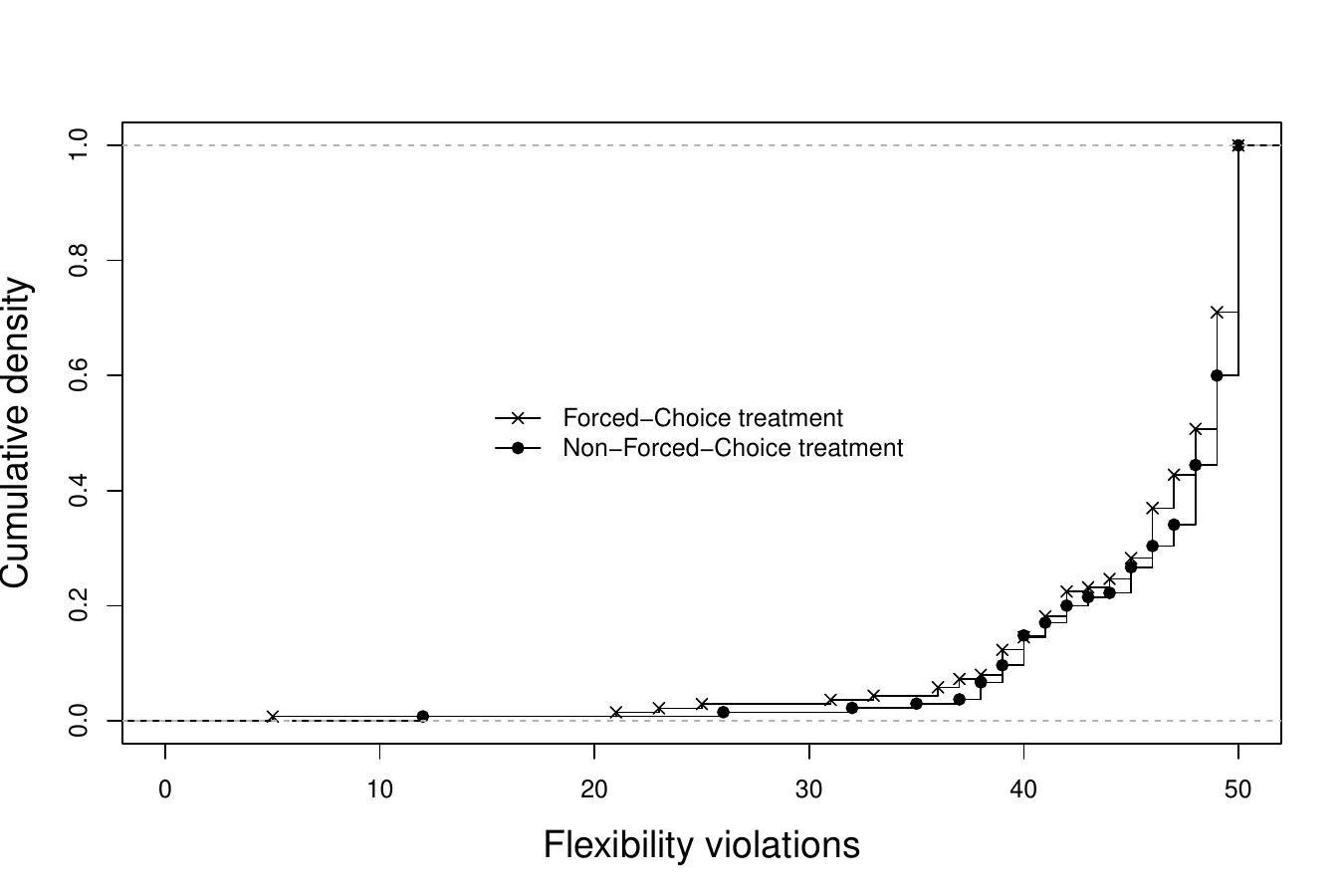}
\label{fig:flex_commit}
\end{figure}

However, the basic aggregate-level analysis in the main part of the paper
readily suggests that few --if any-- subjects in our sample 
of eligible subjects are likely to exhibit a systematic 
preference for flexibility.\footnote{Four subjects in the original 
sample could indeed be thought of as showing --trivially-- such a consistent 
preference for flexibility throughout the experiment by always choosing 
all alternatives at every menu. As noted in the main part of the paper,
those subjects were excluded from the subsequent analysis  
because their behaviour was completely uninformative about 
their preferences over gift-card pairs.} 
This point is validated in Figure \ref{fig:flex_commit}, 
which depicts the subjects' cumulative densities of 
flexibility violations.
More specifically, 5 and 12 are the smallest 
numbers of menus where this principle was violated, 
in each case by a single subject, while 
the means/medians in the two treatments
stand at 45.8/48 (Forced Choice) and 46.6/49 (Free Choice). 
Therefore, in line with the caveats expressed at the beginning 
of this subsection, looking at our experimental data 
through this lens does not seem to add any notable new insights
into subjects' behaviour. This provides some further reassurance toward 
the validity of the results presented earlier in this section.

\section{Motivation for Strongly Symmetric Menu Collections}

Recall that subjects were presented with the 50 menus that comprised all those 
with two (15), three (20) and four (15) alternatives. This excludes all singleton 
menus and all those with 5 or 6 alternatives. Obtaining decision data from the 
full collection of menus may be desirable at some levels (e.g. because they 
provide additional information) but undesirable at others 
(e.g. because choice fatigue 
could negatively affect decision quality) or even impractical (for example, when 
the grand choice set comprises 6, 7 or 8 alternatives the full collection includes 
63, 127 or 255 menus). In situations such as this the researcher may be inclined 
to limit the menus seen by subjects to some manageable number. 
How should this be done? 

We argue that the collection of menus presented to subjects should satisfy 
\textit{strong symmetry} in the sense that the distribution of menu sizes 
where an alternative is feasible is the same for \textit{every} alternative. 
This requirement in turn implies the \textit{weak symmetry} 
condition whereby all alternatives are feasible at some menu 
the same number of times. For example, a collection that comprises menus 
$P=\{x,y,z\}$, $Q=\{w,z\}$ and $R=\{x,y,w\}$ satisfies 
weak but not strong symmetry 
because, although each alternative appears twice in the dataset, $x$ and $y$ 
appear only in ternary menus while $w$ and $z$ do so in one binary and 
one ternary menu instead. By contrast, a collection consisting of 
$P=\{x,y,z\}$, $R=\{x,y,w\}$, $S=\{w,y,z\}$ and $T=\{w,x,z\}$ satisfies 
strong symmetry, with each alternative appearing three times in as many 
menus of the same size. 

The motivation for the strong symmetry requirement, which is indeed satisfied 
in our experimental dataset, is intuitive: if the menu-size distributions differ 
across alternatives, this means that at least one of them is feasible in at least 
one larger/smaller menu compared to at least one other alternative. This could 
then pave the way for a potential bias in favour of or against that option 
in the ensuing analysis. In the first menu collection above, for example, 
suppose $x$ is (uniquely) chosen at $P$, $z$ at $Q$ and $w$ at $R$. Clearly, 
this is a revealed-preference cycle. Moreover, removing any of the three 
observations from this dataset breaks that cycle. Which one should be removed? 
One might be tempted to keep the choice at $Q$ as the potentially more accurate 
of the three because it is derived from a binary menu. But doing so and 
removing $P$, for example, would amount to giving a possibly unfair 
(dis)advantage to alternative $z$ ($x$). Indeed, $z$ ($x$) would then 
appear first (last) in the inferred revealed preference ordering even though 
$x$ ($z$) would have been second (third) if $Q$ had been removed instead. 
Yet even though it could be that $x$ would continue to be chosen over $z$ 
also at the binary menu $\{x,z\}$, the (symmetry-breaking) fact that 
this observation is unavailable works against that alternative.

\pagebreak

\section{Choice Probabilities in Simulations}

\begin{figure}[!htbp]
\centering
\caption{\centering Choice/deferral probabilities under 
multi-valued choice simulations at various menu sizes.\vspace{-3pt}}
\caption*{\centering \scriptsize 
(a) Every feasible alternative is chosen with $\approx0.5$ 
probability as menu size increases, 
while deferral becomes less likely.\vspace{-15pt}}
\includegraphics[width=0.51\textwidth]{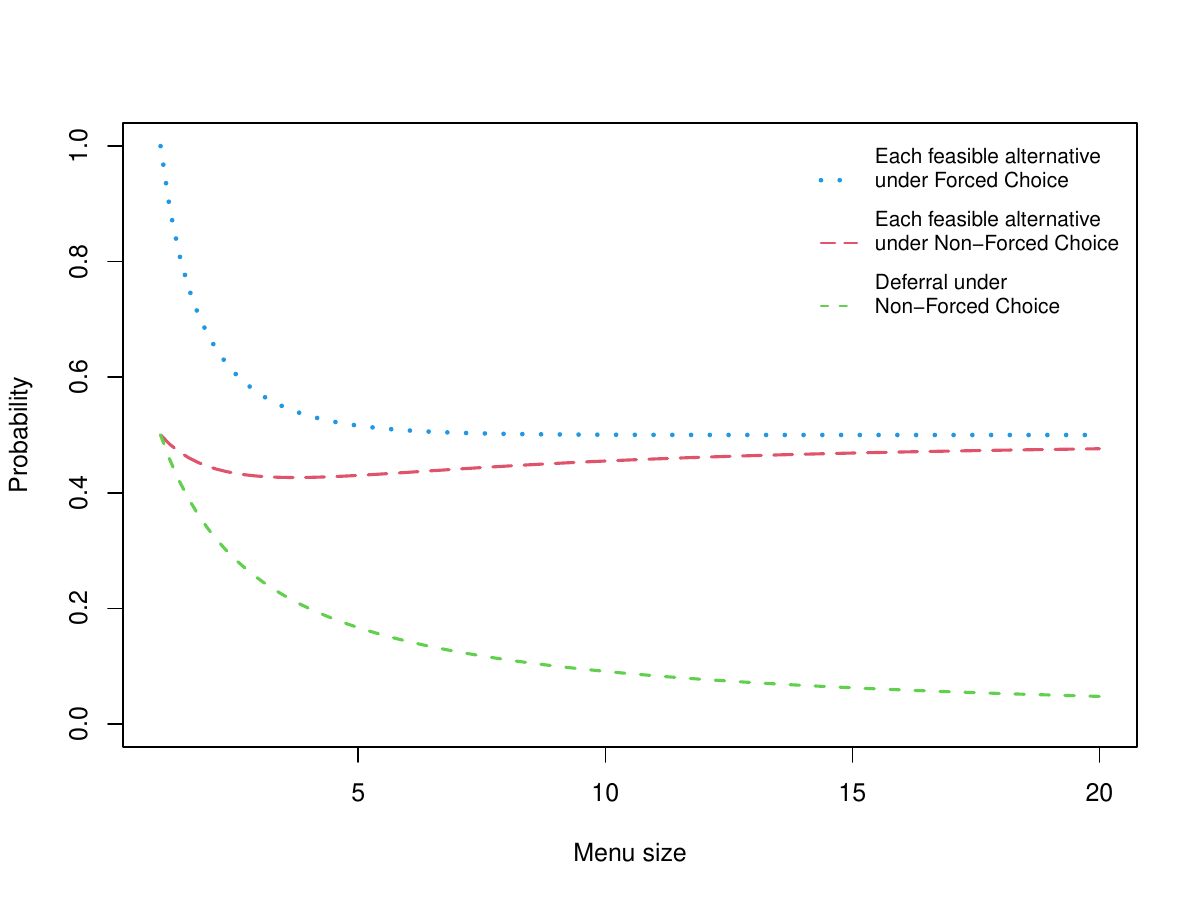}
\vspace{-20pt}
\caption*{\centering \scriptsize 
(b) The probability of choosing any non-empty 
submenu decreases as menu size increases.\vspace{-15pt}}
\includegraphics[width=0.51\textwidth]{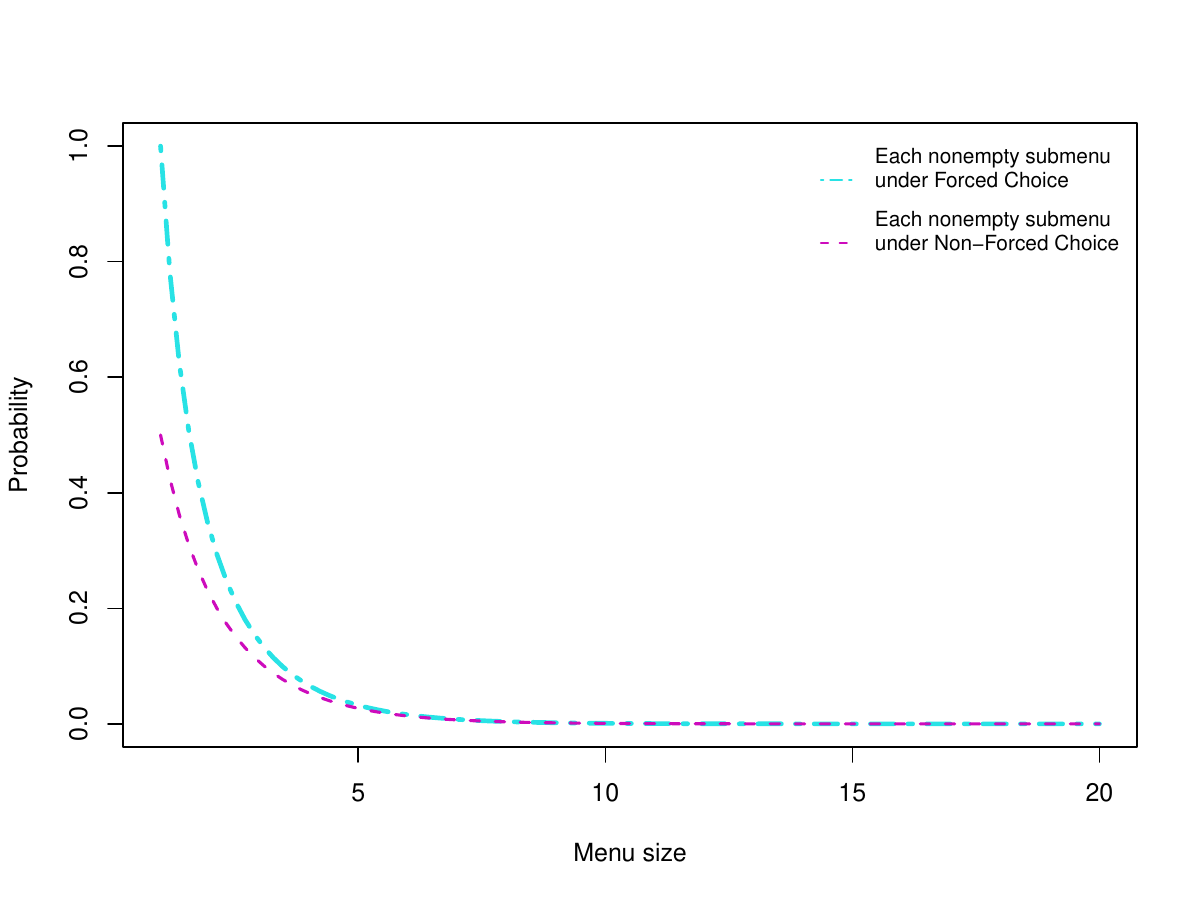}
\label{fig:simulations2}
\label{fig:simulations}
\end{figure}

{\small
The probability of an alternative being chosen at 
a menu under multi-valued choice simulations 
is interpreted here as the probability that this alternative 
belongs to the chosen submenu of that menu. 
Assuming Forced Choice first, and considering an arbitrary 
menu $A$ with $k$ alternatives, 
every non-empty submenu $A'\subseteq A$ is equally likely to be chosen, 
and is therefore chosen with 
probability $\frac{1}{2^k-1}$. Since each of the $k$ feasible alternatives 
belongs to exactly 
$\frac{2^k}{2}$ of these submenus, it follows that each of them is 
chosen in the above sense with 
probability $\frac{2^k}{2(2^k-1)}$. Under Non-Forced Choice now, 
since some non-empty submenu of 
$A$ is chosen with probability $\frac{k}{k+1}$ because choice 
is deferred with probability 
$\frac{1}{k+1}$, the corresponding probability for each of 
the $k$ active-choice alternatives 
is adjusted accordingly. This results to the probabilities 
described in Table \ref{tab:simulations} 
and Figure \ref{fig:simulations}.}

\begin{table}[!htbp]
\centering
\footnotesize
\caption{Choice/deferral probabilities at a menu with $k$ alternatives 
under multi-valued choice simulations.}
\setlength{\tabcolsep}{6pt} 
\renewcommand{\arraystretch}{1.3} 
\makebox[\textwidth][c]{
\begin{tabular}{|l|c|c|c|}
	\hline
	& \textbf{Probability of each}	
	& \textbf{Probability of each} 
	& \textbf{Probability of} \\
	& \textbf{non-empty submenu} 
	& \textbf{active-choice alternative} 
	& \textbf{deferral}\\
	& \textbf{being chosen}	
	& \textbf{being chosen} 
	& \\
	\hline
	\textbf{Forced-Choice simulations} 
	& $\displaystyle\frac{1}{2^k-1}$ 
	& $\displaystyle\frac{1}{2}\frac{2^k}{2^k-1}$ 
	& NA \\
	\hline
	\textbf{Free-Choice simulations}
	& $\displaystyle\frac{1}{2^k-1}\frac{k}{k+1}$ 
	& $\displaystyle\frac{1}{2}\frac{2^k}{2^k-1}\frac{k}{k+1}$ 
	& $\displaystyle\frac{1}{k+1}$ \\
	\hline
\end{tabular}
}
\caption*{\raggedright \scriptsize }
\label{tab:simulations}
\end{table}

\vspace{-15pt}

\pagebreak

\section{Graphs of Model-Optimal Revealed Preference Orderings}

\vspace{-10pt}

\noindent \textit{\textbf{Note: arrows point to dominated options/indifference classes}}

\captionsetup[figure]{labelformat=empty}
\captionsetup[subfigure]{labelformat=empty}

\subsection{Rational Choice/Utility Maximization}

\begin{figure}[!htbp]
\centering
\begin{subfigure}[b]{0.22\textwidth}
\centering
\includegraphics[width=0.4\textwidth]{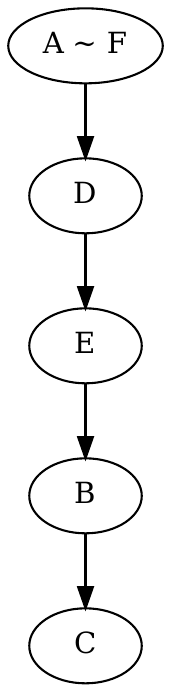}
\caption{\centering
	ID=1140, JHM=0,\\ 1 optimal relation}
	\end{subfigure}\hspace{15pt}
	\begin{subfigure}[b]{0.22\textwidth}
\centering
\includegraphics[width=0.3\textwidth]{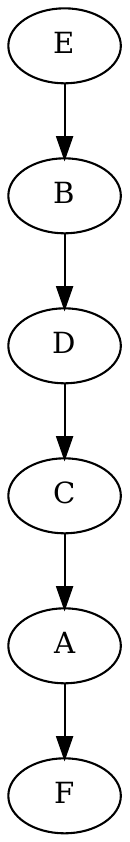}
\caption{\centering
	ID=2582, JHM=0,\\ 1 optimal relation}
	\end{subfigure}
	\begin{subfigure}[b]{0.22\textwidth}
\centering
\includegraphics[width=0.4\textwidth]{/graphs/NFC-UM-0-3843-1.pdf}
\caption{\centering
	ID=3843, JHM=0,\\ 1 optimal relation}
	\end{subfigure}
	\begin{subfigure}[b]{0.22\textwidth}
\centering
\includegraphics[width=0.4\textwidth]{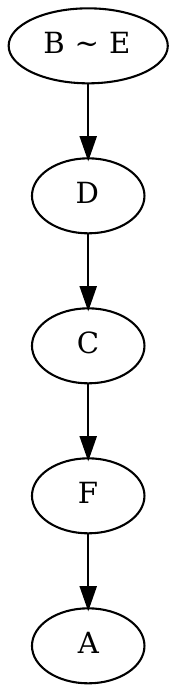}
\caption{\centering
	ID=7228, JHM=0,\\ 1 optimal relation}
	\end{subfigure}
\end{figure}

\begin{figure}[!htbp]
\centering
\begin{subfigure}[b]{0.22\textwidth}
\centering
\includegraphics[width=0.3\textwidth]{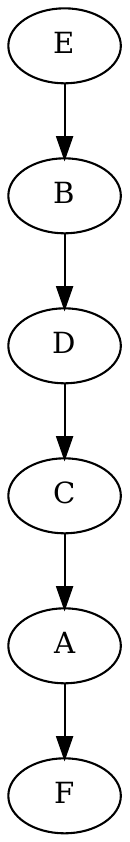}
\caption{\centering
	ID=7538, JHM=0,\\ 1 optimal relation}
	\end{subfigure}
	\begin{subfigure}[b]{0.22\textwidth}
\centering
\includegraphics[width=0.3\textwidth]{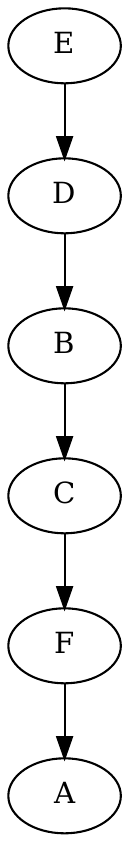}
\caption{\centering
	ID=9139, JHM=0,\\ 1 optimal relation}
	\end{subfigure}
	\begin{subfigure}[b]{0.22\textwidth}
\centering
\includegraphics[width=0.85\textwidth]{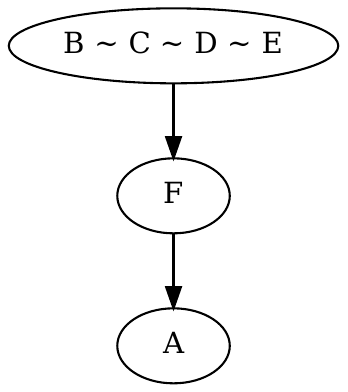}
\caption{\centering
	ID=9425, JHM=0,\\ 1 optimal relation}
	\end{subfigure}
	\begin{subfigure}[b]{0.22\textwidth}
\centering
\includegraphics[width=0.3\textwidth]{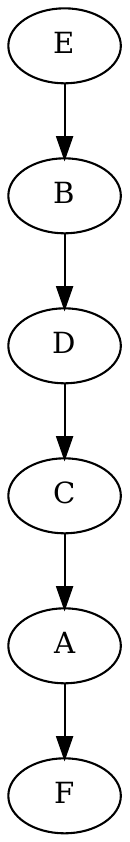}
\caption{\centering
	ID=9507, JHM=0,\\ 1 optimal relation}
	\end{subfigure}
\end{figure}

\begin{figure}[!htbp]
\centering
\begin{subfigure}[b]{0.22\textwidth}
\centering
\includegraphics[width=0.7\textwidth]{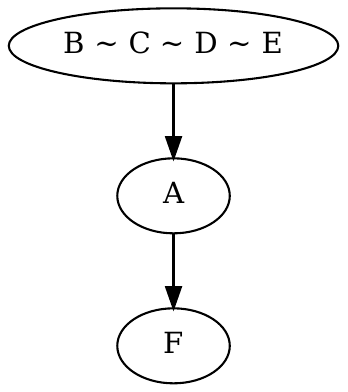}
\caption{\centering
	ID=9946, JHM=0.333,\\ 1 optimal relation}
	\end{subfigure}
	\begin{subfigure}[b]{0.22\textwidth}
\centering
\includegraphics[width=0.45\textwidth]{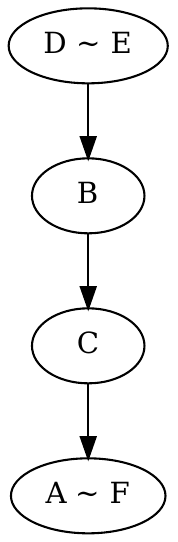}
\caption{\centering
	ID=1411, JHM=0.5,\\ 1 optimal relation}
	\end{subfigure}
	\begin{subfigure}[b]{0.22\textwidth}
\centering
\includegraphics[width=0.8\textwidth]{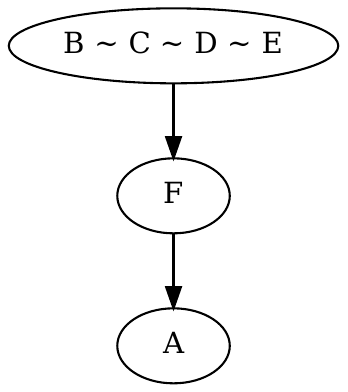}
\caption{\centering
	ID=1967, JHM=0.5,\\ 1 optimal relation}
	\end{subfigure}
	\begin{subfigure}[b]{0.22\textwidth}
\centering
\includegraphics[width=0.8\textwidth]{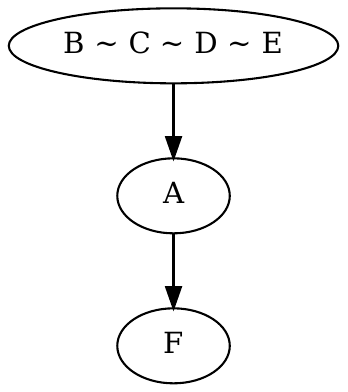}
\caption{\centering
	ID=2895, JHM=0.5,\\ 1 optimal relation}
	\end{subfigure}
\end{figure}

\begin{figure}[!htbp]
\begin{subfigure}[b]{0.22\textwidth}
\centering
\includegraphics[width=0.9\textwidth]{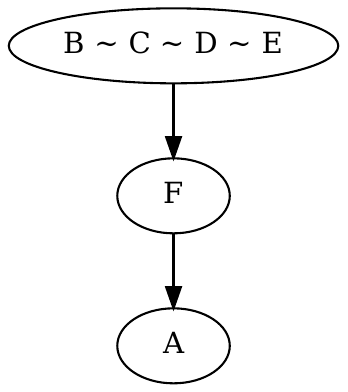}
\caption{\centering
	ID=6861, JHM=0.5,\\ 1 optimal relation}
	\end{subfigure}
	\begin{subfigure}[b]{0.22\textwidth}
\centering
\includegraphics[width=0.45\textwidth]{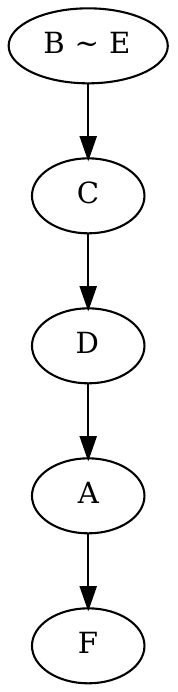}
\caption{\centering
	ID=8360, JHM=0.5,\\ 1 optimal relation}
	\end{subfigure}
	\begin{subfigure}[b]{0.22\textwidth}
\centering
\includegraphics[width=0.65\textwidth]{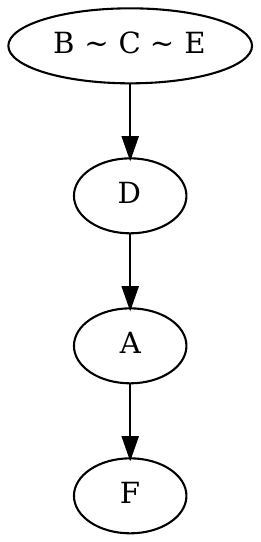}
\caption{\centering
	ID=6969, JHM=0.5,\\ 1 optimal relation}
	\end{subfigure}
	\begin{subfigure}[b]{0.22\textwidth}
\centering
\includegraphics[width=0.7\textwidth]{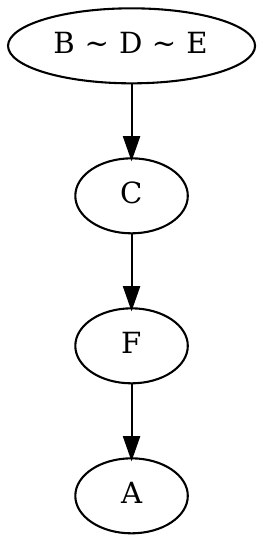}
\caption{\centering
	ID=8712, JHM=0.5,\\ 1 optimal relation}
	\end{subfigure}
\end{figure}

\begin{figure}[!htbp]
\centering
\begin{subfigure}[b]{0.22\textwidth}
\centering
\includegraphics[width=0.9\textwidth]{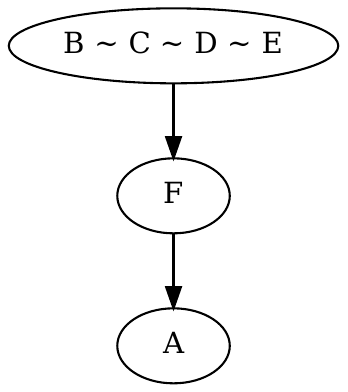}
\caption{\centering
	ID=4430, JHM=0.5,\\ 1 optimal relation}
	\end{subfigure}
	\begin{subfigure}[b]{0.22\textwidth}
\centering
\includegraphics[width=0.45\textwidth]{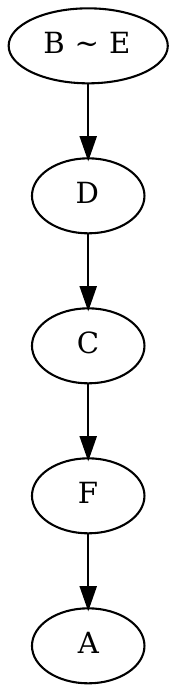}
\caption{\centering
	ID=4540, JHM=0.5,\\ 1 optimal relation}
	\end{subfigure}
	\begin{subfigure}[b]{0.22\textwidth}
\centering
\includegraphics[width=0.9\textwidth]{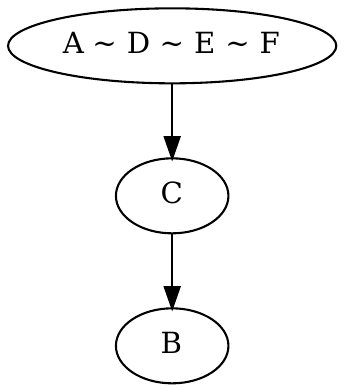}
\caption{\centering
	ID=3797, JHM=0.75,\\ 1 optimal relation}
	\end{subfigure}
	\begin{subfigure}[b]{0.22\textwidth}
\centering
\includegraphics[width=0.9\textwidth]{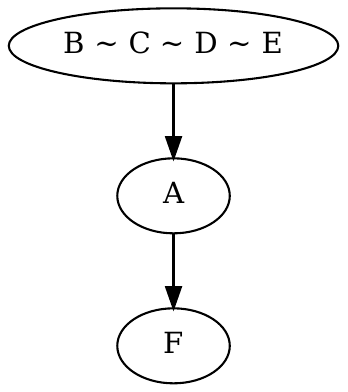}
\caption{\centering
	ID=6679, JHM=0.833,\\ 1 optimal relation}
	\end{subfigure}
\end{figure}

\begin{figure}[!htbp]
\centering
\begin{subfigure}[b]{0.22\textwidth}
\centering
\includegraphics[width=0.9\textwidth]{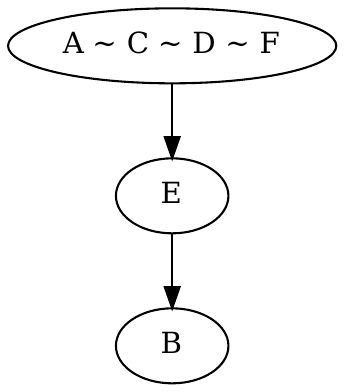}
\caption{\centering 
	ID=8722, JHM=0.833,\\ 1 optimal relation}
	\end{subfigure}
	\begin{subfigure}[b]{0.22\textwidth}
\centering
\includegraphics[width=0.35\textwidth]{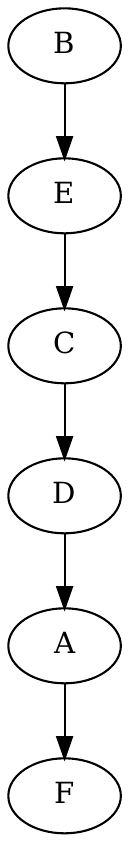}
\caption{\centering 
	ID=4542, JHM=1,\\ 1 optimal relation}
	\end{subfigure}
	\begin{subfigure}[b]{0.22\textwidth}
\centering
\includegraphics[width=0.25\textwidth]{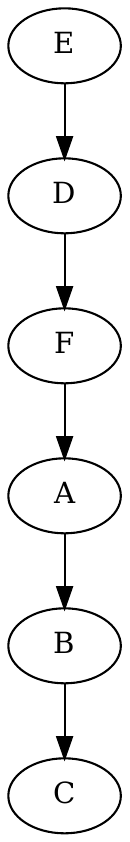}
\includegraphics[width=0.25\textwidth]{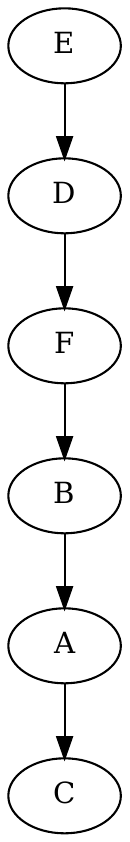}
\includegraphics[width=0.3\textwidth]{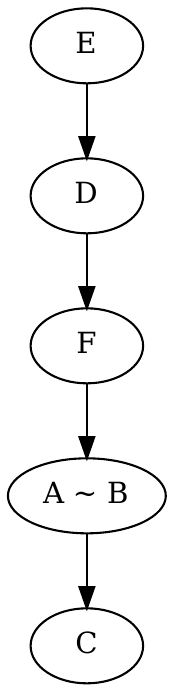}
\caption{\centering
	ID=6890, JHM=1,\\ 3 optimal relations}
	\end{subfigure}
	\begin{subfigure}[b]{0.22\textwidth}
\centering
\includegraphics[width=0.27\textwidth]{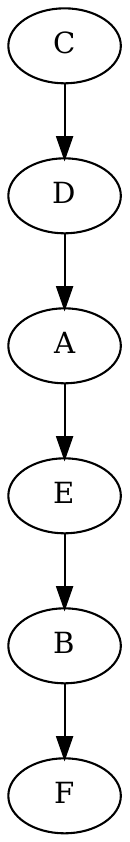}
\caption{\centering 
	ID=7081, JHM=1,\\ 1 optimal relation}
	\end{subfigure}	
\end{figure}

\begin{figure}[!htbp]
\begin{subfigure}[b]{0.22\textwidth}
\centering
\includegraphics[width=0.45\textwidth]{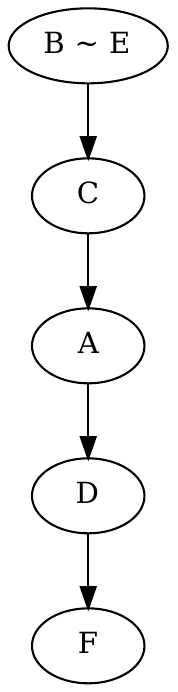}
\caption{\centering 
	ID=3622, JHM=1,\\ 1 optimal relation}
	\end{subfigure}
	\begin{subfigure}[b]{0.22\textwidth}
\centering
\includegraphics[width=0.45\textwidth]{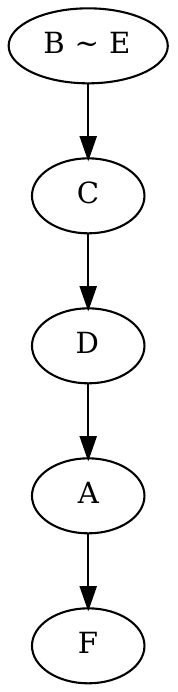}
\caption{\centering
	ID=3826, JHM=1,\\ 1 optimal relation}
	\end{subfigure}
	\begin{subfigure}[b]{0.22\textwidth}
\centering
\includegraphics[width=0.45\textwidth]{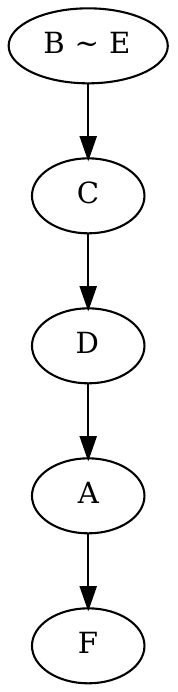}
\caption{\centering
	ID=6778, JHM=1,\\ 1 optimal relation}
	\end{subfigure}	
	\begin{subfigure}[b]{0.22\textwidth}
\centering
\includegraphics[width=0.3\textwidth]{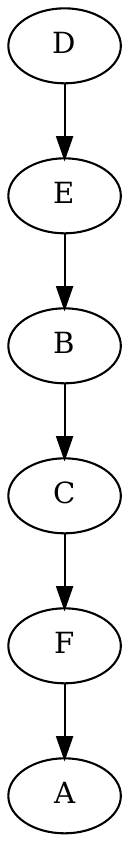}
\caption{\centering
	ID=7464, JHM=1,\\ 1 optimal relation}
	\end{subfigure}
\end{figure}

\begin{figure}[!htbp]
\centering
\begin{subfigure}[b]{0.22\textwidth}
\centering
\includegraphics[width=0.9\textwidth]{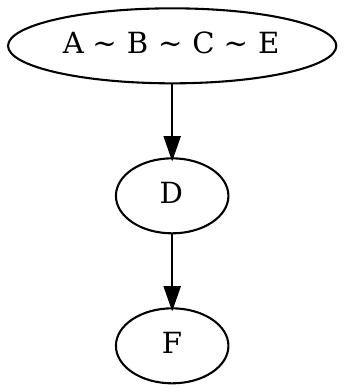}
\caption{\centering
	ID=7752, JHM=1,\\ 1 optimal relation}
	\end{subfigure}
	\begin{subfigure}[b]{0.22\textwidth}
\centering
\includegraphics[width=0.9\textwidth]{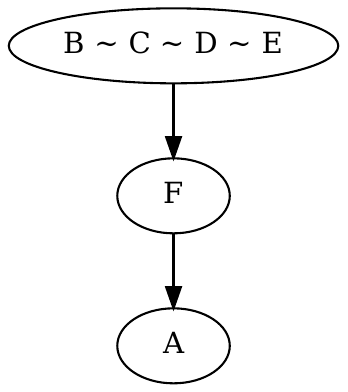}
\caption{\centering
	ID=3986, JHM=1.167,\\ 1 optimal relation}
	\end{subfigure}
	\begin{subfigure}[b]{0.25\textwidth}
\centering
\includegraphics[width=0.99\textwidth]{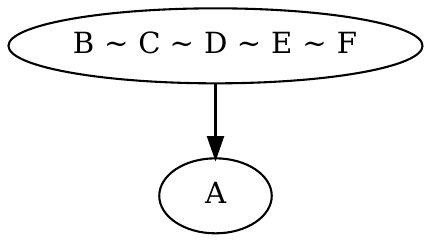}
\caption{\centering
	ID=6112, JHM=1.167,\\ 1 optimal relation}
	\end{subfigure}
	\begin{subfigure}[b]{0.22\textwidth}
\centering
\includegraphics[width=0.7\textwidth]{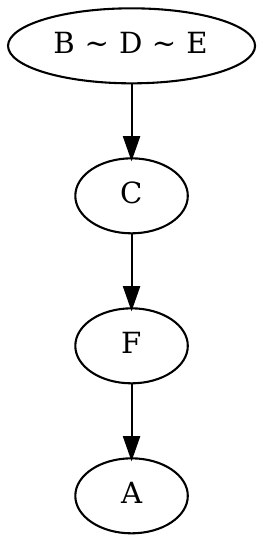}
\caption{\centering
	ID=3452, Score=1.333,\\ 1 optimal relation}
	\end{subfigure}
\end{figure}

\begin{figure}[!htbp]
\centering
\begin{subfigure}[b]{0.22\textwidth}
\centering
\includegraphics[width=0.9\textwidth]{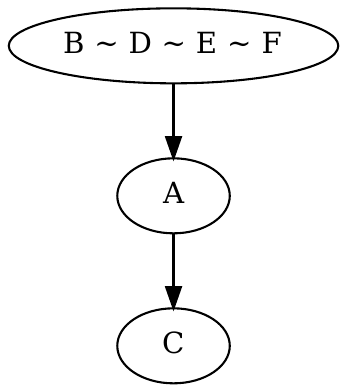}
\caption{\centering
	ID=8855, JHM=1.333,\\ 1 optimal relation}
	\end{subfigure}
	\begin{subfigure}[b]{0.22\textwidth}
\centering
\includegraphics[width=0.9\textwidth]{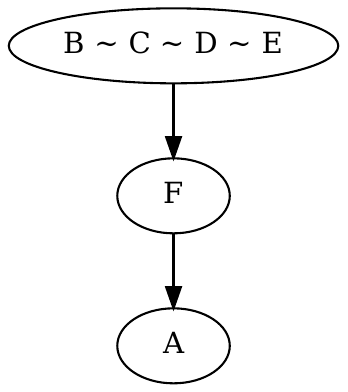}
\caption{\centering
	ID=1230, JHM=1.333,\\ 1 optimal relation}
	\end{subfigure}\hspace{10pt}
	\begin{subfigure}[b]{0.22\textwidth}
\centering
\includegraphics[width=0.3\textwidth]{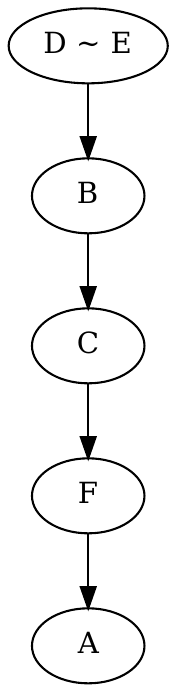}
\includegraphics[width=0.3\textwidth]{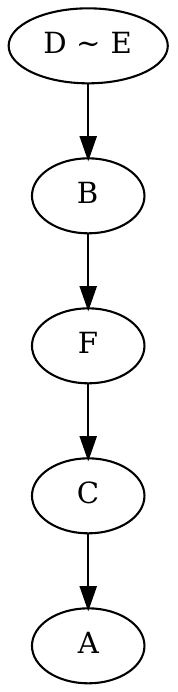}
\includegraphics[width=0.3\textwidth]{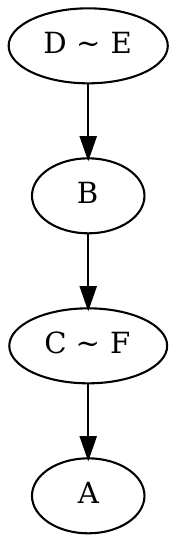}		
\caption{\centering
	ID=5134, JHM=1.5,\\ 3 optimal relations}
	\end{subfigure}	
	\begin{subfigure}[b]{0.22\textwidth}
\centering
\includegraphics[width=0.3\textwidth]{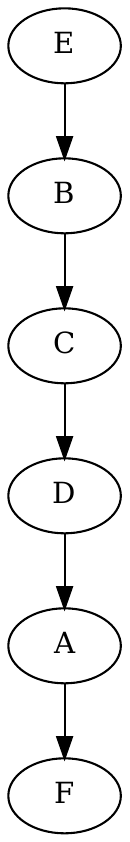}
\caption{\centering 
	ID=8553, JHM=1.5,\\ 1 optimal relation}
	\end{subfigure}
\end{figure}

\begin{figure}[!htbp]
\begin{subfigure}[b]{0.22\textwidth}
\centering
\includegraphics[width=0.3\textwidth]{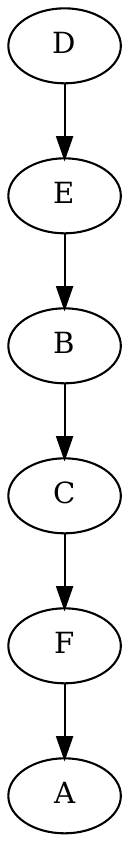}
\caption{\centering
	ID=1827, JHM=1.5,\\ 1 optimal relation}
	\end{subfigure}	
	\begin{subfigure}[b]{0.22\textwidth}
\centering
\includegraphics[width=0.3\textwidth]{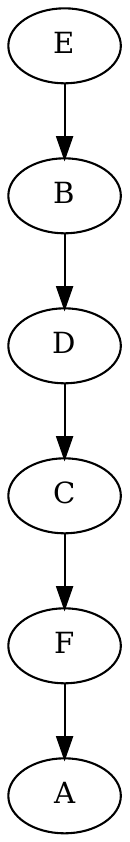}
\caption{\centering
	ID=3377, JHM=1.5,\\ 1 optimal relation}
	\end{subfigure}
	\begin{subfigure}[b]{0.22\textwidth}
\centering
\includegraphics[width=0.9\textwidth]{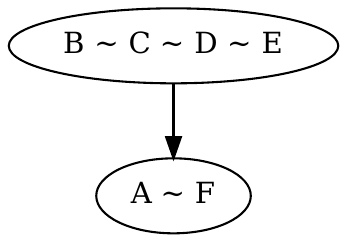}
\caption{\centering
	ID=3473, JHM=1.5,\\ 1 optimal relation}
	\end{subfigure}
	\begin{subfigure}[b]{0.21\textwidth}
\centering
\includegraphics[width=0.5\textwidth]{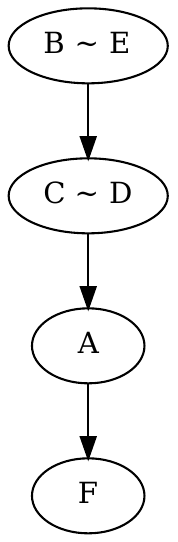}
\caption{\centering
	ID=5337, JHM=1.5,\\ 1 optimal relation}
	\end{subfigure}
\end{figure}

\begin{figure}[!htbp]
\centering
\begin{subfigure}[b]{0.22\textwidth}
\centering
\includegraphics[width=0.3\textwidth]{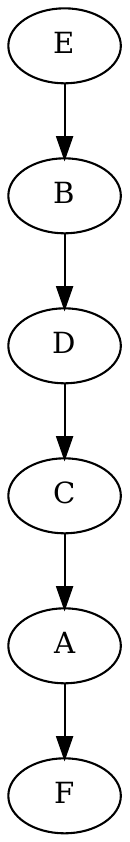}
\caption{\centering
	ID=8773, JHM=1.5,\\ 1 optimal relation}
	\end{subfigure}
	\begin{subfigure}[b]{0.22\textwidth}
\centering
\includegraphics[width=0.5\textwidth]{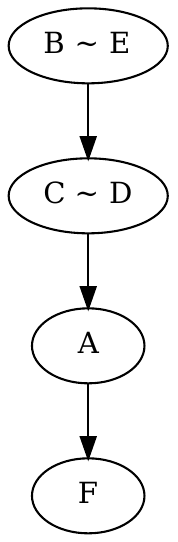}
\caption{\centering
	ID=7677, JHM=1.5,\\ 1 optimal relation}
	\end{subfigure}
	\begin{subfigure}[b]{0.23\textwidth}
\centering
\includegraphics[width=0.45\textwidth]{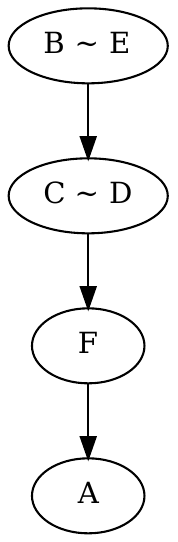}
\caption{\centering
	ID=1557, Score=1.667, 1 optimal relation}
	\end{subfigure}
	\begin{subfigure}[b]{0.22\textwidth}
\centering
\includegraphics[width=0.9\textwidth]{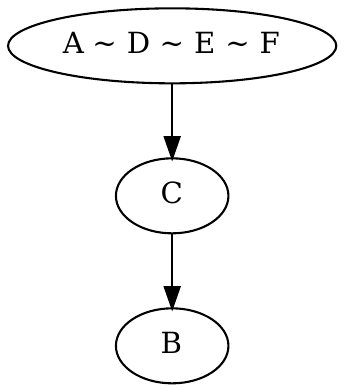}
\caption{\centering
	ID=9039, JHM=1.667,\\ 1 optimal relation}
	\end{subfigure}
\end{figure}

\begin{figure}[!htbp]
\centering
\begin{subfigure}[b]{0.22\textwidth}
\centering
\includegraphics[width=0.99\textwidth]{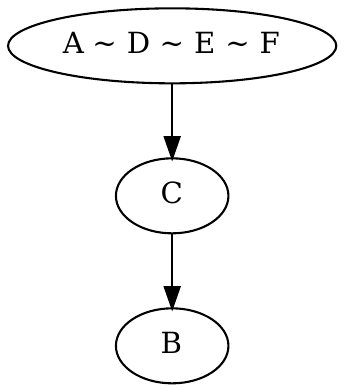}
\caption{\centering
	ID=4753, JHM=1.667,\\ 1 optimal relation}
	\end{subfigure}
	\begin{subfigure}[b]{0.31\textwidth}
\centering
\includegraphics[width=0.3\textwidth]{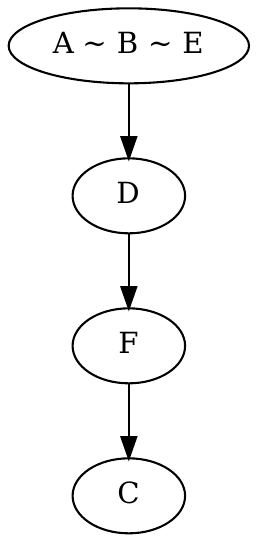}
\includegraphics[width=0.3\textwidth]{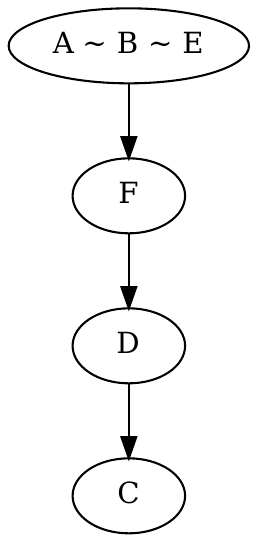}
\includegraphics[width=0.3\textwidth]{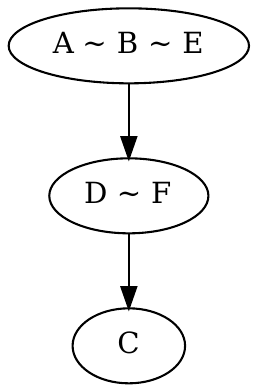}		
\caption{\centering
	ID=5061, JHM=1.833,\\ 3 optimal relations}
	\end{subfigure}
	\begin{subfigure}[b]{0.22\textwidth}
\centering
\includegraphics[width=0.45\textwidth]{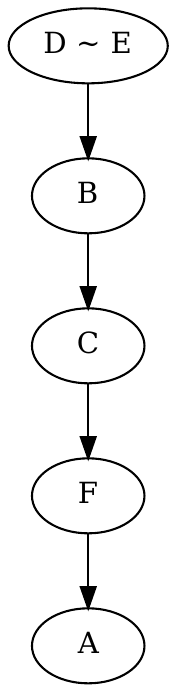}
\caption{\centering
	ID=6923, JHM=1.833,\\ 1 optimal relation}
	\end{subfigure}	
	\begin{subfigure}[b]{0.22\textwidth}
\centering
\includegraphics[width=0.44\textwidth]{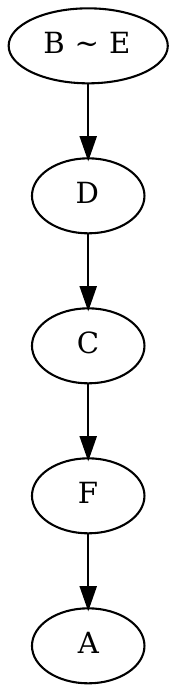}
\caption{\centering
	ID=2952, JHM=1.833,\\ 1 optimal relation}
	\end{subfigure}
\end{figure}

\begin{figure}[!htbp]
\centering
\begin{subfigure}[b]{0.22\textwidth}
\centering
\includegraphics[width=0.45\textwidth]{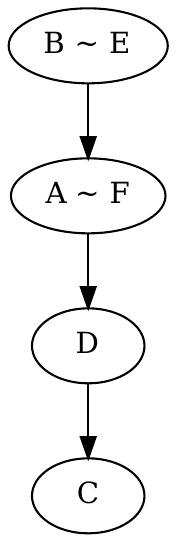}
\caption{\centering
	ID=9766, JHM=2,\\ 1 optimal relation}
	\end{subfigure}
	\begin{subfigure}[b]{0.22\textwidth}
\centering
\includegraphics[width=0.45\textwidth]{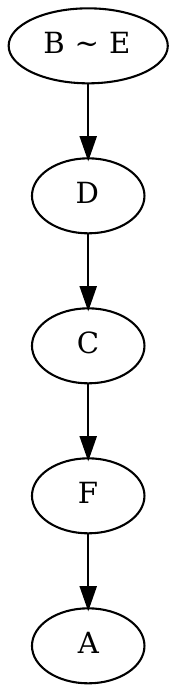}
\caption{\centering
	ID=9934, JHM=2,\\ 1 optimal relation}
	\end{subfigure}
	\begin{subfigure}[b]{0.22\textwidth}
\centering
\includegraphics[width=0.45\textwidth]{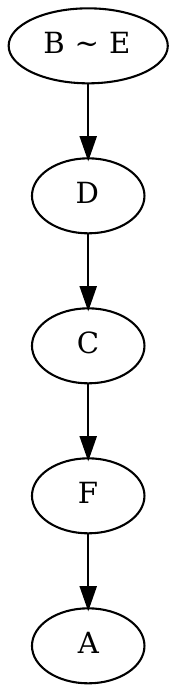}
\includegraphics[width=0.45\textwidth]{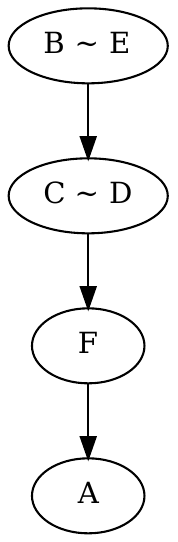}
\caption{\centering
	ID=1068, JHM=2,\\ 2 optimal relations}
	\end{subfigure}\hspace{10pt}
	\begin{subfigure}[b]{0.22\textwidth}
\centering
\includegraphics[width=0.44\textwidth]{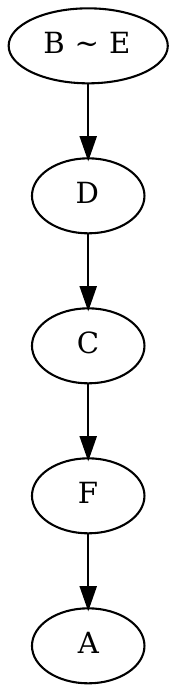}
\caption{\centering
	ID=1609, JHM=2,\\ 1 optimal relation}
	\end{subfigure}
\end{figure}

\begin{figure}[!htbp]
\centering
\begin{subfigure}[b]{0.22\textwidth}
\centering
\includegraphics[width=0.9\textwidth]{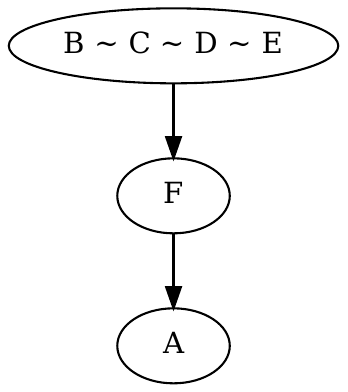}
\caption{\centering
	ID=6222, JHM=2,\\ 1 optimal relation}
	\end{subfigure}
	\begin{subfigure}[b]{0.22\textwidth}
\centering
\includegraphics[width=0.43\textwidth]{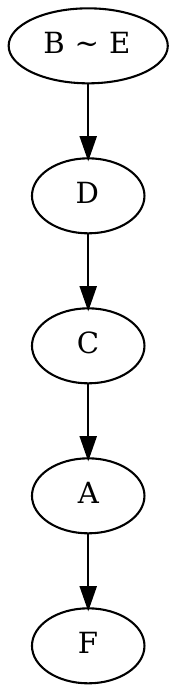}
\includegraphics[width=0.45\textwidth]{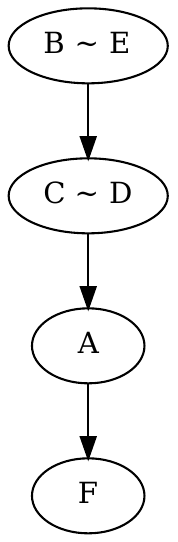}		
\caption{\centering
	ID=8468, JHM=2,\\ 2 optimal relations}
	\end{subfigure}
	\begin{subfigure}[b]{0.22\textwidth}
\centering
\includegraphics[width=0.45\textwidth]{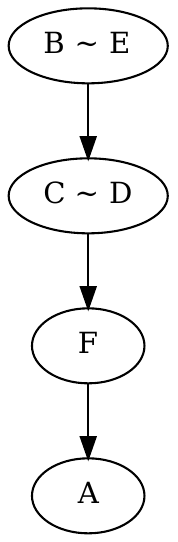}
\caption{\centering
	ID=8989, JHM=2,\\ 1 optimal relation}
	\end{subfigure}
	\begin{subfigure}[b]{0.24\textwidth}
\centering
\includegraphics[width=0.9\textwidth]{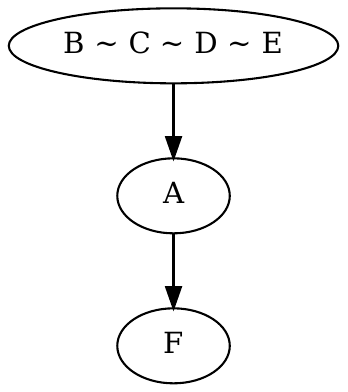}
\caption{\centering
	ID=5820, JHM=2.167,\\ 1 optimal relation}
	\end{subfigure}
\end{figure}

\begin{figure}[!htbp]
\begin{subfigure}[b]{0.22\textwidth}
\centering
\includegraphics[width=0.9\textwidth]{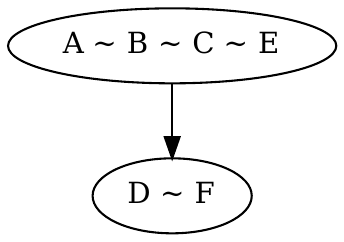}
\caption{\centering
	ID=4597, JHM=2.25,\\ 1 optimal relation}
	\end{subfigure}
	\begin{subfigure}[b]{0.23\textwidth}
\centering
\includegraphics[width=0.99\textwidth]{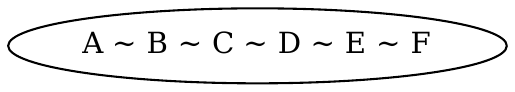}
\caption{\centering
	ID=4261, JHM=2.25,\\ 1 optimal relation}
	\end{subfigure}
	\begin{subfigure}[b]{0.22\textwidth}
\centering
\includegraphics[width=0.99\textwidth]{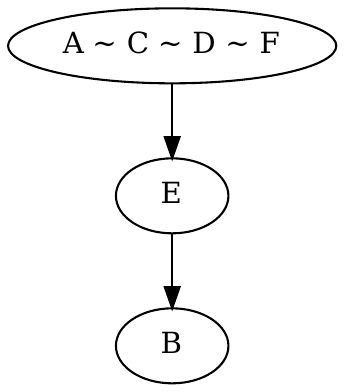}
\caption{\centering
	ID=9460, JHM=2.417,\\ 1 optimal relation}
	\end{subfigure}
	\begin{subfigure}[b]{0.23\textwidth}
\centering
\includegraphics[width=0.3\textwidth]{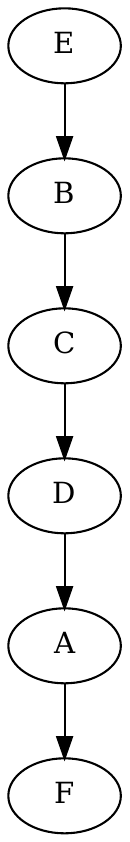}
\caption{\centering
	ID=5878, JHM=2.5,\\ 1 optimal relation}
	\end{subfigure}
\end{figure}

\begin{figure}[!htbp]
\begin{subfigure}[b]{0.26\textwidth}
\centering
\includegraphics[width=0.3\textwidth]{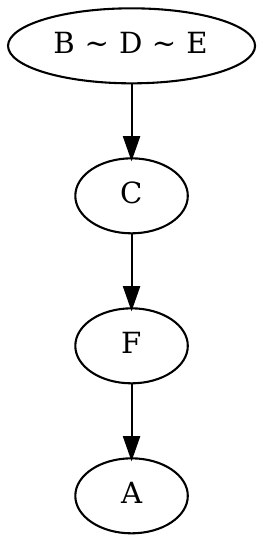}
\includegraphics[width=0.3\textwidth]{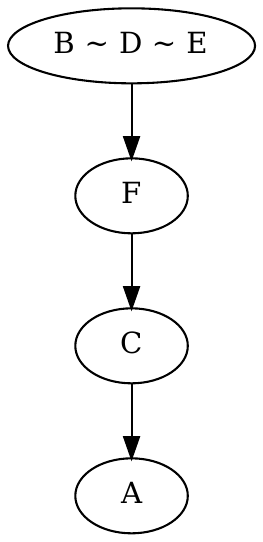}
\includegraphics[width=0.3\textwidth]{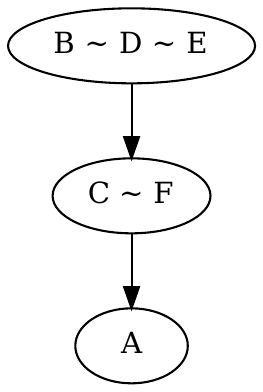}
\caption{\centering ID=2437, JHM=2.5,\\ 3 optimal relations}
\end{subfigure}
\begin{subfigure}[b]{0.22\textwidth}
\centering
\includegraphics[width=0.45\textwidth]{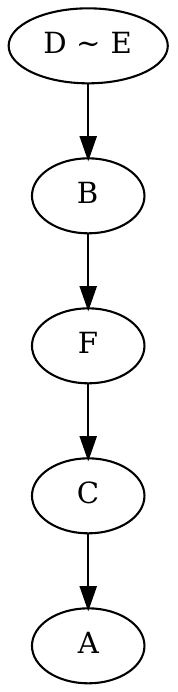}
\includegraphics[width=0.45\textwidth]{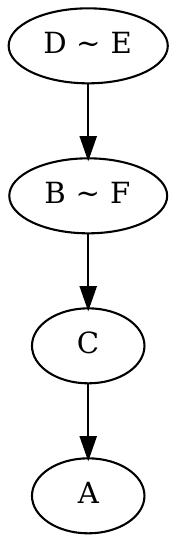}
\caption{\centering
	ID=4152, JHM=2.5,\\ 2 optimal relations}
	\end{subfigure}
	\begin{subfigure}[b]{0.22\textwidth}
\centering
\includegraphics[width=0.45\textwidth]{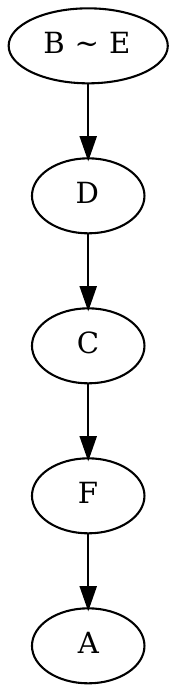}
\caption{\centering
	ID=8892, JHM=2.667,\\ 1 optimal relation}
	\end{subfigure}
	\begin{subfigure}[b]{0.22\textwidth}
\centering
\includegraphics[width=0.35\textwidth]{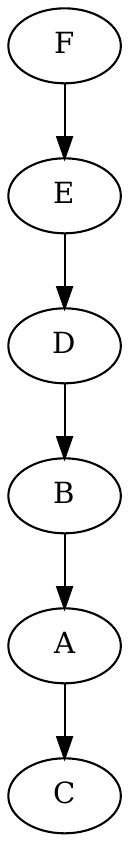}
\caption{\centering
	ID=5323, JHM=3,\\ 1 optimal relation}
	\end{subfigure}
\end{figure}

\begin{figure}[!htbp]
\begin{subfigure}[b]{0.22\textwidth}
\centering
\includegraphics[width=0.3\textwidth]{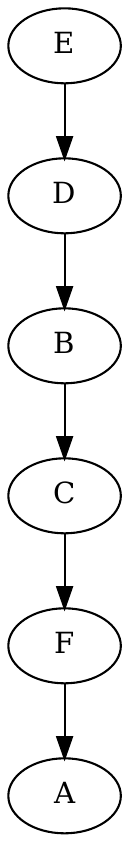}
\caption{\centering
	ID=7996, JHM=3,\\ 1 optimal relations}
	\end{subfigure}
	\begin{subfigure}[b]{0.22\textwidth}
\centering
\includegraphics[width=0.3\textwidth]{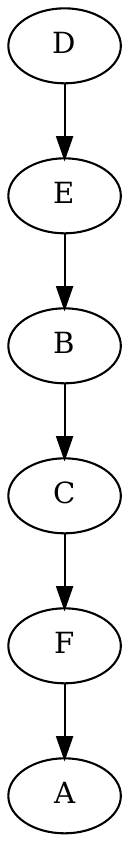}
\caption{\centering
	ID=2155, JHM=3,\\ 1 optimal relation}
	\end{subfigure}
	\begin{subfigure}[b]{0.22\textwidth}
\centering
\includegraphics[width=0.45\textwidth]{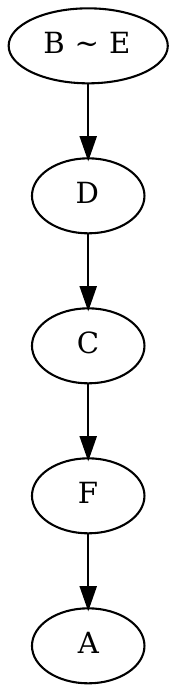}
\caption{\centering
	ID=3788, JHM=3,\\ 1 optimal relation}
	\end{subfigure}
	\begin{subfigure}[b]{0.22\textwidth}
\centering
\includegraphics[width=0.3\textwidth]{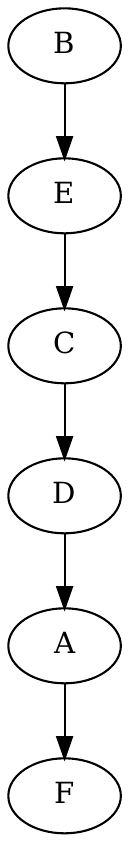}
\caption{\centering
	ID=4924, JHM=3,\\ 1 optimal relation}
	\end{subfigure}
\end{figure}

\begin{figure}[!htbp]
\centering
\begin{subfigure}[b]{0.22\textwidth}
\centering
\includegraphics[width=0.45\textwidth]{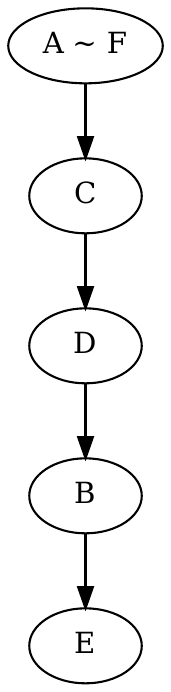}
\caption{\centering
	ID=8732, JHM=3,\\ 1 optimal relation}
	\end{subfigure}
	\begin{subfigure}[b]{0.22\textwidth}
\centering
\includegraphics[width=0.45\textwidth]{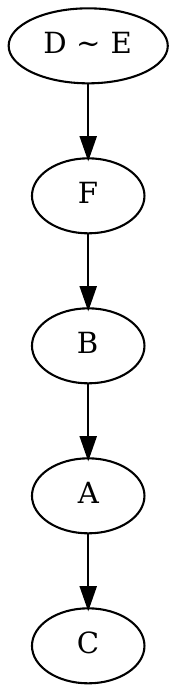}
\caption{\centering
	ID=9899, JHM=3,\\ 1 optimal relation}
	\end{subfigure}	
	\begin{subfigure}[b]{0.25\textwidth}
\centering
\includegraphics[width=0.45\textwidth]{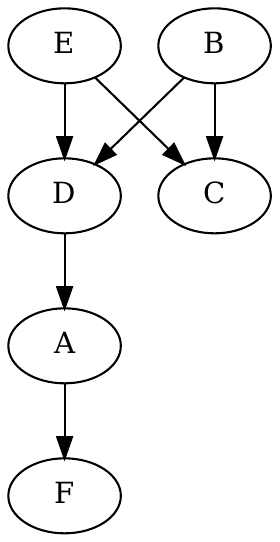}
\includegraphics[width=0.45\textwidth]{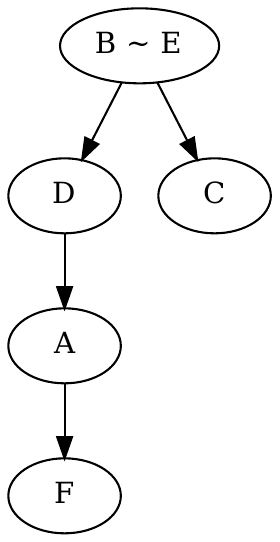}
\caption{\centering
	ID=9292, JHM=3.167,\\ 2 optimal relations}
	\end{subfigure}
	\begin{subfigure}[b]{0.22\textwidth}
\centering
\includegraphics[width=0.7\textwidth]{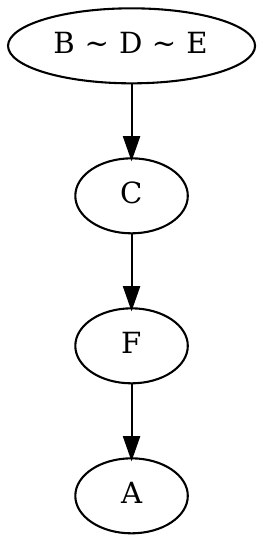}
\caption{\centering
	ID=4652, JHM=3.167,\\ 1 optimal relation}
	\end{subfigure}
\end{figure}

\begin{figure}[!htbp]
\centering
\begin{subfigure}[b]{0.22\textwidth}
\centering
\includegraphics[width=0.9\textwidth]{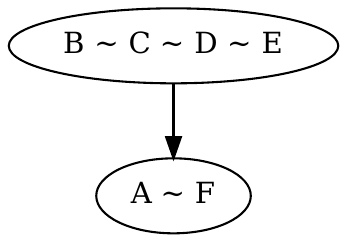}
\caption{\centering
	ID=6384, JHM=3.167,\\ 1 optimal relation}
	\end{subfigure}
	\begin{subfigure}[b]{0.22\textwidth}
\centering
\includegraphics[width=0.3\textwidth]{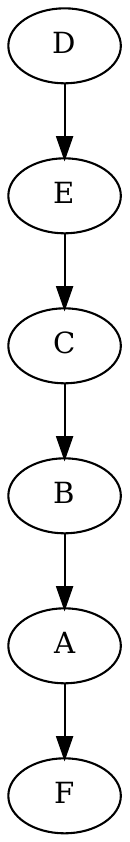}
\caption{\centering
	ID=7215, JHM=3.5,\\ 1 optimal relation}
	\end{subfigure}
	\begin{subfigure}[b]{0.22\textwidth}
\centering
\includegraphics[width=0.9\textwidth]{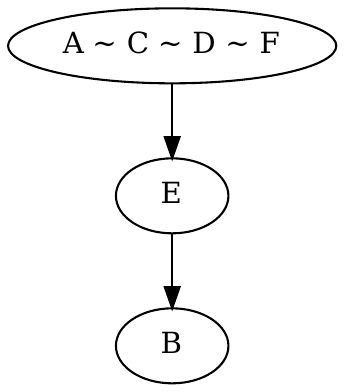}
\caption{\centering
	ID=9134, JHM=3.583,\\ 1 optimal relation}
	\end{subfigure}
	\begin{subfigure}[b]{0.22\textwidth}
\centering
\includegraphics[width=0.45\textwidth]{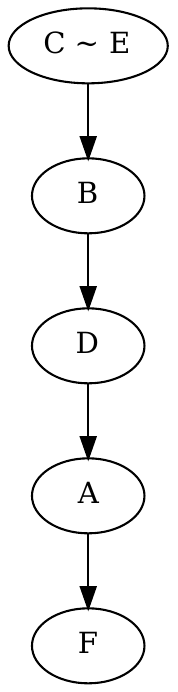}
\caption{\centering
	ID=2864, JHM=3.667,\\ 1 optimal relation}
	\end{subfigure}
\end{figure}

\begin{figure}[!htbp]
\centering
\begin{subfigure}[b]{0.22\textwidth}
\centering
\includegraphics[width=0.7\textwidth]{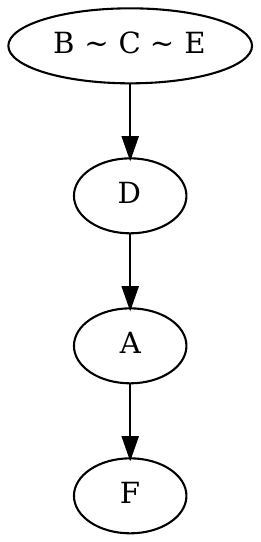}
\caption{\centering
	ID=9985, JHM=3.667,\\ 1 optimal relation}
	\end{subfigure}
	\begin{subfigure}[b]{0.3\textwidth}
\centering
\includegraphics[width=0.31\textwidth]{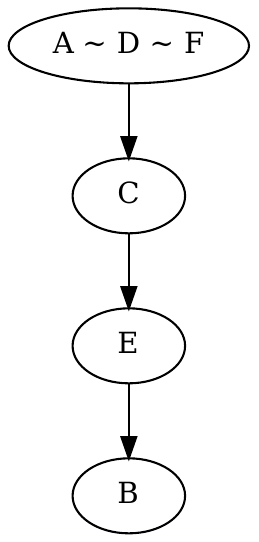}
\includegraphics[width=0.31\textwidth]{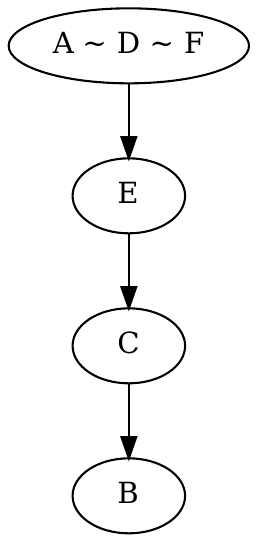}
\includegraphics[width=0.31\textwidth]{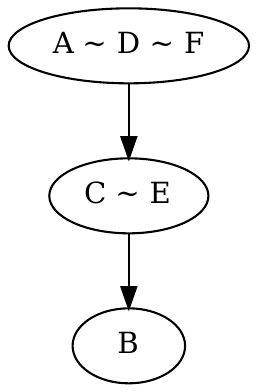}
\caption{\centering
	ID=1873, JHM=3.667,\\ 3 optimal relations}
	\end{subfigure}	
	\begin{subfigure}[b]{0.22\textwidth}
\centering
\includegraphics[width=0.9\textwidth]{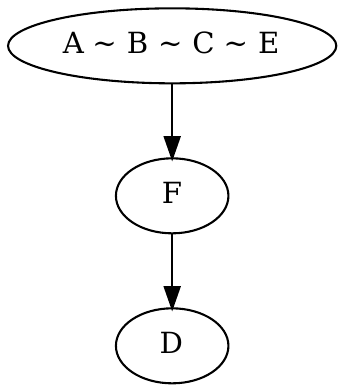}
\caption{\centering
	ID=4587, JHM=3.75,\\ 1 optimal relation}
	\end{subfigure}
	\begin{subfigure}[b]{0.23\textwidth}
\centering
\includegraphics[width=0.4\textwidth]{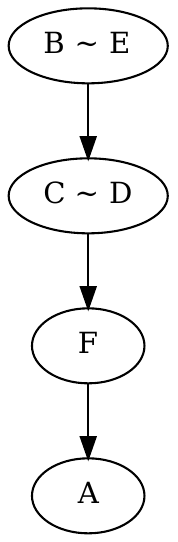}
\caption{\centering
	ID=5646, JHM=4,\\ 1 optimal relation}
	\end{subfigure}
\end{figure}

\begin{figure}[!htbp]
\centering
\begin{subfigure}[b]{0.22\textwidth}
\centering
\includegraphics[width=0.45\textwidth]{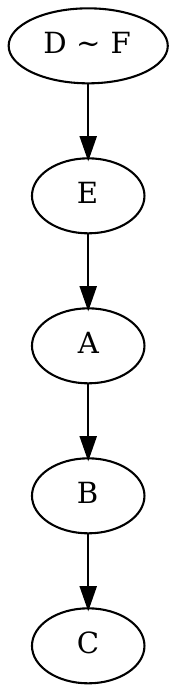}
\includegraphics[width=0.45\textwidth]{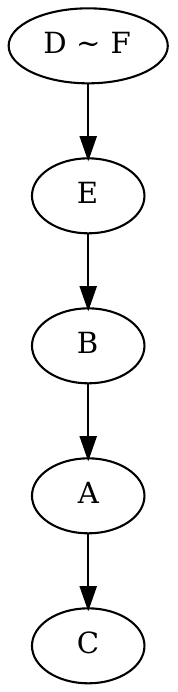}
\caption{\centering
	ID=2753, JHM=4,\\ 2 optimal relations}
	\end{subfigure}
	\begin{subfigure}[b]{0.22\textwidth}
\centering
\includegraphics[width=0.45\textwidth]{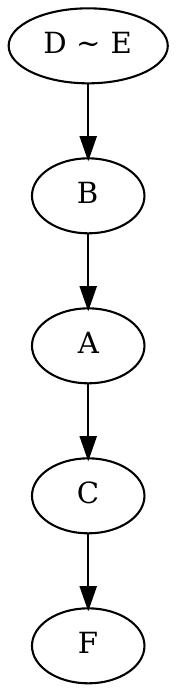}
\caption{\centering
	ID=1727, JHM=4,\\ 1 optimal relation}
	\end{subfigure}
	\begin{subfigure}[b]{0.22\textwidth}
\centering
\includegraphics[width=0.3\textwidth]{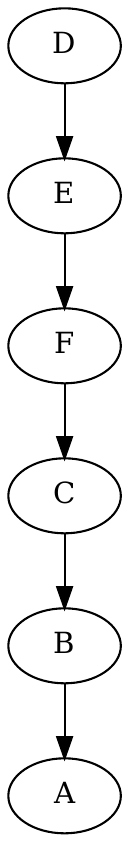}
\caption{\centering
	ID=6715, JHM=4.5,\\ 1 optimal relation}
	\end{subfigure}
	\begin{subfigure}[b]{0.22\textwidth}
\centering
\includegraphics[width=0.45\textwidth]{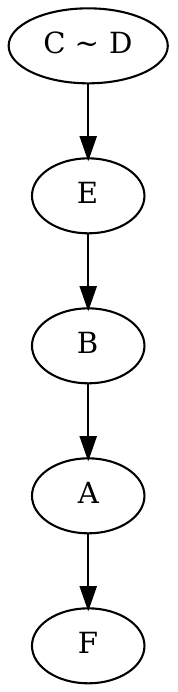}
\caption{\centering
	ID=6657, JHM=4.833,\\ 1 optimal relation}
	\end{subfigure}
\end{figure}

\begin{figure}[!htbp]
\centering
\begin{subfigure}[b]{0.32\textwidth}
\centering
\includegraphics[width=0.48\textwidth]{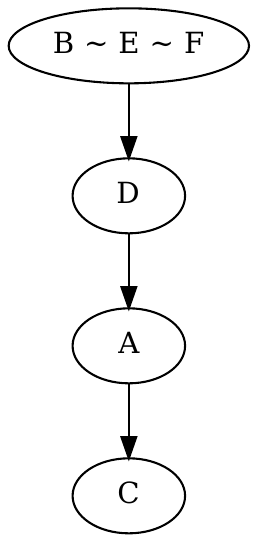}
\includegraphics[width=0.48\textwidth]{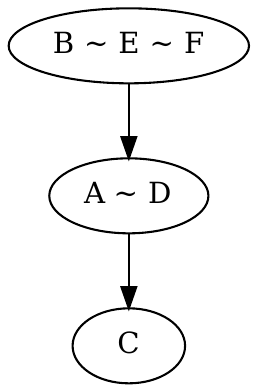}
\caption{\centering
	ID=4285, JHM=4.833,\\ 2 optimal relations}
	\end{subfigure}
	\begin{subfigure}[b]{0.32\textwidth}
\centering
\includegraphics[width=0.5\textwidth]{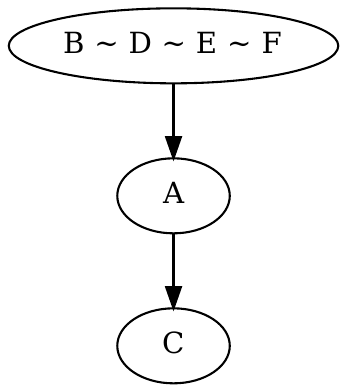}
\caption{\centering 
	ID=6535, JHM=4.833,\\ 1 optimal relation}
	\end{subfigure}
	\begin{subfigure}[b]{0.32\textwidth}
\centering
\includegraphics[width=0.5\textwidth]{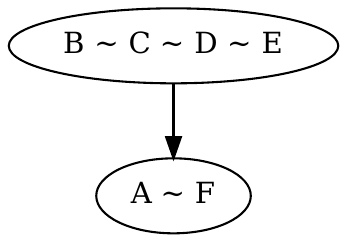}
\caption{\centering
	ID=7735, JHM=4.833,\\ 1 optimal relation}
	\end{subfigure}
\end{figure}

\begin{figure}[!htbp]
\begin{subfigure}[b]{0.22\textwidth}
\centering
\includegraphics[width=0.9\textwidth]{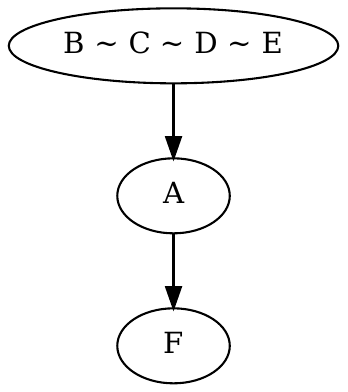}
\caption{\centering
	ID=8542, JHM=4.833,\\ 1 optimal relation}
	\end{subfigure}
	\begin{subfigure}[b]{0.22\textwidth}
\centering
\includegraphics[width=0.9\textwidth]{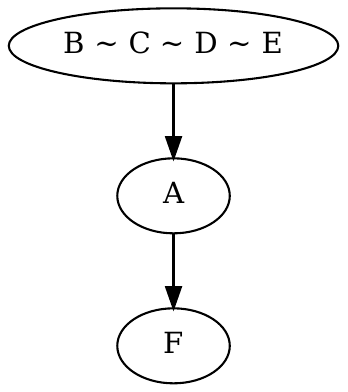}
\caption{\centering
	ID=9294, JHM=4.917,\\ 1 optimal relation}
	\end{subfigure}
	\begin{subfigure}[b]{0.22\textwidth}
\centering
\includegraphics[width=0.45\textwidth]{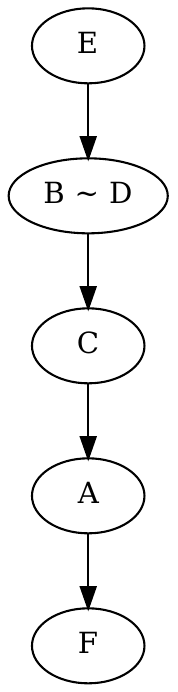}
\caption{\centering
	ID=7763, JHM=5,\\ 1 optimal relation}
	\end{subfigure}
	\begin{subfigure}[b]{0.22\textwidth}
\centering
\includegraphics[width=0.7\textwidth]{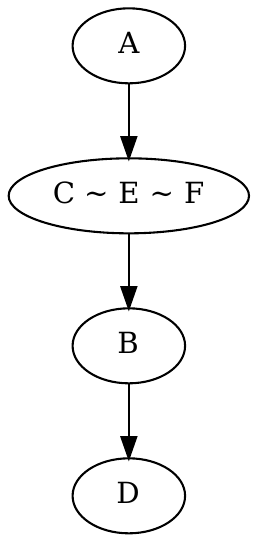}
\caption{\centering
	ID=7332, JHM=5,\\ 1 optimal relation}
	\end{subfigure}
\end{figure}

\pagebreak

\subsection{Undominated Choice with Incomplete Preferences}

\begin{figure}[!htbp]
\centering
\includegraphics[width=0.25\textwidth]{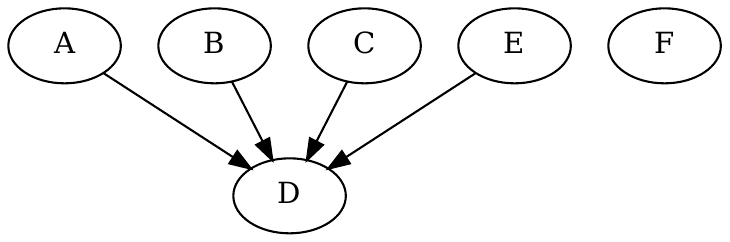}
\includegraphics[width=0.22\textwidth]{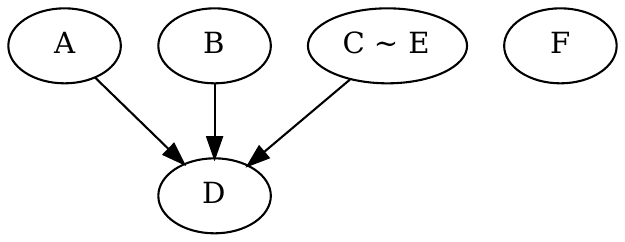}
\includegraphics[width=0.22\textwidth]{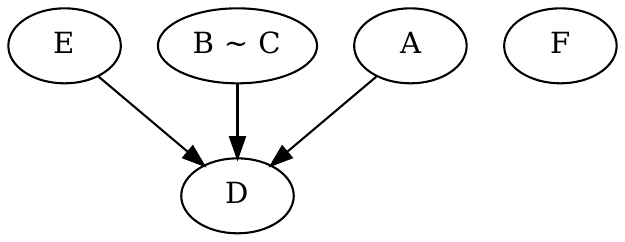}
\includegraphics[width=0.22\textwidth]{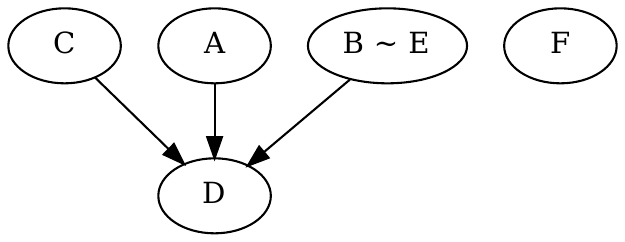}
\includegraphics[width=0.22\textwidth]{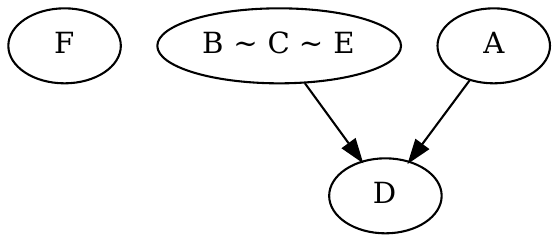}
\includegraphics[width=0.22\textwidth]{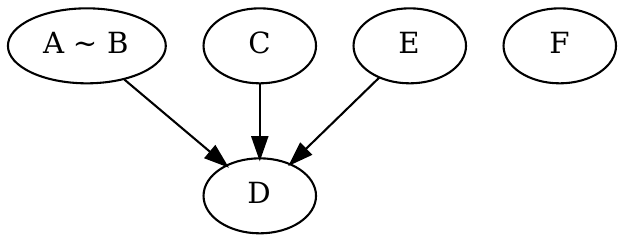}
\includegraphics[width=0.22\textwidth]{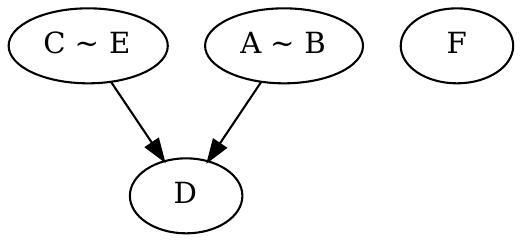}
\includegraphics[width=0.22\textwidth]{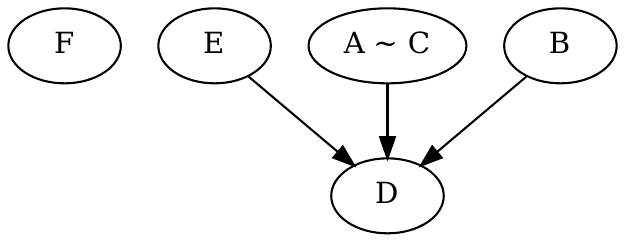}
\includegraphics[width=0.22\textwidth]{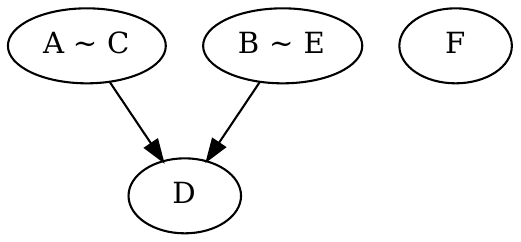}
\includegraphics[width=0.22\textwidth]{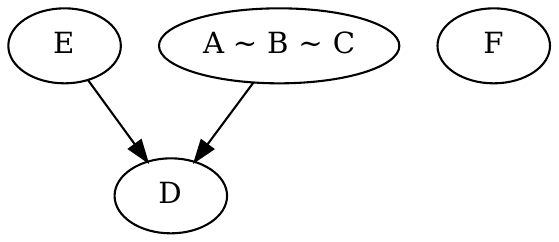}
\includegraphics[width=0.22\textwidth]{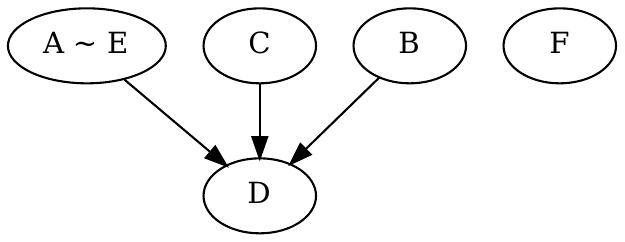}
\includegraphics[width=0.22\textwidth]{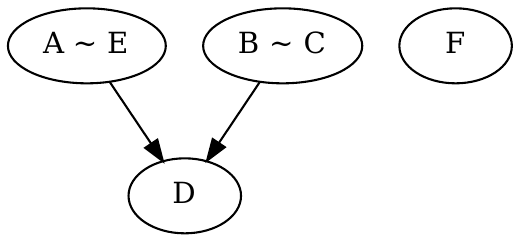}
\includegraphics[width=0.22\textwidth]{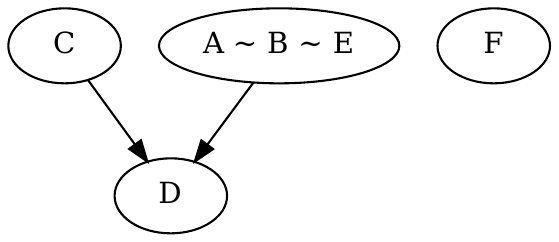}
\includegraphics[width=0.22\textwidth]{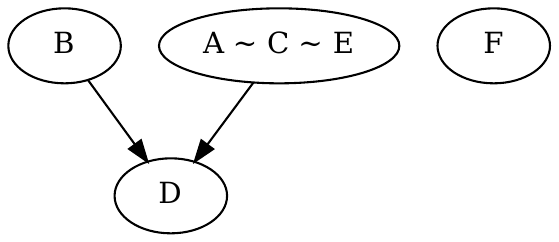}
\includegraphics[width=0.22\textwidth]{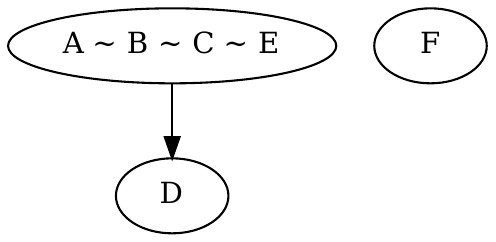}
\caption{\centering ID=1889, JHM=0.667, 15 optimal relations}
\end{figure}

\begin{figure}[!htbp]
\centering
\includegraphics[width=0.25\textwidth]{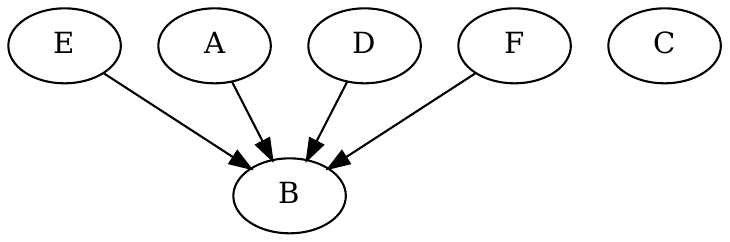}
\includegraphics[width=0.22\textwidth]{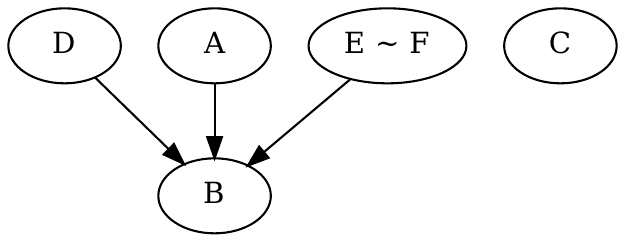}
\includegraphics[width=0.22\textwidth]{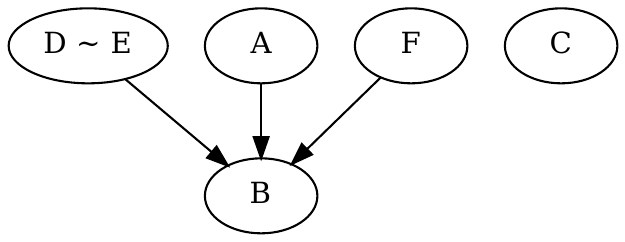}
\includegraphics[width=0.22\textwidth]{/graphs/FC-UC-1-9001-3.pdf}
\includegraphics[width=0.22\textwidth]{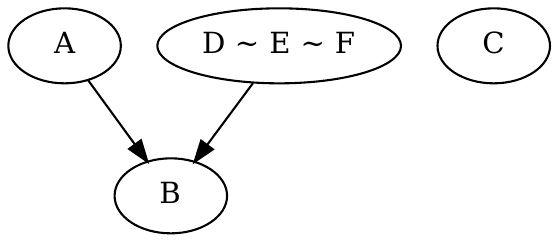}
\includegraphics[width=0.22\textwidth]{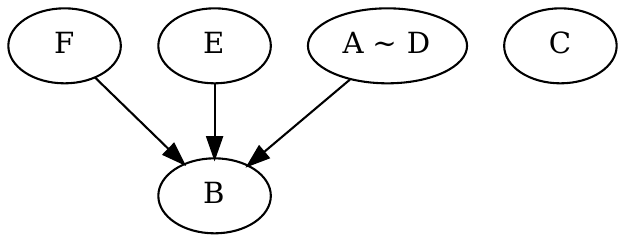}
\includegraphics[width=0.22\textwidth]{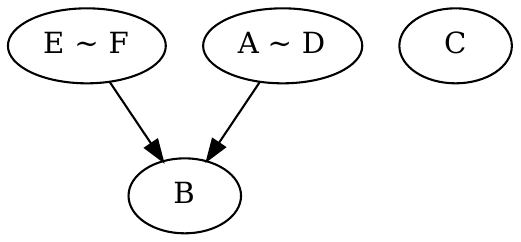}
\includegraphics[width=0.22\textwidth]{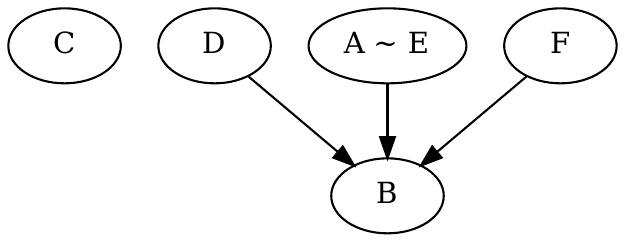}
\includegraphics[width=0.22\textwidth]{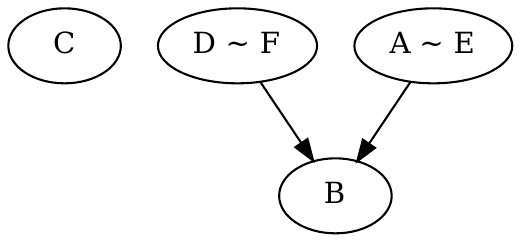}
\includegraphics[width=0.22\textwidth]{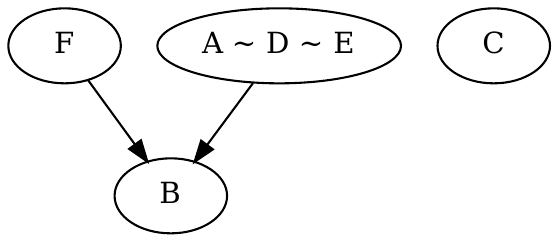}
\includegraphics[width=0.22\textwidth]{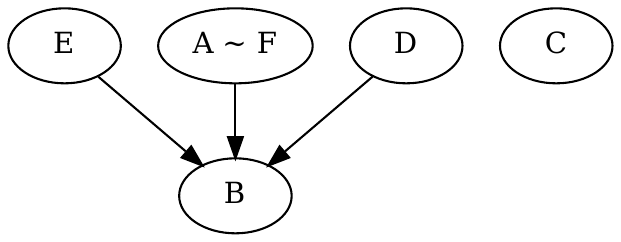}
\includegraphics[width=0.22\textwidth]{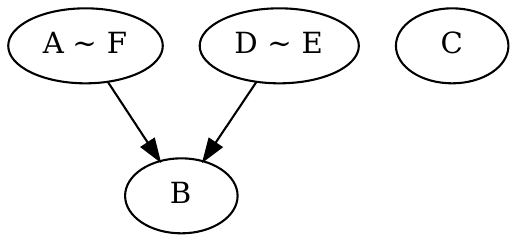}
\includegraphics[width=0.22\textwidth]{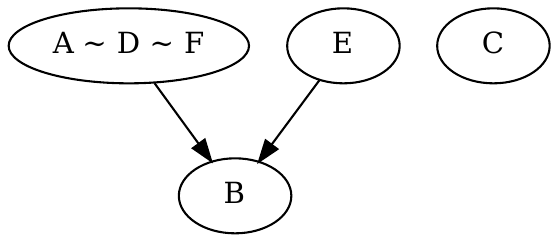}
\includegraphics[width=0.22\textwidth]{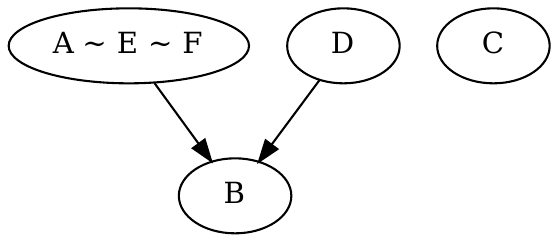}
\includegraphics[width=0.22\textwidth]{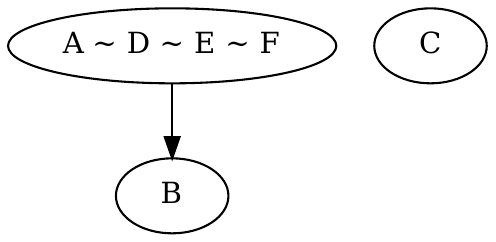}
\caption{\centering ID=9001, JHM=1, 15 optimal relations}
\end{figure}

\begin{figure}[!htbp]
\begin{subfigure}[b]{0.22\textwidth}
\centering
\includegraphics[width=0.75\textwidth]{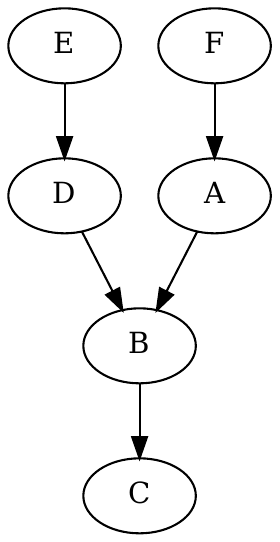}
\caption{\centering
	ID=8945, JHM=1.333,\\ 1 optimal relation}
	\end{subfigure}
	\begin{subfigure}[b]{0.77\textwidth}
\centering
\includegraphics[width=0.25\textwidth]{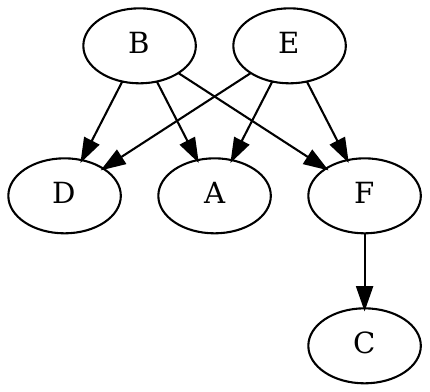}
\includegraphics[width=0.25\textwidth]{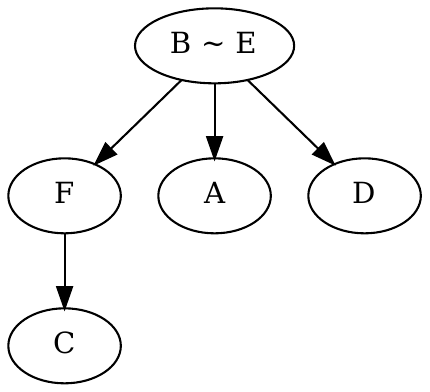}
\includegraphics[width=0.20\textwidth]{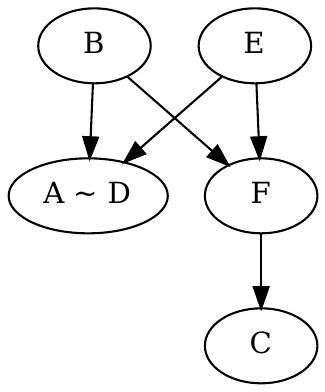}
\includegraphics[width=0.20\textwidth]{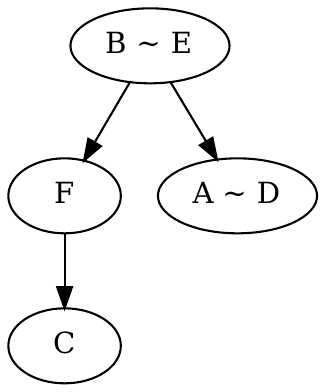}
\caption{\centering
	ID=6948, JHM=1.5, 4 optimal relations}
	\end{subfigure}
\end{figure}

\begin{figure}[!htbp]
\begin{subfigure}[b]{0.22\textwidth}
\centering
\includegraphics[width=0.75\textwidth]{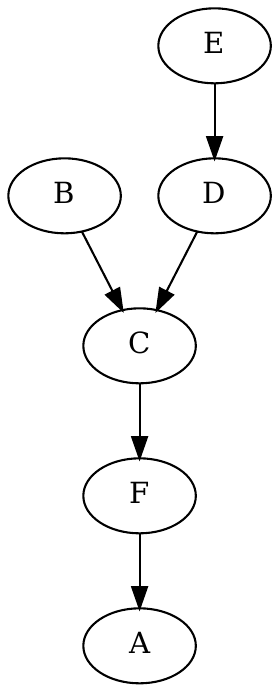}
\caption{\centering
	ID=8721, JHM=1.5,\\ 1 optimal relation}
	\end{subfigure}
	\begin{subfigure}[b]{0.77\textwidth}
\centering
\includegraphics[width=0.26\textwidth]{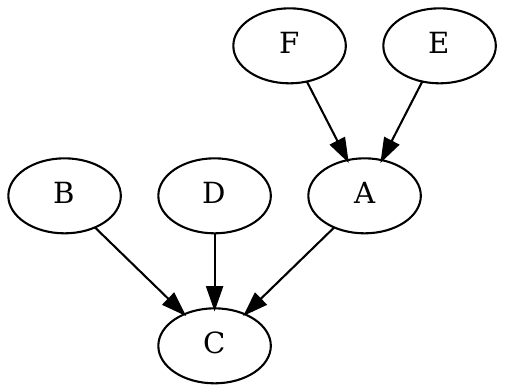}
\includegraphics[width=0.25\textwidth]{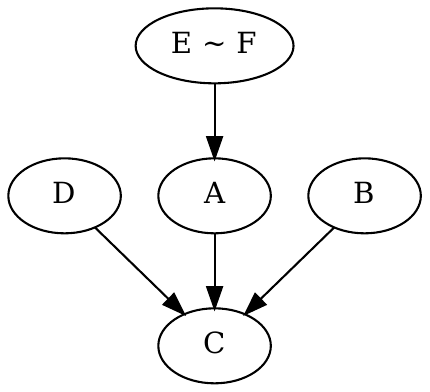}
\includegraphics[width=0.22\textwidth]{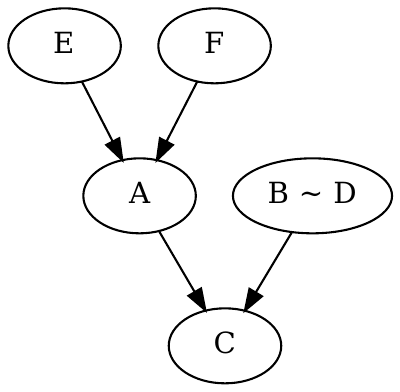}
\includegraphics[width=0.20\textwidth]{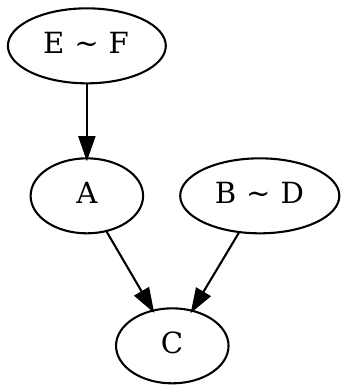}
\caption{\centering
	ID=6521, JHM=2.083, 4 optimal relations}
	\end{subfigure}
\end{figure}

\begin{figure}[!htbp]
\begin{subfigure}[b]{0.48\textwidth}
\centering
\includegraphics[width=0.54\textwidth]{/graphs/FC-UC-2.167-6140-1.pdf}
\includegraphics[width=0.42\textwidth]{/graphs/FC-UC-2.167-6140-2.pdf}
\caption{\centering
	ID=6140, JHM=2.167, 2 optimal relations}
	\end{subfigure}
	\begin{subfigure}[b]{0.48\textwidth}
\centering
\includegraphics[width=0.54\textwidth]{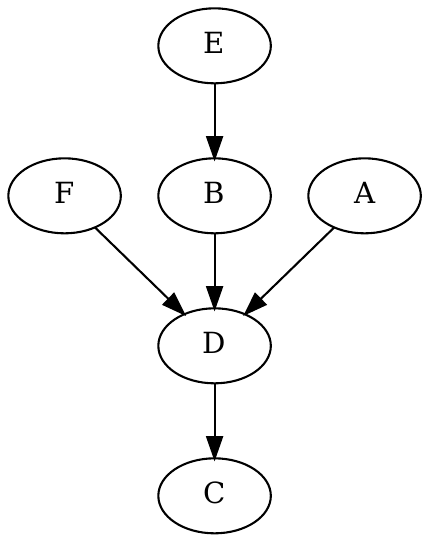}
\includegraphics[width=0.42\textwidth]{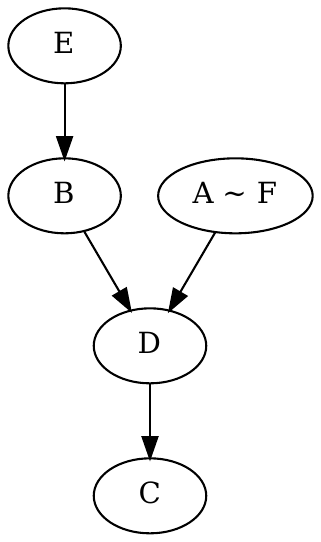}
\caption{\centering
	ID=2432, JHM=3, 2 optimal relations}
	\end{subfigure}	
\end{figure}

\begin{figure}
\centering
\includegraphics[width=0.25\textwidth]{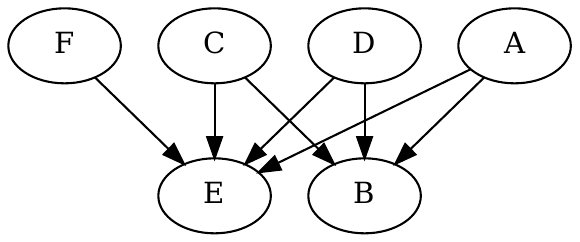}
\includegraphics[width=0.25\textwidth]{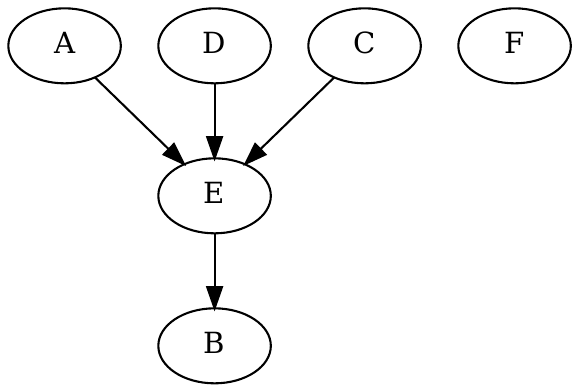}
\includegraphics[width=0.22\textwidth]{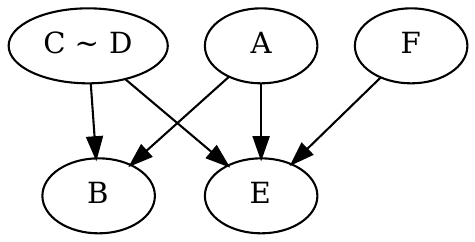}
\includegraphics[width=0.22\textwidth]{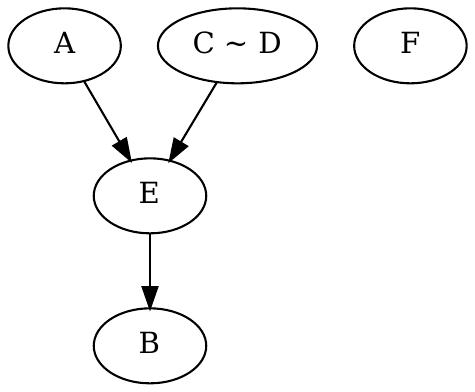}
\includegraphics[width=0.25\textwidth]{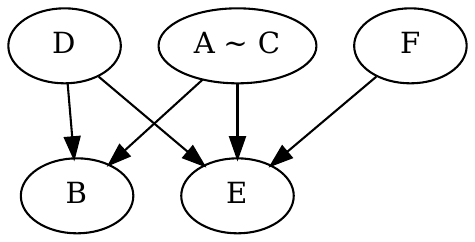}
\includegraphics[width=0.25\textwidth]{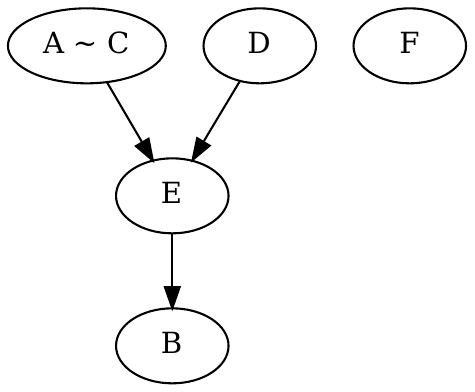}
\includegraphics[width=0.25\textwidth]{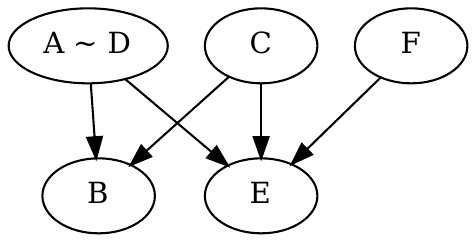}
\includegraphics[width=0.25\textwidth]{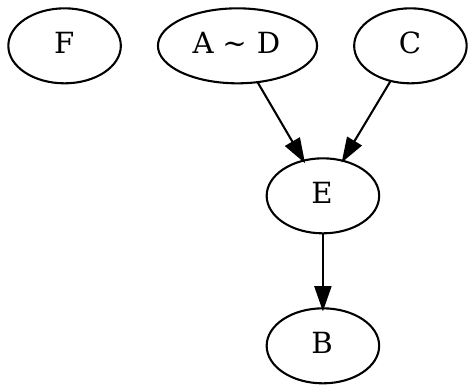}
\includegraphics[width=0.22\textwidth]{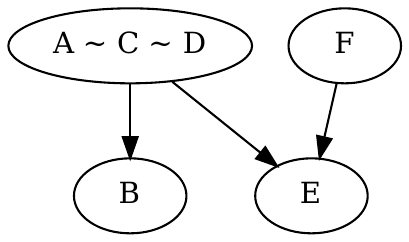}
\includegraphics[width=0.22\textwidth]{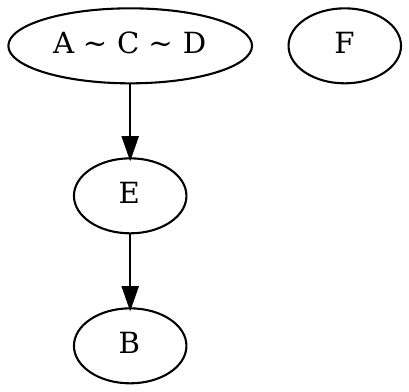}
\caption{\centering ID=7017, JHM=3.083, 10 optimal relations}
\end{figure}

\begin{figure}
\centering
\begin{subfigure}[b]{0.3\textwidth}
\centering
\includegraphics[width=0.48\textwidth]{/graphs/NFC-UC-3.167-9292-1.pdf}
\includegraphics[width=0.48\textwidth]{/graphs/NFC-UC-3.167-9292-2.pdf}
\caption{\centering
	ID=9292, JHM=3.167,\\ 2 optimal relations}
	\end{subfigure}
	\begin{subfigure}[b]{0.3\textwidth}
\centering
\includegraphics[width=0.48\textwidth]{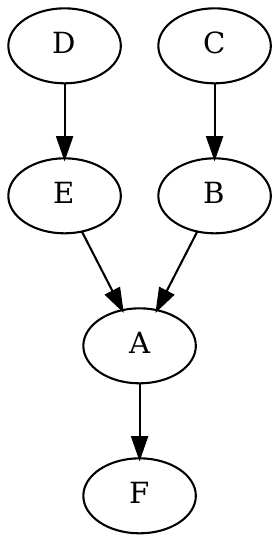}
\caption{\centering
	ID=9222, JHM=3.333,\\ 1 optimal relation}
	\end{subfigure}
	\begin{subfigure}[b]{0.35\textwidth}
\centering
\includegraphics[width=0.54\textwidth]{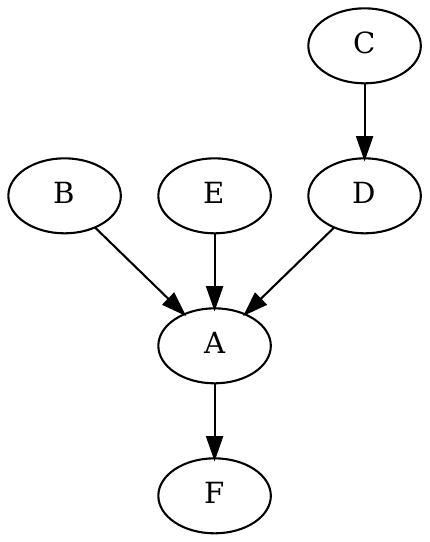}
\includegraphics[width=0.43\textwidth]{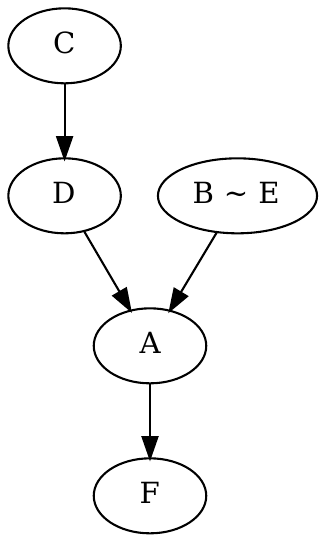}
\caption{\centering
	ID=8075, JHM=3.5,\\ 2 optimal relations}
	\end{subfigure}	
\end{figure}

\begin{figure}
\centering
\begin{subfigure}[b]{0.3\textwidth}
\centering
\includegraphics[width=0.48\textwidth]{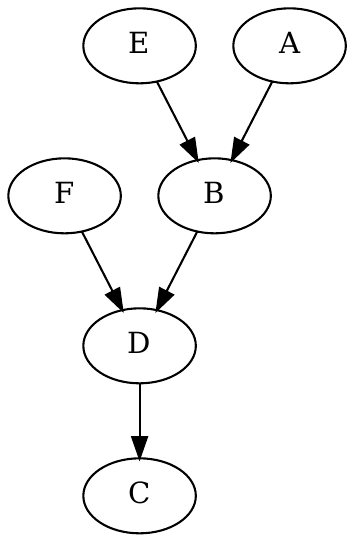}
\includegraphics[width=0.48\textwidth]{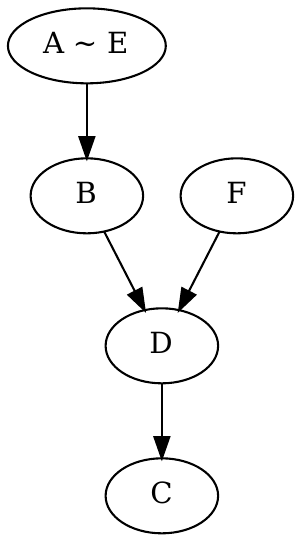}
\caption{\centering
	ID=8719, JHM=3.667,\\ 2 optimal relations}
	\end{subfigure}
	\begin{subfigure}[b]{0.3\textwidth}
\centering
\includegraphics[width=0.48\textwidth]{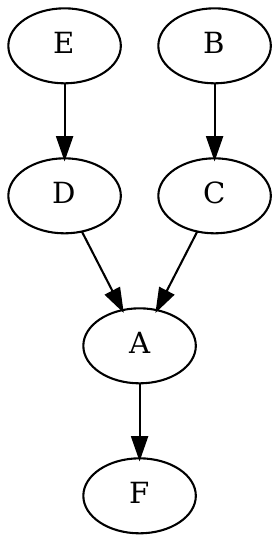}
\caption{\centering
	ID=3574, JHM=4,\\ 1 optimal relation}
	\end{subfigure}
	\begin{subfigure}[b]{0.3\textwidth}
\centering
\includegraphics[width=0.48\textwidth]{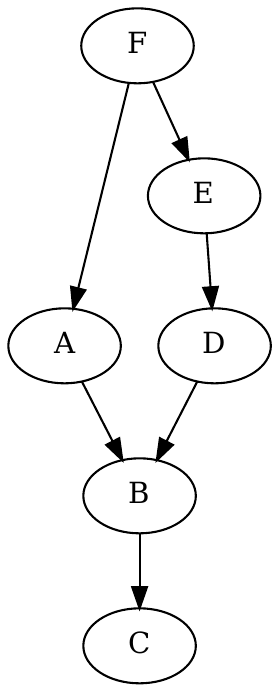}
\caption{\centering
	ID=7423, JHM=4,\\ 1 optimal relation}
	\end{subfigure}
\end{figure}

\begin{figure}
\centering
\begin{subfigure}[b]{0.3\textwidth}
\centering
\includegraphics[width=0.48\textwidth]{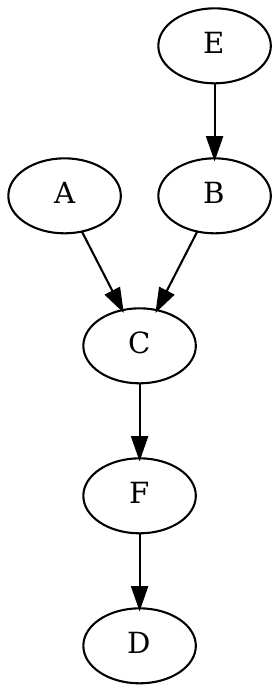}
\caption{\centering
	ID=9819, JHM=4,\\ 1 optimal relation}
	\end{subfigure}
	\begin{subfigure}[b]{0.3\textwidth}
\centering
\includegraphics[width=0.48\textwidth]{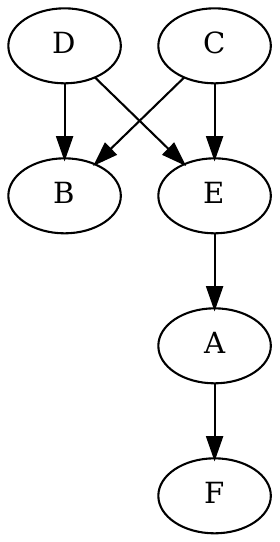}
\includegraphics[width=0.48\textwidth]{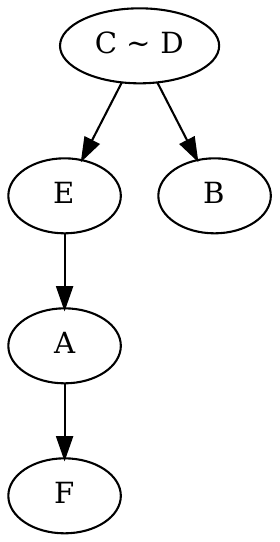}
\caption{\centering
	ID=5049, JHM=4.167,\\ 2 optimal relations}
	\end{subfigure}	
\end{figure}

\begin{figure}
\centering
\includegraphics[width=0.19\textwidth]{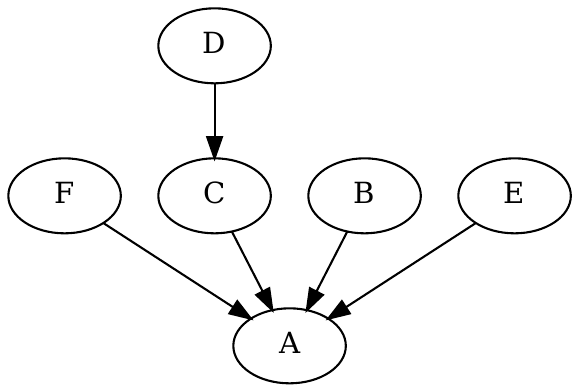}
\includegraphics[width=0.19\textwidth]{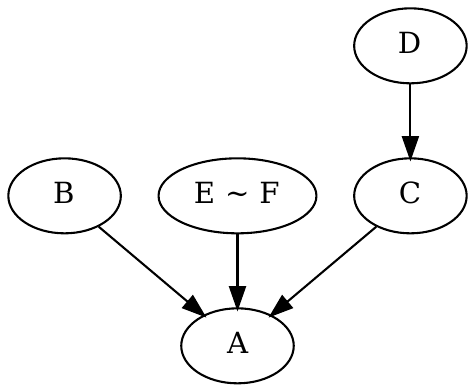}
\includegraphics[width=0.19\textwidth]{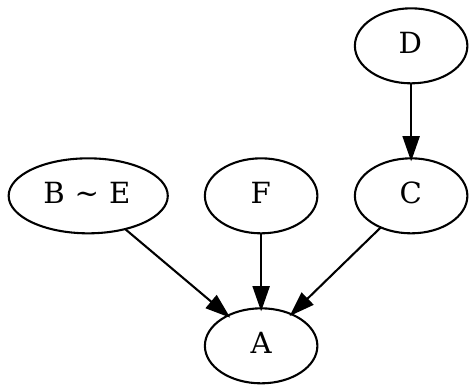}
\includegraphics[width=0.19\textwidth]{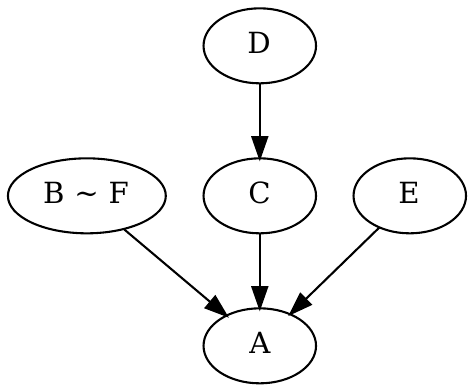}
\includegraphics[width=0.19\textwidth]{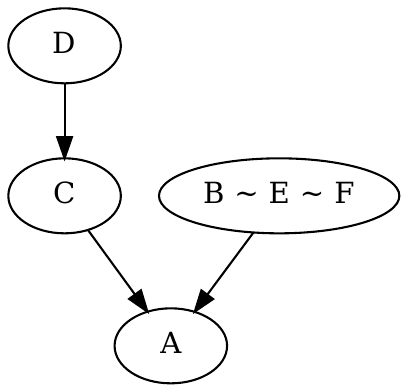}
\caption{\centering ID=5283, JHM=4.167, 5 optimal relations}
\end{figure}

\begin{figure}
\includegraphics[width=0.22\textwidth]{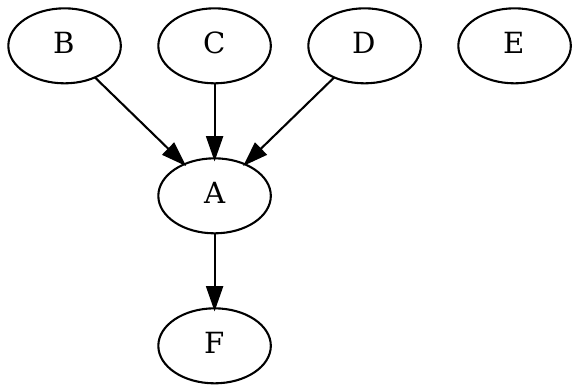}
\includegraphics[width=0.22\textwidth]{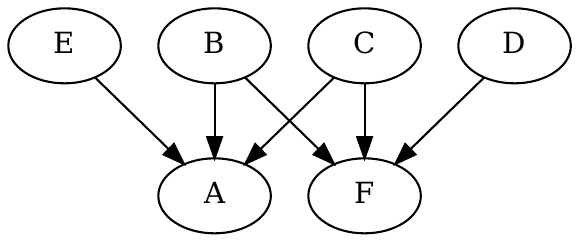}
\includegraphics[width=0.22\textwidth]{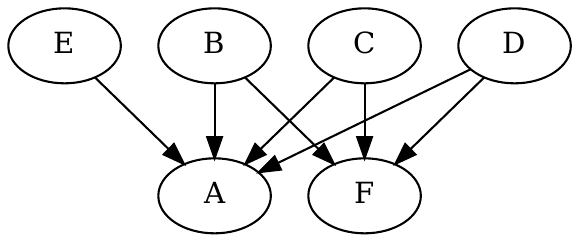}
\includegraphics[width=0.22\textwidth]{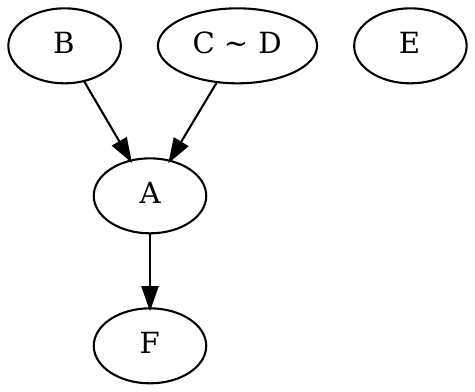}
\includegraphics[width=0.22\textwidth]{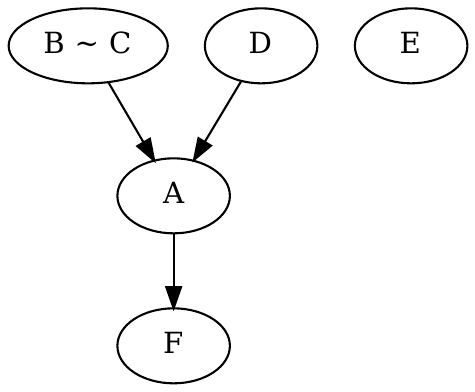}
\includegraphics[width=0.22\textwidth]{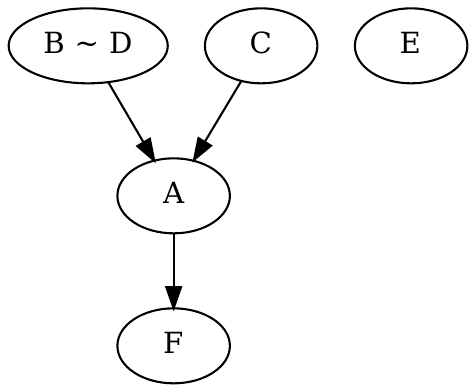}
\includegraphics[width=0.22\textwidth]{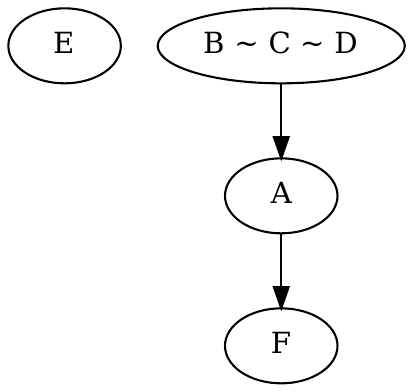}
\includegraphics[width=0.22\textwidth]{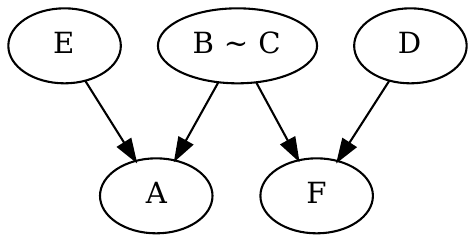}
\includegraphics[width=0.22\textwidth]{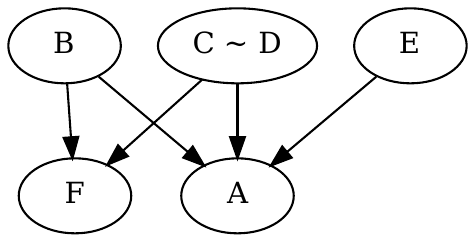}
\includegraphics[width=0.22\textwidth]{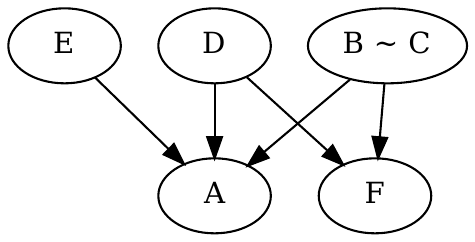}
\includegraphics[width=0.22\textwidth]{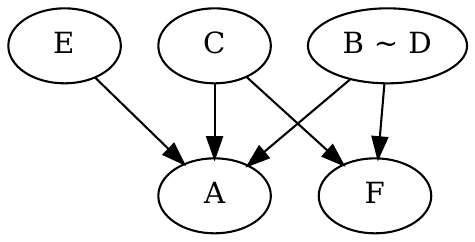}
\includegraphics[width=0.22\textwidth]{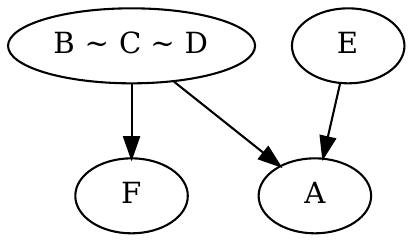}
\caption{\centering ID=8049, JHM=4.25, 12 optimal relations}
\end{figure}

\begin{figure}
\centering
\begin{subfigure}[b]{0.3\textwidth}
\centering
\includegraphics[width=0.48\textwidth]{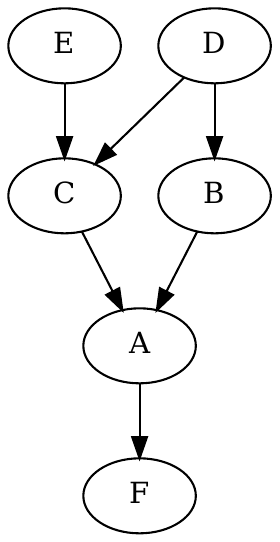}
\caption{\centering
	ID=2038, JHM=4.667,\\ 1 optimal relation}
	\end{subfigure}
	\begin{subfigure}[b]{0.3\textwidth}
\centering
\includegraphics[width=0.53\textwidth]{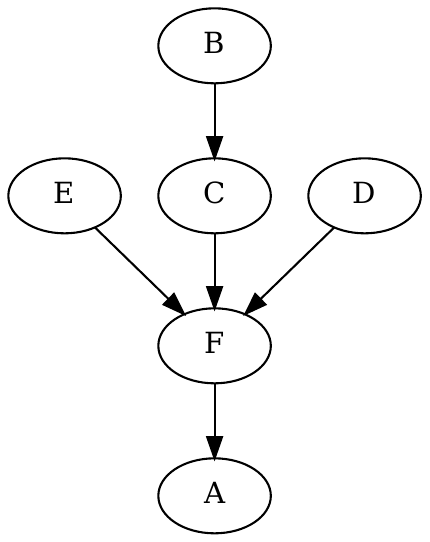}
\includegraphics[width=0.43\textwidth]{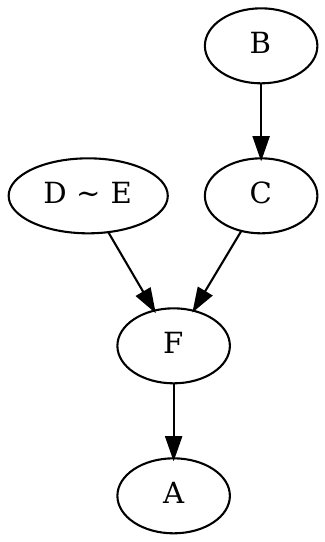}
\caption{\centering
	ID=2375, JHM=4.667,\\ 2 optimal relations}
	\end{subfigure}	
\end{figure}

\begin{figure}
\centering
\includegraphics[width=0.27\textwidth]{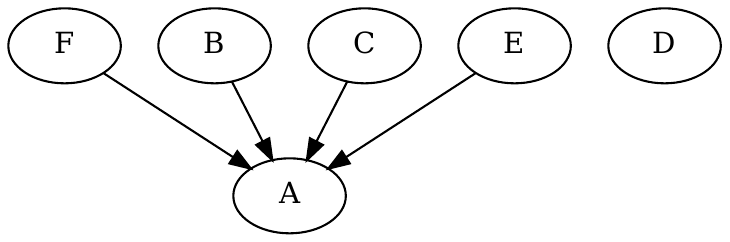}
\includegraphics[width=0.24\textwidth]{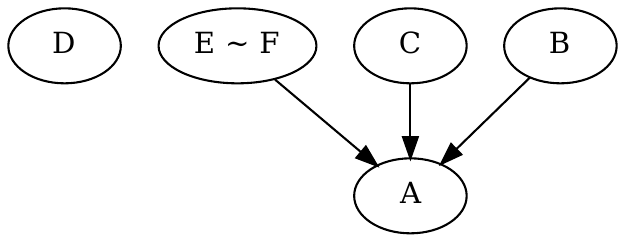}
\includegraphics[width=0.24\textwidth]{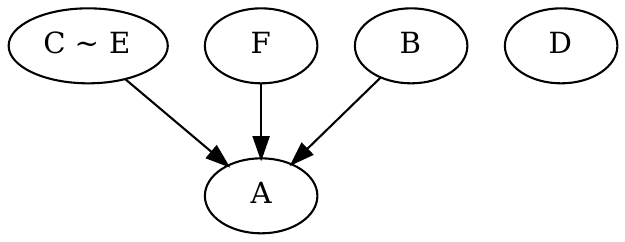}
\includegraphics[width=0.24\textwidth]{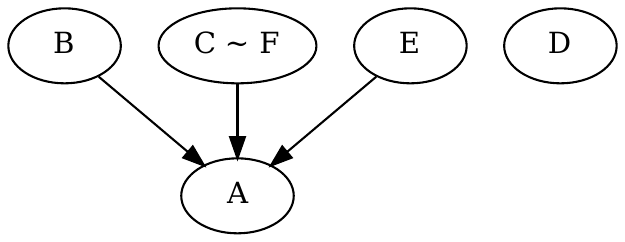}
\includegraphics[width=0.24\textwidth]{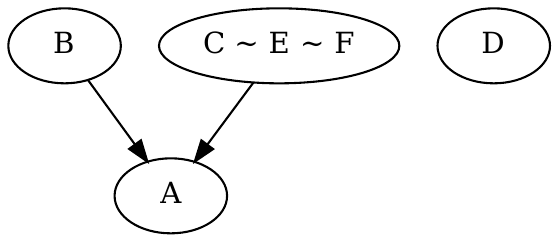}
\includegraphics[width=0.24\textwidth]{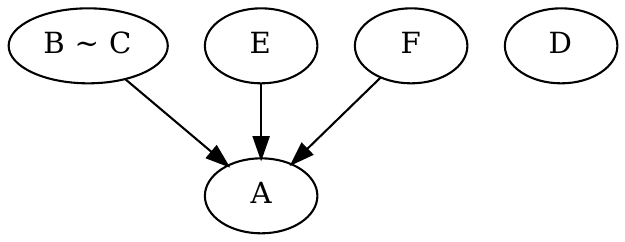}
\includegraphics[width=0.24\textwidth]{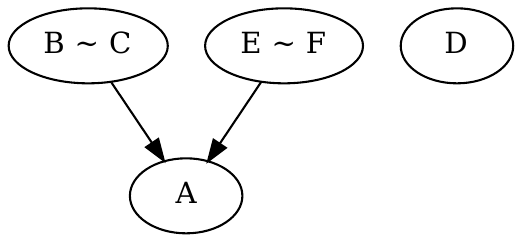}
\includegraphics[width=0.24\textwidth]{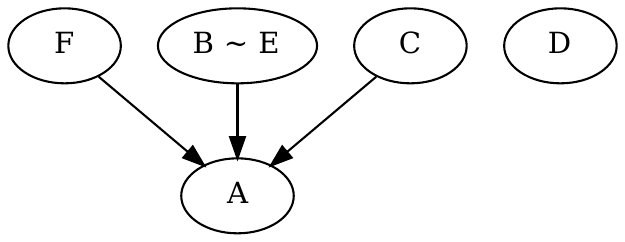}
\includegraphics[width=0.24\textwidth]{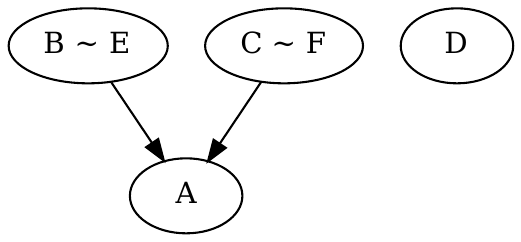}
\includegraphics[width=0.24\textwidth]{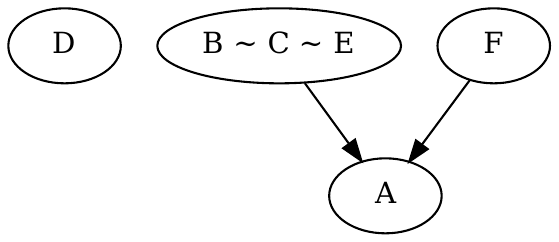}
\includegraphics[width=0.24\textwidth]{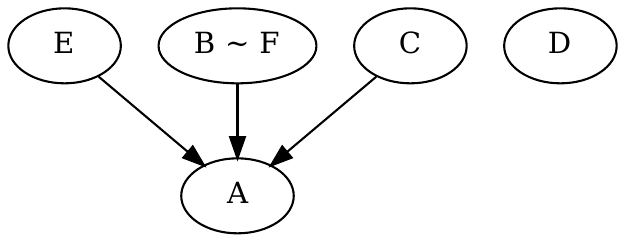}
\includegraphics[width=0.24\textwidth]{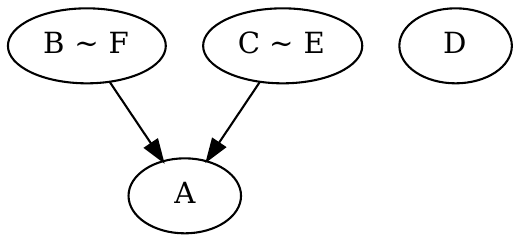}
\includegraphics[width=0.24\textwidth]{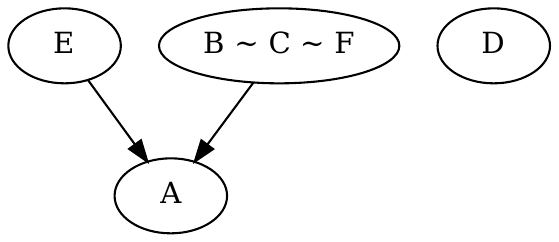}
\includegraphics[width=0.24\textwidth]{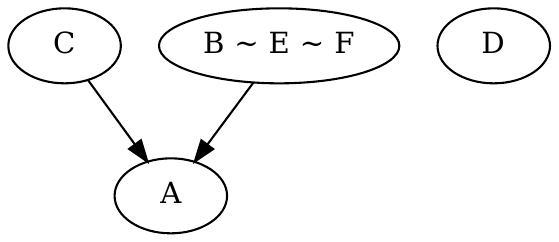}
\includegraphics[width=0.24\textwidth]{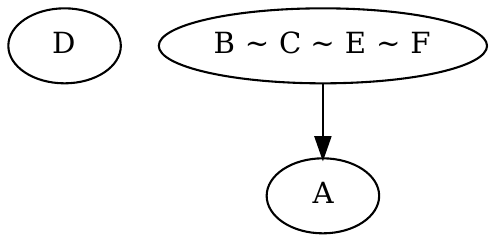}
\caption{\centering ID=1308, JHM=4.833, 15 optimal relations}
\end{figure}

\begin{figure}
\centering
\includegraphics[width=0.2\textwidth]{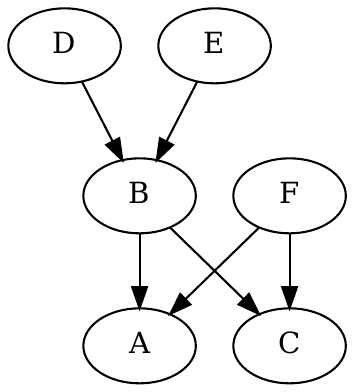}
\includegraphics[width=0.2\textwidth]{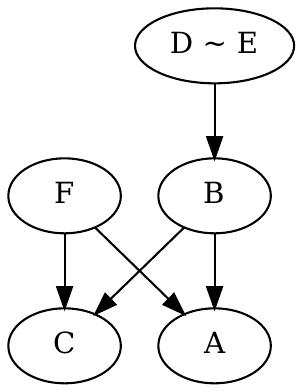}
\includegraphics[width=0.2\textwidth]{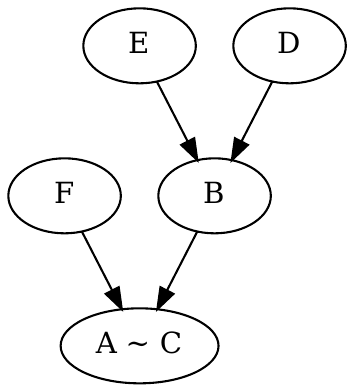}
\includegraphics[width=0.2\textwidth]{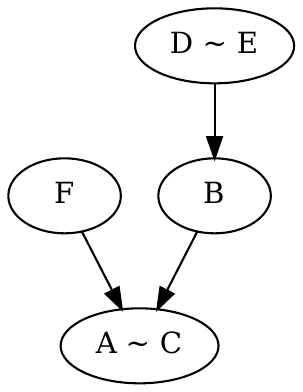}
\caption{\centering ID=3645, JHM=4.833, 4 optimal relations}
\end{figure}

\pagebreak

\subsection{Dominant Choice with Incomplete Preferences}

\begin{figure}[!htbp]
\centering
\begin{subfigure}[b]{0.3\textwidth}
\centering
\includegraphics[width=0.9\textwidth]{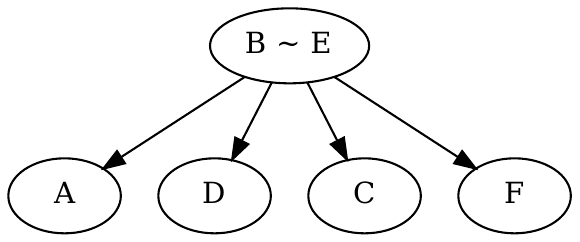}
\caption{\centering
	ID=1716, JHM=0,\\ 1 optimal relation}
	\end{subfigure}
	\begin{subfigure}[b]{0.3\textwidth}
\centering
\includegraphics[width=0.7\textwidth]{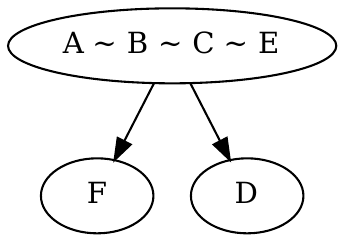}
\caption{\centering
	ID=3487, JHM=0,\\ 1 optimal relation}
	\end{subfigure}
	\begin{subfigure}[b]{0.3\textwidth}
\centering
\includegraphics[width=0.9\textwidth]{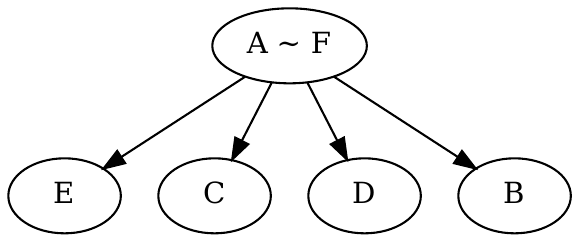}
\caption{\centering
	ID=3746, JHM=0,\\ 1 optimal relation}
	\end{subfigure}
\end{figure}

\begin{figure}[!htbp]
\centering
\begin{subfigure}[b]{0.3\textwidth}
\centering
\includegraphics[width=0.9\textwidth]{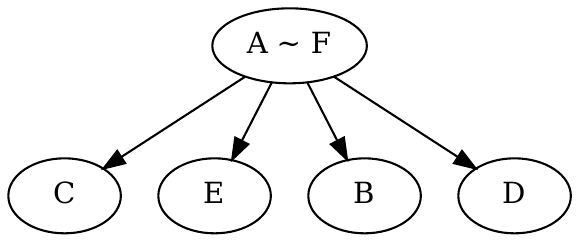}
\caption{\centering
	ID=4486, JHM=0,\\ 1 optimal relation}
	\end{subfigure}
	\begin{subfigure}[b]{0.3\textwidth}
\centering
\includegraphics[width=0.9\textwidth]{/graphs/NFC-MDC-0-5649-1.pdf}
\caption{\centering
	ID=5649, JHM=0,\\ 1 optimal relation}
	\end{subfigure}	
	\begin{subfigure}[b]{0.3\textwidth}
\centering
\includegraphics[width=0.9\textwidth]{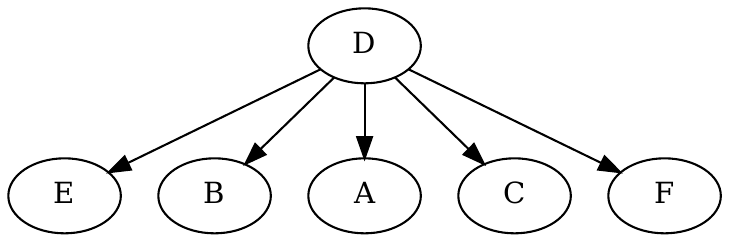}
\caption{\centering
	ID=5813, JHM=0,\\ 1 optimal relation}
	\end{subfigure}
\end{figure}

\begin{figure}[!htbp]
\centering
\begin{subfigure}[b]{0.36\textwidth}
\centering
\includegraphics[width=0.99\textwidth]{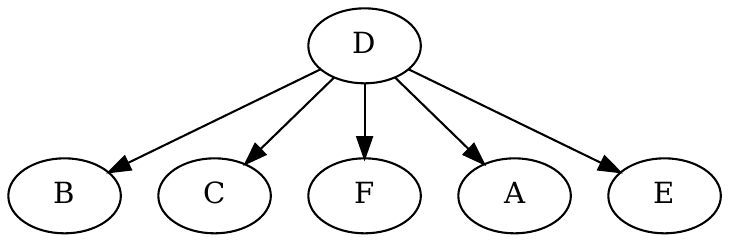}
\caption{\centering
	ID=9142, JHM=0,\\ 1 optimal relation}
	\end{subfigure}
	\begin{subfigure}[b]{0.3\textwidth}
\centering
\includegraphics[width=0.7\textwidth]{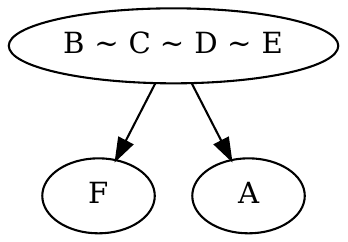}
\caption{\centering
	ID=2558, JHM=0.333,\\ 1 optimal relation}
	\end{subfigure}
	\begin{subfigure}[b]{0.3\textwidth}
\centering
\includegraphics[width=0.7\textwidth]{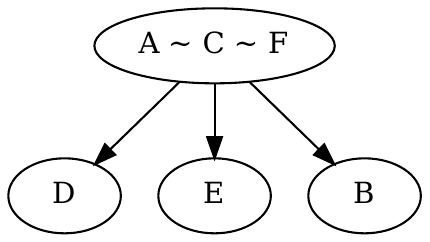}
\caption{\centering
	ID=7128, JHM=0.5,\\ 1 optimal relation}
	\end{subfigure}
\end{figure}

\begin{figure}[!htbp]
\centering
\begin{subfigure}[b]{0.3\textwidth}
\centering
\includegraphics[width=0.7\textwidth]{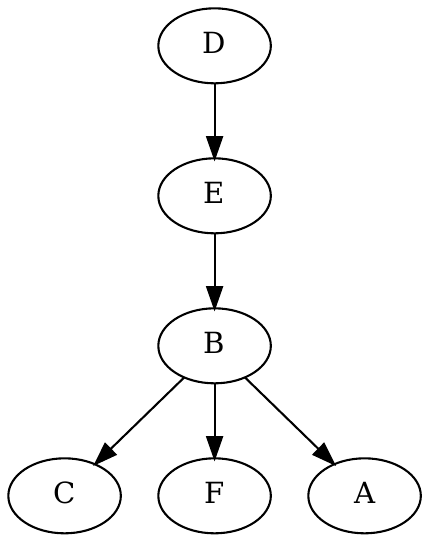}
\caption{\centering
	ID=8407, JHM=0.5,\\ 1 optimal relation}
	\end{subfigure}
	\begin{subfigure}[b]{0.3\textwidth}
\centering
\includegraphics[width=0.9\textwidth]{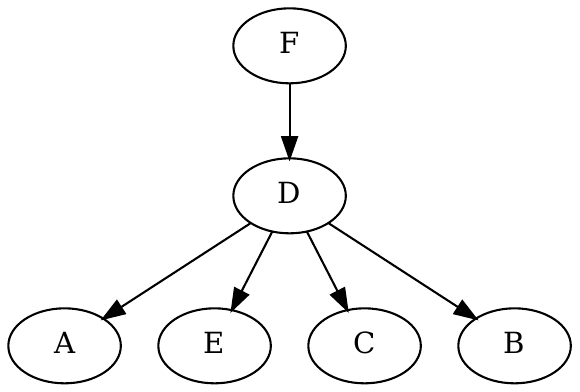}
\caption{\centering
	ID=4112, JHM=1,\\ 1 optimal relation}
	\end{subfigure}
	\begin{subfigure}[b]{0.3\textwidth}
\centering
\includegraphics[width=0.9\textwidth]{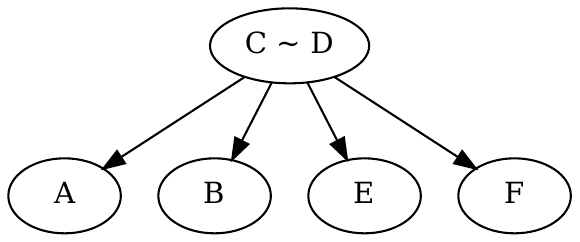}
\caption{\centering
	ID=5317, JHM=1,\\ 1 optimal relation}
	\end{subfigure}
\end{figure}

\begin{figure}[!htbp]
\centering
\begin{subfigure}[b]{0.3\textwidth}
\centering
\includegraphics[width=0.5\textwidth]{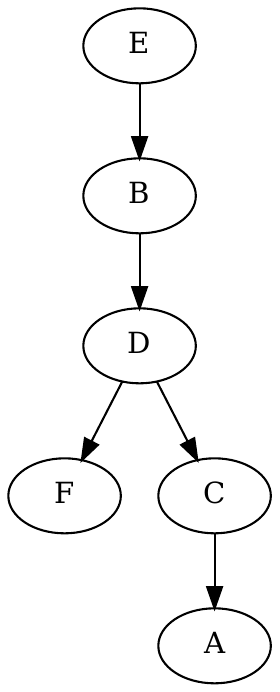}
\caption{\centering
	ID=8555, JHM=1,\\ 1 optimal relation}
	\end{subfigure}
	\begin{subfigure}[b]{0.3\textwidth}
\centering
\includegraphics[width=0.5\textwidth]{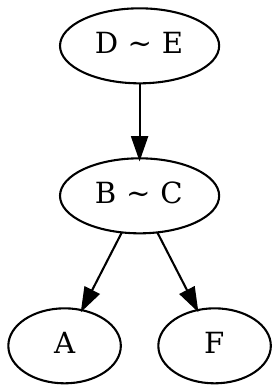}
\caption{\centering
	ID=2882, JHM=1.5,\\ 1 optimal relation}
	\end{subfigure}	
	\begin{subfigure}[b]{0.3\textwidth}
\centering
\includegraphics[width=0.7\textwidth]{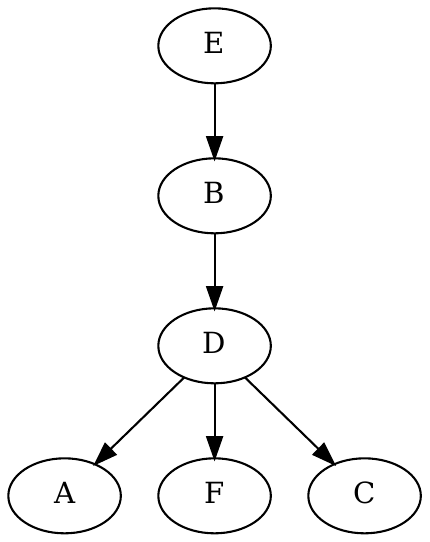}
\caption{\centering
	ID=5591, JHM=1.5,\\ 1 optimal relation}		
	\end{subfigure}
\end{figure}

\begin{figure}[!htbp]
\centering
\begin{subfigure}[b]{0.3\textwidth}
\centering
\includegraphics[width=0.7\textwidth]{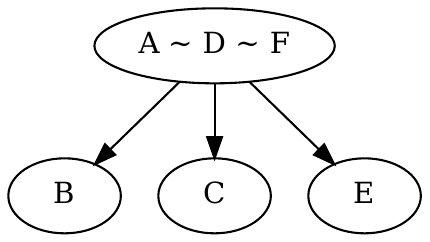}
\caption{\centering
	ID=2855, JHM=2,\\ 1 optimal relation}		
	\end{subfigure}
	\begin{subfigure}[b]{0.3\textwidth}
\centering
\includegraphics[width=0.5\textwidth]{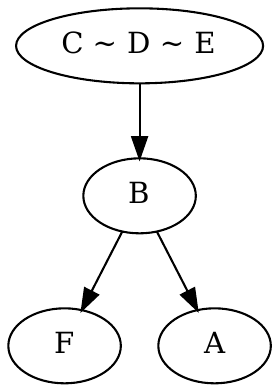}
\caption{\centering
	ID=3116, JHM=2,\\ 1 optimal relation}		
	\end{subfigure}
	\begin{subfigure}[b]{0.3\textwidth}
\centering
\includegraphics[width=0.7\textwidth]{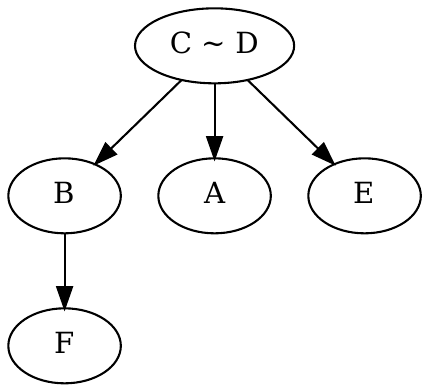}
\caption{\centering
	ID=7286, JHM=2,\\ 1 optimal relation}
	\end{subfigure}
\end{figure}

\begin{figure}[!htbp]
\centering
\begin{subfigure}[b]{0.3\textwidth}
\centering
\includegraphics[width=0.9\textwidth]{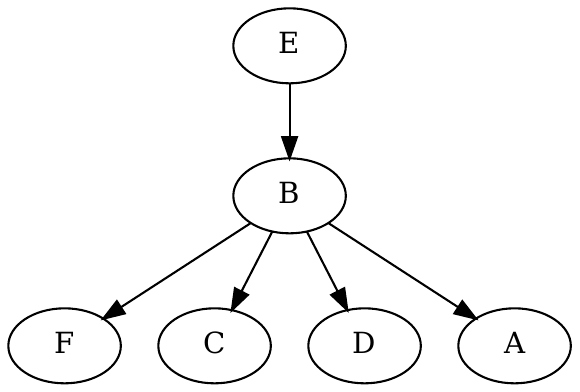}
\caption{\centering
	ID=7621, JHM=2,\\ 1 optimal relation}
	\end{subfigure}
	\begin{subfigure}[b]{0.3\textwidth}
\centering
\includegraphics[width=0.75\textwidth]{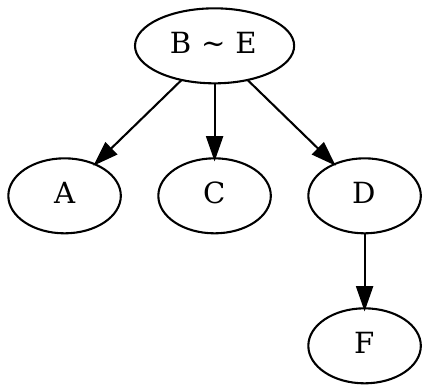}
\caption{\centering
	ID=9072, JHM=2,\\ 1 optimal relation}
	\end{subfigure}
	\begin{subfigure}[b]{0.3\textwidth}
\centering
\includegraphics[width=0.75\textwidth]{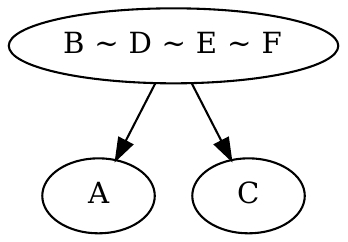}
\caption{\centering
	ID=9650, JHM=2.083,\\ 1 optimal relation}
	\end{subfigure}
\end{figure}

\begin{figure}[!htbp]
\centering
\begin{subfigure}[b]{0.3\textwidth}
\centering
\includegraphics[width=0.5\textwidth]{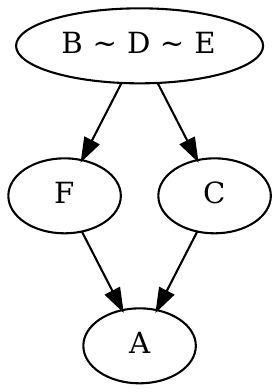}
\caption{\centering
	ID=1124, JHM=2.167,\\ 1 optimal relation}
	\end{subfigure}
	\begin{subfigure}[b]{0.3\textwidth}
\centering
\includegraphics[width=0.55\textwidth]{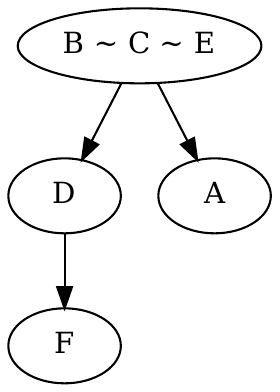}
\caption{\centering
	ID=5064, JHM=2.333,\\ 1 optimal relation}
	\end{subfigure}
	\begin{subfigure}[b]{0.3\textwidth}
\centering
\includegraphics[width=0.7\textwidth]{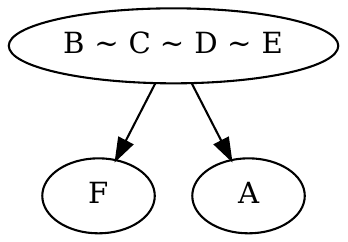}
\caption{\centering
	ID=5860, JHM=2.333,\\ 1 optimal relation}
	\end{subfigure}
\end{figure}

\begin{figure}[!htbp]
\centering
\begin{subfigure}[b]{0.3\textwidth}
\centering
\includegraphics[width=0.55\textwidth]{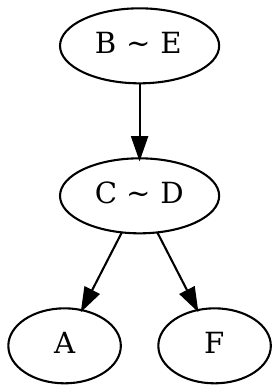}
\caption{\centering
	ID=7481, JHM=2.333,\\ 1 optimal relation}
	\end{subfigure}
	\begin{subfigure}[b]{0.3\textwidth}
\centering
\includegraphics[width=0.75\textwidth]{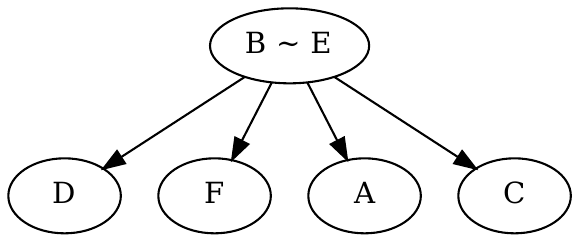}
\caption{\centering
	ID=4155, JHM=2.5,\\ 1 optimal relation}
	\end{subfigure}
	\begin{subfigure}[b]{0.3\textwidth}
\centering
\includegraphics[width=0.5\textwidth]{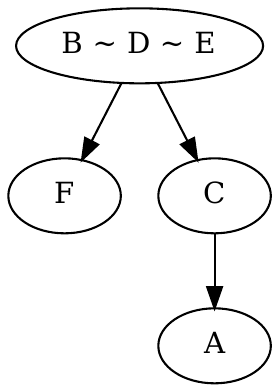}
\caption{\centering
	ID=8801, JHM=2.5,\\ 1 optimal relation}
	\end{subfigure}
	\begin{subfigure}[b]{0.35\textwidth}
\centering
\includegraphics[width=0.95\textwidth]{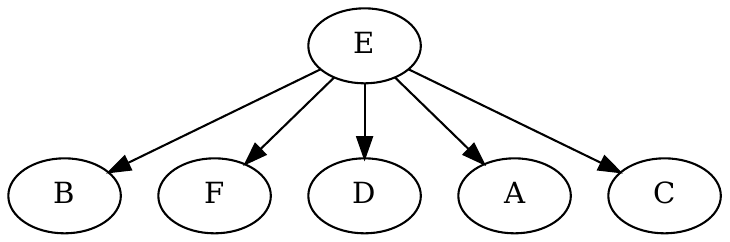}
\caption{\centering
	ID=5131, JHM=3,\\ 1 optimal relation}
	\end{subfigure}
	\begin{subfigure}[b]{0.3\textwidth}
\centering
\includegraphics[width=0.65\textwidth]{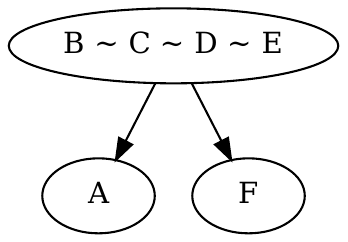}
\caption{\centering
	ID=5440, JHM=3,\\ 1 optimal relation}
	\end{subfigure}
\end{figure}

\begin{figure}[!htbp]
\centering
\begin{subfigure}[b]{0.3\textwidth}
\centering
\includegraphics[width=0.95\textwidth]{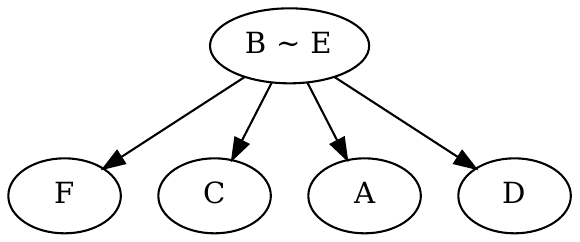}
\caption{\centering
	ID=6055, JHM=3,\\ 1 optimal relation}
	\end{subfigure}
	\begin{subfigure}[b]{0.3\textwidth}
\centering
\includegraphics[width=0.75\textwidth]{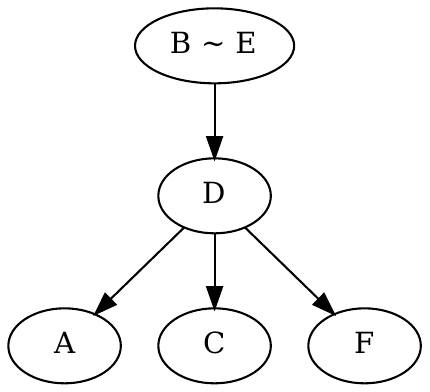}
\caption{\centering
	ID=7869, JHM=3.5,\\ 1 optimal relation}
	\end{subfigure}
	\begin{subfigure}[b]{0.3\textwidth}
\centering
\includegraphics[width=0.75\textwidth]{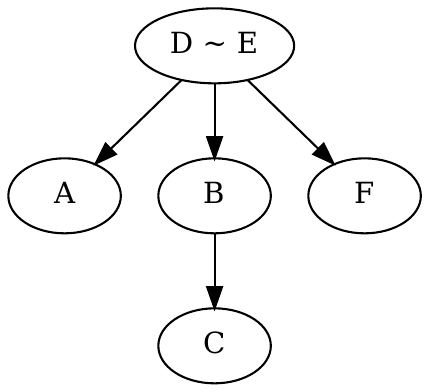}
\caption{\centering
	ID=8133, JHM=3.5,\\ 1 optimal relation}
	\end{subfigure}
\end{figure}

\begin{figure}[!htbp]
\centering
\begin{subfigure}[b]{0.6\textwidth}
\centering
\includegraphics[width=0.3\textwidth]{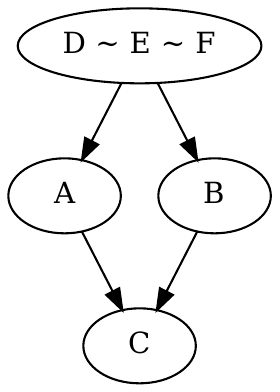}
\includegraphics[width=0.3\textwidth]{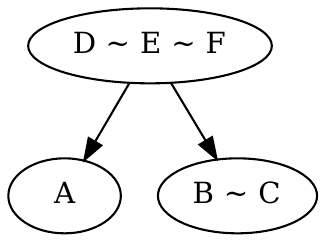}
\includegraphics[width=0.3\textwidth]{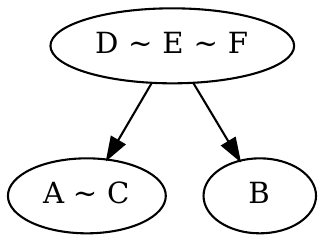}
\caption{\centering ID=8781, JHM=3.667, 3 optimal relations}
\end{subfigure}
\begin{subfigure}[b]{0.37\textwidth}
\centering
\includegraphics[width=0.7\textwidth]{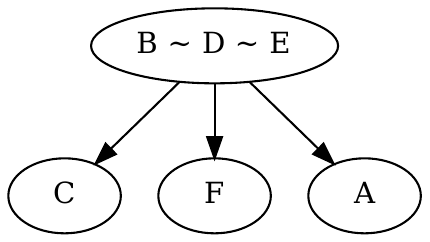}
\caption{\centering
	ID=9305, JHM=4,\\ 1 optimal relation}
	\end{subfigure}	
\end{figure}

\begin{figure}[!htbp]
\centering
\begin{subfigure}[b]{0.3\textwidth}
\centering
\includegraphics[width=0.7\textwidth]{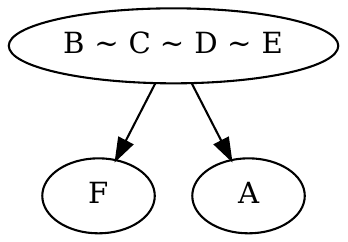}
\caption{\centering
	ID=5580, JHM=4.5,\\ 1 optimal relation}
	\end{subfigure}
	\begin{subfigure}[b]{0.3\textwidth}
\centering
\includegraphics[width=0.5\textwidth]{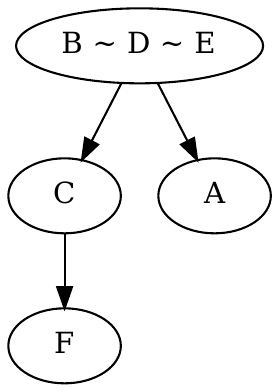}
\caption{\centering
	ID=7071, JHM=4.833,\\ 1 optimal relation}
	\end{subfigure}
	\begin{subfigure}[b]{0.3\textwidth}
\centering
\includegraphics[width=0.99\textwidth]{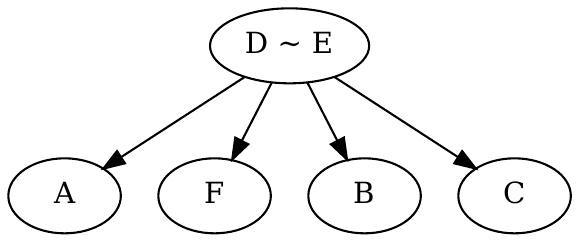}
\caption{\centering
	ID=7577, JHM=5,\\ 1 optimal relation}
	\end{subfigure}
\end{figure}

\pagebreak

\bibliographystyle{ecta}
\bibliography{MultiValued}

\end{document}